\documentclass[11pt,letterpaper]{article}
\usepackage{amsmath,amssymb,amsthm}
\usepackage[numbers]{natbib}
\usepackage{graphicx,booktabs}
\usepackage{tikz}
\usetikzlibrary{calc,arrows.meta}
\definecolor{cA}{HTML}{2B6CD6}
\definecolor{cB}{HTML}{E8692E}
\usepackage{algorithm}
\usepackage{algpseudocode}
\usepackage[hypertexnames=false]{hyperref}
\hypersetup{colorlinks=true,linkcolor=blue,citecolor=blue,urlcolor=blue}

\newtheorem{theorem}{Theorem}[section]
\newtheorem{proposition}[theorem]{Proposition}
\newtheorem{lemma}[theorem]{Lemma}
\newtheorem{corollary}[theorem]{Corollary}
\newtheorem{remark}[theorem]{Remark}

\numberwithin{equation}{section}

\newcommand{\R}{\mathbb{R}}
\newcommand{\E}{\mathbb{E}}
\newcommand{\M}{\mathsf{M}}
\newcommand{\T}{\top}
\newcommand{\KL}{\mathrm{KL}}

\makeatletter
\newcommand{\tagsleft}{\tagsleft@true}
\makeatother
\begin{document}

\begin{center}
{\Large \textbf{Multilevel Filtering on Manifolds with Light Schr\"odinger Bridges}}

\vspace{0.5cm}

JASON BEH$^{1}$, PIERRE DEL MORAL$^{2}$ \& AJAY JASRA$^{1}$

{\footnotesize $^{1}$Computing \& Mathematical Sciences Division, Mohamed Bin Zayed University of Artificial Intelligence, Abu Dhabi, UAE.}\\
{\footnotesize $^{2}$Centre de Recherche Inria Bordeaux Sud-Ouest, Talence, FR.}\\
{\footnotesize E-Mail:\,}\texttt{\emph{\footnotesize {Jason.KaiXun,Ajay.Jasra}@mbzuai.ac.ae, pierre.del\_moral@inria.fr}}

\vspace{0.5cm}

\textbf{Abstract}

\vspace{0.2cm}
\begin{minipage}{0.9\textwidth}
In this article we consider the filtering of partially observed diffusions when the observations, made at discrete times, are noiseless. Each observation then confines the hidden state to a manifold defined by the data, whose type, for instance a sphere or an affine subspace, may change from one observation time to the next. In practice, the filter is approximated by particle filters, which require the diffusion to be discretized in time. Multilevel Monte Carlo reduces the cost of this discretization only if the particle approximations at consecutive discretization levels stay close, which standard particle filters do not ensure. We develop a new particle filter based on light Schr\"odinger bridges, which works between changing manifolds and can be coupled across discretization levels. A key insight is that the previous state enters the light Schr\"odinger bridge only through a few features, so that the bridge only has to be fitted on the new manifold, regardless of the type of the previous one. We bound the bias of this particle filter and the gap between its approximations at consecutive levels, and show numerically, on spheres and on manifolds that change type, that this gap is much smaller than with standard multilevel particle filters, and that, with the cost of the fits included, a hybrid estimator that uses it at the fine levels reduces the cost at small tolerances.

\vspace{0.2cm}
\end{minipage}
\end{center}

\section{Introduction}\label{sec:related}

Filtering is the problem of estimating the current state of a system that evolves randomly over time from a stream of partial measurements, for instance the position of a moving target from sensor readings. At each observation time, the filter is the conditional law of the current state given the measurements so far; it is needed in tracking, navigation, the geosciences and econometrics \citep{bain2009fundamentals,cappe2005inference,delmoral2004feynman}. Except in special cases, such as linear Gaussian models, it has no closed form. Particle filters are a class of numerical methods that approximate the filter by a cloud of weighted samples, called particles, which are moved with the dynamics, reweighted at each new measurement and then resampled \citep{gordon1993novel,delmoral2004feynman,delmoral2013mean}. When the state follows a diffusion, in many cases of practical interest, its motion between two observations must also be approximated, by a time discretization.

This section first describes the model: the filtering problem with noiseless observations, and the manifolds on which the filter then lives. It then turns to the numerical approximation of the filter: the time discretization of the dynamics, particle filters and why noiseless observations make them difficult, and multilevel Monte Carlo. It then states the two properties that we want from a new particle filter, reviews related work, and ends with our contributions. The reader is referred to Appendix~\ref{app:notation} for the notation used throughout the paper.

\subsection{Filtering with noiseless observations}\label{sec:setting-filter}

We consider a partially observed diffusion, that is a state-space model whose latent signal is a diffusion. The signal solves the stochastic differential equation $dX_t=\mu(X_t)\,dt+\sigma\,dW_t$ on $\R^d$, where $W$ is a $d$-dimensional Brownian motion, the drift $\mu:\R^d\to\R^d$ gives the deterministic part of the motion, the velocity of the state in the absence of noise, and the diffusion coefficient $\sigma>0$, a constant, scales the random part. The signal is started from a known law $\pi_0$, and it is observed without noise at the integer times through
\[
  Y_n = H_n(X_n),\qquad n\ge1 ,
\]
for known maps $H_n:\R^d\to\R^{d_y}$. Once $y_{1:n}$ is recorded, $X_n$ must belong to $\M_n:=\{x\in\R^d: H_n(x)=y_n\}$. Under appropriate mathematical assumptions,
to be discussed later, the filter $\pi_n$, the conditional law of $X_n$ given $Y_{1:n}=y_{1:n}$, is supported by $\M_n$. The filter is $\pi_n(\varphi):=\int\varphi(x)\,\pi_n(dx)$ for suitable test functions $\varphi$ and again, we shall define
when this probability measure is well-defined in some sense.

Write $f_n(x\,|\,x')$ for the transition density of $X$ over one observation interval. That is, the density of the state $X_n$ at time $n$, given that the state was $X_{n-1}=x'$ at time $n-1$. The density exists under standard assumptions on
$\mu$ such as being Lipschitz and satisfying a growth condition.
Since the observations are noiseless, the filter is obtained, for the observation maps of this paper (Appendix~\ref{app:coarea}), by restricting the predicted law to $\M_n$:
\[
  \pi_n(dx) \propto \int f_n(x\,|\,x')\,\pi_{n-1}(dx')\,\lambda_{\M_n}(dx),
\]
with $\lambda_{\M_n}$ the natural measure on $\M_n$, for instance Lebesgue measure on an affine subspace and the surface measure on a sphere. Note that the types of manifolds
that we consider are not `simple' in that there are only a small number of countable solutions $y_n=H_n(x)$.

\subsection{Changing manifolds}\label{sec:setting-manifolds}

\begin{figure}[!t]
\centering
\begin{tikzpicture}[scale=0.62,
  tA/.style={draw=cA,line width=0.9pt},
  tB/.style={draw=cB,line width=0.9pt},
  sens/.style={fill=black,inner sep=1.2pt,circle},
  targ/.style={inner sep=1.6pt,circle}]
\def\panel#1#2#3{%
  \begin{scope}[xshift=#1cm]
    \draw[gray!60,line width=0.4pt] (-2.1,-2.1) rectangle (2.1,2.1);
    \begin{scope}\clip (-2.1,-2.1) rectangle (2.1,2.1);
    #2
    \end{scope}
    \node[below,font=\small,align=center] at (0,-2.2) {#3};
  \end{scope}}
\panel{0}{%
  \node[sens,label={[font=\scriptsize]left:$o_1$}] (o1) at (-1.1,0.6) {};
  \node[sens,label={[font=\scriptsize]right:$o_2$}] (o2) at (1.2,-0.7) {};
  \draw[tA] (o1) circle (0.85);
  \draw[tB] (o2) circle (0.7);
  \node[targ,fill=cA] at ($(o1)+(40:0.85)$) {};
  \node[targ,fill=cB] at ($(o2)+(160:0.7)$) {};
}{(a) two ranges\\torus $\mathbb S^1\times\mathbb S^1$}
\panel{5.4}{%
  \node[sens,label={[font=\scriptsize]left:$o_1$}] (o1) at (-1.1,0.6) {};
  \draw[tA] (o1) circle (0.85);
  \node[targ,fill=cA] at ($(o1)+(40:0.85)$) {};
  \node[sens,label={[font=\scriptsize]left:$p_2$}] (p2) at (1.5,-1.6) {};
  \draw[tB] ($(p2)+(-80:0.6)$) -- ($(p2)+(100:4.2)$);
  \node[targ,fill=cB] at ($(p2)+(100:2.3)$) {};
}{(b) range and line\\cylinder $\mathbb S^1\times\mathbb R$}
\panel{10.8}{%
  \node[sens,label={[font=\scriptsize]right:$p_1$}] (p1) at (-1.5,-1.6) {};
  \draw[tA] ($(p1)+(-100:0.6)$) -- ($(p1)+(80:4.2)$);
  \node[targ,fill=cA] at ($(p1)+(80:2.8)$) {};
  \node[sens,label={[font=\scriptsize]left:$p_2$}] (p2) at (1.5,-1.6) {};
  \draw[tB] ($(p2)+(-80:0.6)$) -- ($(p2)+(100:4.2)$);
  \node[targ,fill=cB] at ($(p2)+(100:2.3)$) {};
}{(c) two lines\\plane $\mathbb R\times\mathbb R$}
\panel{16.2}{%
  \node[sens,label={[font=\scriptsize]left:$o_1$}] (o1) at (-1.1,0.6) {};
  \node[sens,label={[font=\scriptsize]right:$o_2$}] (o2) at (1.2,-0.7) {};
  \node[targ,fill=cA] (x1) at (-0.3,1.3) {};
  \node[targ,fill=cB] (x2) at (0.5,-0.2) {};
  \draw[cA,densely dashed] (o1) -- (x1);
  \draw[cB,densely dashed] (o2) -- (x2);
}{(d) one range for both\\three-sphere $\mathbb S^3$}
\end{tikzpicture}
\caption{Manifolds for two targets in the plane, $X_n=(X_{n,(1)},X_{n,(2)})\in\R^4$. Blue: first target; orange: second target; black: sensors at known positions. In (a)--(c), each target is confined to a circle or a line, and the state lies on their product. In (d), a single range for the pair, $|x_{(1)}-o_1|^2+|x_{(2)}-o_2|^2=y_n$, confines the state to a three-sphere in $\R^4$, which cannot be drawn in the plane; the dashed segments show the two distances it combines.}\label{fig:sets}
\end{figure}
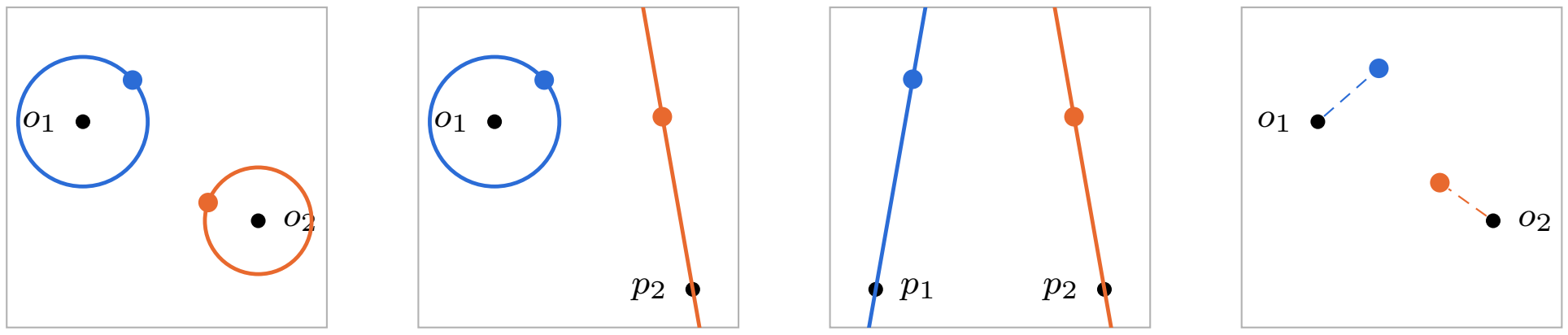

It is the hidden state $X_n$, not the observation, that lives on a manifold. The observation $Y_n=H_n(X_n)\in\R^{d_y}$ is a measured value, a fixed vector once it is recorded as $y_n$. The set $\M_n=\{x\in\R^d:H_n(x)=y_n\}$ is the level set of the observation map at that value, a subset of the state space. Since the observation is noiseless, the state must satisfy $H_n(X_n)=y_n$, that is $X_n\in\M_n$, and the filter $\pi_n$ is supported by $\M_n$. The data $y_n$ thus determine which manifold the state is confined to, and the map $H_n$ determines its shape.

Consider a target moving in the plane, $X_n\in\R^2$. A range sensor at a known position $o$ measures $H_n(x)=|x-o|^2$, so that $\M_n$ is the circle of center $o$ and radius $y_n^{1/2}$. A linear sensor measures $H_n(x)=\langle e,x\rangle$ for a known unit vector $e$, so that $\M_n$ is the line $\{x:\langle e,x\rangle=y_n\}$. In higher dimension the same observations give spheres and affine subspaces, such as hyperplanes, which move with the data.

With several targets, the sets combine. For two targets in the plane (Figure~\ref{fig:sets}), the state $X_n\in\R^4$ lies on a product: two ranges give a torus, a circle times a circle; a range and a linear sensor give a cylinder, a circle times a line; and two linear sensors give a plane. When the two targets are too close to be resolved, a single range for the pair is observed, and the state lies on a three-sphere in $\R^4$. If the sensors change with time, because targets move in and out of coverage or are resolved or not, the manifold $\M_n$ changes in dimension and in type from one observation to the next. Our examples are of this kind: affine subspaces, spheres, and two targets whose manifold switches between a torus, a cylinder, a plane and a three-sphere.

\subsection{Time discretization}\label{sec:setting-disc}

A filter needs the transition density $f_n$, or at least samples from it, to predict where the state goes between two observations. It is rarely available in closed form, since it requires solving the stochastic differential equation over the whole interval, and exact samples of $X_n$ given $X_{n-1}$ are rarely available either. The standard remedy is to cut the interval into small time steps and to approximate the motion over each step, where the drift is nearly constant and the noise nearly Gaussian. This is a time discretization. It introduces an error, which decreases as the step decreases, while the computing time increases with the number of steps. We fix a finest level $L\ge1$ and use the time steps $\Delta_l:=2^{-l}$, indexed by the level $l\in\{0,\dots,L\}$. Level $0$ takes the whole observation interval in one step, and each level halves the step of the previous one, so that larger levels are finer and more accurate, but take more time; level $L$ has the smallest time step $\Delta_L$. For the Euler scheme, one observation interval contains $\Delta_l^{-1}$ steps,
\[
  X^{[l]}_{k+1} = X^{[l]}_k+\mu\big(X^{[l]}_k\big)\Delta_l+\sigma\,\Delta W^{[l]}_k,\qquad \Delta W^{[l]}_k\sim\mathcal N(0,\Delta_l I),
\]
with independent Brownian increments $\Delta W^{[l]}_k$
and $\mathcal N(0,\Delta_l I)$ the $d$-dimensional Gaussian
distribution of mean 0 and covariance $\Delta_l I$. Let $f_n^{[l]}(x\,|\,x')$ be the density of the level-$l$ transition over one interval. The level-$l$ filter is obtained in the same way, with $f_n^{[l]}$ in place of $f_n$:
\[
  \pi_n^{[l]}(dx) \propto \int f_n^{[l]}(x\,|\,x')\,\pi_{n-1}^{[l]}(dx')\,\lambda_{\M_n}(dx),\qquad \pi_0^{[l]}:=\pi_0,
\]
and (under assumptions) $\pi_n^{[l]}(\varphi)\to\pi_n(\varphi)$ as the time step $\Delta_l$ decreases to $0$.

\begin{figure}[t]
\centering
\begin{tikzpicture}[scale=0.8,
  man/.style={draw=black!70,line width=0.8pt},
  pt/.style={circle,fill=black!75,inner sep=0pt,minimum size=4pt},
  ar/.style={-{Stealth[length=4pt]},line width=0.5pt}]
\def\frame#1{%
  \draw[gray!60,line width=0.4pt] (-1.5,-2.0) rectangle (4.0,2.0);
  \draw[man] (0,0) circle (1.0);
  \draw[man] (3.0,-1.8) -- (3.0,1.8);
  \node[font=\scriptsize] at (-0.9,-1.65) {$\M_{n-1}$};
  \node[font=\scriptsize] at (3.45,-1.65) {$\M_n$};
  \node[below,font=\small,align=center] at (1.25,-2.1) {#1};}
\def\frameN#1{%
  \draw[gray!60,line width=0.4pt] (-1.5,-2.0) rectangle (4.0,2.0);
  \fill[cA!12] (2.55,-1.95) rectangle (3.45,1.95);
  \draw[cA!60,densely dotted,line width=0.6pt] (3.0,-1.95) -- (3.0,1.95);
  \node[font=\scriptsize,cA] at (3.0,-1.7) {$y_n$};
  \node[below,font=\small,align=center] at (1.25,-2.1) {#1};}
\begin{scope}[yshift=5.6cm]
\frameN{1. move with\\the dynamics}
\foreach \x/\y/\u/\v in {-0.6/1.2/2.45/1.35,-0.9/0.5/3.6/0.7,-0.3/0.1/2.2/0.25,-1.0/-0.4/3.15/-0.3,-0.5/-0.9/2.85/-0.85,-0.8/-1.4/3.75/-1.3}{
  \node[pt] (s) at (\x,\y) {};
  \draw[ar,gray,densely dashed] (s) -- (\u-0.08,\v);
  \fill[black!60] (\u,\v) circle (2pt);}
\end{scope}
\begin{scope}[xshift=6.2cm,yshift=5.6cm]
\frameN{2. weight by\\the likelihood}
\foreach \u/\v/\r in {2.45/1.35/2.0,3.6/0.7/1.4,2.2/0.25/1.0,3.15/-0.3/5.6,2.85/-0.85/4.6,3.75/-1.3/0.9}{
  \fill[cA] (\u,\v) circle (\r pt);}
\end{scope}
\begin{scope}[xshift=12.4cm,yshift=5.6cm]
\frameN{3. resample:\\equal weights, copies}
\foreach \u/\v/\c in {3.15/-0.3/3,2.85/-0.85/2,2.45/1.35/1}{
  \fill[cA] (\u,\v) circle (2.2pt);
  \node[font=\scriptsize,cA,anchor=east] at (\u-0.1,\v) {$\times\c$};}
\foreach \u/\v in {3.6/0.7,2.2/0.25,3.75/-1.3}{
  \draw[gray!70] (\u,\v) circle (2.2pt);}
\end{scope}
\begin{scope}
\frame{(a) moved by the dynamics:\\every weight is zero\\with probability one}
\foreach \a/\x/\y in {25/2.45/1.3,75/3.55/0.85,125/2.3/0.3,190/3.45/-0.05,250/2.6/-0.7,310/3.6/-1.15}{
  \node[pt] (s) at (\a:1.0) {};
  \draw[ar,gray,densely dashed] (s) -- (\x,\y);
  \draw[cB,line width=0.7pt] (\x-0.09,\y-0.09) -- (\x+0.09,\y+0.09) (\x-0.09,\y+0.09) -- (\x+0.09,\y-0.09);}
\end{scope}
\begin{scope}[xshift=6.2cm]
\frame{(b) drawn on $\M_n$:\\unequal weights}
\foreach \a/\y/\r in {25/1.25/2.2,75/0.75/5.0,125/0.35/1.6,190/-0.1/6.4,250/-0.65/2.8,310/-1.2/1.2}{
  \node[pt] (s) at (\a:1.0) {};
  \draw[ar,gray,densely dashed] (s) -- (2.88,\y);
  \fill[cA] (3.0,\y) circle (\r pt);}
\end{scope}
\begin{scope}[xshift=12.4cm]
\frame{(c) resampled:\\equal weights, copies}
\foreach \y/\c in {0.75/2,-0.1/3,-0.65/1}{
  \fill[cA] (3.0,\y) circle (2.2pt);
  \node[font=\scriptsize,cA,anchor=west] at (3.12,\y) {$\times\c$};}
\foreach \y in {1.25,0.35,-1.2}{
  \draw[gray!70] (3.0,\y) circle (2.2pt);}
\node[font=\scriptsize,gray,anchor=west,align=left] at (-1.4,1.55) {light weights dropped,\\heavy ones copied};
\end{scope}
\end{tikzpicture}
\caption{One step of a particle filter. Top: with a noisy observation $y_n$, whose likelihood is shown by the shaded band, the three steps work: the particles are moved, weighted (dot size) and resampled. Bottom: with a noiseless observation, from $\M_{n-1}$ (here a circle) to $\M_n$ (here a line). (a) Particles moved by the dynamics land off $\M_n$ with probability one, so that every weight vanishes. (b) Drawing each particle on $\M_n$ from the dynamics restricted to $\M_n$ gives unequal weights (dot size). (c) Resampling discards the particles with small weights and copies those with large weights.}\label{fig:pf}
\end{figure}

\subsection{Particle filters}\label{sec:setting-pf}

In the case of noisy observations, 
a particle filter \citep{cappe2005inference,delmoral2004feynman,doucet2000sequential,gordon1993novel} approximates $\pi_n$ (or $\pi_n^{[l]}$) by the empirical measure $N^{-1}\sum_{i=1}^N\delta_{X_n^i}$ of $N$ particles $X_n^1,\dots,X_n^N$. Given particles $X_{n-1}^1,\dots,X_{n-1}^N$ that approximate $\pi_{n-1}$, one step of the filter has three parts (Figure~\ref{fig:pf}, top).
\begin{enumerate}
  \item[1.] Move. Each particle is moved with the dynamics, $\hat X_n^i\sim f_n(\cdot\,|\,X_{n-1}^i)$. The moved particles approximate the predicted law $\int f_n(\cdot\,|\,x')\,\pi_{n-1}(dx')$, the law of $X_n$ before $y_n$ is observed.
  \item[2.] Weight. Each moved particle receives a weight $W_n^i$ proportional to the likelihood of the observation $y_n$ at $\hat X_n^i$, with $\sum_iW_n^i=1$. With noisy observations, $Y_n=H_n(X_n)+\xi_n$, the likelihood is the density of the noise $\xi_n$ at $y_n-H_n(\hat X_n^i)$, so that the particles that explain the data well receive large weights. The weighted particles approximate $\pi_n$.
  \item[3.] Resample. $N$ new particles $X_n^1,\dots,X_n^N$ are drawn among the moved ones, each $\hat X_n^j$ being chosen with probability $W_n^j$. The particles with small weights tend to disappear and those with large weights are copied, so that the new particles again carry equal weights and approximate $\pi_n$.
\end{enumerate}
Without resampling, the weights would degenerate over time, a single particle carrying almost all the weight after a few steps \citep{doucet2000sequential}. At level $l$, the particles are moved with $f_n^{[l]}$ in place of $f_n$, and they approximate $\pi_n^{[l]}$ \citep{jasra2017multilevel}.

With noiseless observations, the likelihood in step~2 is the indicator that $H_n(\hat X_n^i)=y_n$, and the weighting step fails; this is what makes the problem hard. A moved particle lands exactly on $\M_n$ with probability zero, so that, with probability one, every weight vanishes (Figure~\ref{fig:pf}, bottom, (a)). One can instead draw the new particle directly on $\M_n$, from the dynamics restricted to $\M_n$, which is the optimal proposal \citep{doucet2000sequential} written on $\M_n$ \citep{zhumekenov2026}, but the particles then carry unequal weights and must be resampled (Figure~\ref{fig:pf}, bottom, (b)--(c)). \citet{zhumekenov2026} develop a particle
filter on linear manifolds.

\subsection{Multilevel Monte Carlo}\label{sec:setting-mlmc}

The time discretization forces a choice between accuracy and cost: a fine time step gives a small error but a long computation. Multilevel Monte Carlo \citep{beskos2017multilevel,giles2008multilevel,giles2015multilevel} turns the hierarchy of levels into a saving. It rests on the identity
\begin{equation}\label{eq:mlmc}
  \pi_n^{[L]}(\varphi) = \pi_n^{[0]}(\varphi) + \sum_{l=1}^L\big\{\pi_n^{[l]}(\varphi)-\pi_n^{[l-1]}(\varphi)\big\},
\end{equation}
which writes the filter at the finest level as the filter at the coarsest level plus a sum of corrections between consecutive levels. Each correction is estimated by running two particle filters together, one at step $\Delta_l$ and one at step $\Delta_{l-1}$, driven by the same random variables; we call this a coupling of the two levels. For the Euler scheme, the synchronous coupling \citep{giles2008multilevel,jasra2017multilevel} drives both levels by the same Brownian path, the coarse level using the sums of pairs of fine Brownian increments,
\begin{equation}\label{eq:brownian-agg}
  \Delta W^{[l-1]}_k = \Delta W^{[l]}_{2k}+\Delta W^{[l]}_{2k+1}.
\end{equation}
When the coupling is suitably effective, the two particle approximations almost cancel, so that their difference has a small variance, and few samples of it are needed at the fine levels, which are the expensive ones. \citet{jasra2017multilevel} show how the resampling of particle filters can be coupled in a multilevel context.

\paragraph{The three rates.} The computing time needed to reach a given accuracy depends on how three quantities change as the time step $\Delta_l$ decreases.
\begin{enumerate}
  \item[1.] The variance $V_l$ of one sample of the level-$l$ difference. It measures how closely the two levels are coupled, and it decreases as the levels get finer.
  \item[2.] The computing time $C_l$ of one such sample. It increases, since finer levels take more time steps.
  \item[3.] The bias of the finest level, $b_L:=\pi_n^{[L]}(\varphi)-\pi_n(\varphi)$, the error of the time discretization that the estimator cannot remove. It decreases as $L$ increases.
\end{enumerate}
The complexity theorem of multilevel Monte Carlo, \citep[Thm.~3.1]{giles2008multilevel} for $\gamma=1$ and \citep[Thm.~1]{giles2015multilevel} in general, assumes that there exist rates $\alpha,\beta,\gamma>0$, with $\alpha\ge\frac12\min(\beta,\gamma)$, and constants such that
\[
  V_l\le c_1\,\Delta_l^{\beta},\qquad C_l\le c_2\,\Delta_l^{-\gamma},\qquad |b_L|\le c_3\,\Delta_L^{\alpha}.
\]
For the Euler scheme, $\gamma=1$, since halving the step doubles the number of steps, and $\alpha=1$ \citep[\S2]{giles2015multilevel}. The rate $\beta$, on the other hand, is not given by the scheme: it depends on how the two levels are coupled, that is, on the coupling of the particles
through sampling and resampling.

\paragraph{The computing time.} To reach a root-mean-square error $\epsilon$, the finest level $L$ is chosen so that $|b_L|$ is of order $\epsilon$, and the numbers of samples $N_l$ are chosen to make the variance of the estimator of order $\epsilon^2$ at the smallest total computing time. This gives
\[
  N_l\propto\epsilon^{-2}\left(\frac{V_l}{C_l}\right)^{1/2}\sum_{k\le L}(V_kC_k)^{1/2},
\]
and a total computing time
\[
  \sum_{l\le L}N_lC_l \propto \epsilon^{-2}\left(\sum_{l\le L}(V_lC_l)^{1/2}\right)^2 =
  \begin{cases}
    O(\epsilon^{-2}), & \beta>\gamma,\\
    O(\epsilon^{-2}|\log\epsilon|^2), & \beta=\gamma,\\
    O(\epsilon^{-2-(\gamma-\beta)/\alpha}), & \beta<\gamma.
  \end{cases}
\]
The terms of the sum are of order $2^{l(\gamma-\beta)/2}$: they decrease when the variance of the differences decreases faster than their computing time increases. When $\beta>\gamma$, the coarse levels dominate, and the computing time is that of a Monte Carlo estimate with exact samples of $\pi_n$. The critical case $\beta=\gamma$ is the order obtained for a multilevel filter based on entropic optimal transport \citep[Thm.~4.17]{zhao2026sequential}. When $\beta<\gamma$, the fine levels dominate, but the computing time is still smaller than the time $\epsilon^{-3}$ needed with the finest level alone for the Euler scheme.

\paragraph{What this asks of a particle filter.} A larger $\beta$ is better: the variance of the differences then vanishes faster as the levels get finer, so fewer samples are needed at the expensive fine levels. Since $\gamma=1$ here, the goal is $\beta>1$, which gives a computing time of order $\epsilon^{-2}$; this is optimal for any Monte Carlo method. 
If $\sigma$ is constant, two Euler schemes driven by the same Brownian path stay within $O(\Delta_l)$ of each other, which gives $\beta=2$ for the difference of two coupled particles. A standard particle filter does not keep this property due to the resampling operation.
Even with the same random numbers, resampling, which copies some particles and drops others (Figure~\ref{fig:pf}), occasionally gives a particle and its twin, the particle at the other level driven by the same random numbers, different ancestors, which then differ by a quantity of order one. These rare jumps dominate the variance of the increments and limit $\beta$ to about $1$, the critical case, in which the computing time is of order $\epsilon^{-2}|\log\epsilon|^2$. Section~\ref{sec:theory-resampling} gives a lower bound of order $\Delta_l$ for maximally coupled resampling, proved at the first resampling in a linear Gaussian model and checked numerically on the examples of Section~\ref{sec:numerics}.
There are resampling mechanisms \citep[e.g.][]{ballesio2020wasserstein} which can maintain $\beta=2$, but they often have flaws such as being limited to
diffusions in one dimension.

\subsection{Objective}\label{sec:setting-goal}

We can now state the two properties that we want from a new particle filter in this setting.
\begin{enumerate}
  \item[(i)] It moves the particles from $\M_{n-1}$ to $\M_n$ with equal weights, regardless of the type of each manifold, so that no weighting step is needed.
  \item[(ii)] It does not resample, so that two runs of the particle filter with different time steps, driven by the same random variables, stay close, and multilevel Monte Carlo can reach $\beta>1$ and, with a variance that also decreases like $N^{-1}$, a computing time of order $\epsilon^{-2}$.
\end{enumerate}

\subsection{Related work}\label{sec:related-work}

In the context of the filtering problem,
noiseless or `low noise' observations have received 
recent attention from a computational perspective; see for instance \citep{zhumekenov2025smcmc,zhumekenov2026} for a summary; with noisy observations, particle filters on Riemannian manifolds are developed in \citep{zhang}. Particle filters \citep{zhumekenov2026} and multilevel Markov chain Monte Carlo for parameter inference \citep{saporiti2026multilevel} have been developed for linear observations, whose manifolds are affine subspaces, and sequential Markov chain Monte Carlo for more general observations \citep{zhumekenov2025smcmc}. In multilevel particle filters, the particle filters at two levels are coupled through their resampling steps \citep{jasra2017multilevel,sen2018coupling,jacob2020smoothing}.
The resampling operation has been coupled, for instance, using
maximal coupling
and Wasserstein coupling. The latter relies on sorting the particles, which is why it is limited to one-dimensional states \citep{ballesio2020wasserstein}. 
Another direction is based upon
transport-based (particle) filters, which replace resampling altogether.
Among them, the ensemble transform particle filter and its multilevel version solve discrete transport problems whose cost grows faster than quadratically in the number of particles \citep{reich2013nonparametric,gregory2016multilevel}, entropic transport makes the particle filter differentiable at a quadratic cost \citep{corenflos2021differentiable}, and particle flows and transport maps move the particles deterministically \citep{daum2007nonlinear,spantini2022coupling}. 
In some cases, 
Schr\"odinger bridges and entropic transport are also learned from samples with neural networks, including on Riemannian manifolds \citep{cuturi2013sinkhorn,debortoli2021diffusion,thornton2022riemannian,delmoral2026trends}, and filters on circles and spheres are reviewed in \citep{kurz2016recursive}. Closest to our approach, Schr\"odinger bridges have been fitted within the filtering recursion \citep[\S4.4]{bernton2019schrodinger}, building on controlled sequential Monte Carlo \citep{heng2020controlled}, at a cost quadratic in the number of particles, and light Schr\"odinger bridges couple the levels of a multilevel filter at the final time \citep{zhao2026sequential} and the levels of static Bayesian inverse problems \citep{beh2026sequential}. The bridge-based filters require noisy observations, through the density of the observations or a likelihood bounded away from zero, and do not apply when the filter is supported by a manifold.

\subsection{Contributions and structure}\label{sec:contrib}

We obtain both properties (i)--(ii) of Section~\ref{sec:setting-goal} with one tool, the light Schr\"odinger bridge of \citep{aramayo2026light,gushchin2024light,korotin2024light}.
This is the framework of entropic optimal transport, which we use to move the particles from one manifold to the next instead of resampling them. A key insight is that, when the reference of the light Schr\"odinger bridge is the dynamics restricted to the new manifold, the previous position of a particle enters the bridge only through a few features. The bridge therefore only has to be fitted on the new manifold, regardless of the type of the previous one, which gives property (i). Since no particle is resampled, two runs of the light filter with different time steps can be driven by the same random variables and stay close, which gives the variance rate of property (ii); the computing time of order $\epsilon^{-2}$ remains conditional on a bound that also decreases like $N^{-1}$, which we do not prove.

Our contributions are as follows.
\begin{itemize}
  \item We build a new particle filter that moves its particles between changing manifolds with light Schr\"odinger bridges, on a class of Riemannian manifolds that includes planes, spheres, cylinders and tori, and we couple two runs of it with different time steps for multilevel Monte Carlo (Sections~\ref{sec:general} and~\ref{sec:filter}).
  \item We bound the error of the population version of the filter by how far each fitted bridge is from the exact one. We then show that the light filter inherits the rate at which two particles driven by the same Brownian path stay close, $\beta=2$ with constant $\sigma$ (Section~\ref{sec:setting-mlmc}): under a condition on one step and a condition on the fits, the latter proved in a linear Gaussian model for a fitting procedure adapted to the proof and checked numerically for it, the variance of the difference between two time steps is at most of order $\Delta_l^\beta$. Particle filters with maximally coupled resampling do not inherit it: under a nondegeneracy condition, which we prove at the first resampling in a linear Gaussian model and check numerically, $N$ times the variance of the same difference decreases, for large $N$, no faster than the time step, regardless of that rate. We are not aware of a proof of such a lower bound (Section~\ref{sec:theory}).
  \item We compare the method with multilevel particle filters on planes, on spheres, and on a complex example in which the manifold alternates between a torus, a cylinder, a plane and a three-sphere under nonlinear dynamics. In each case the runs at successive time steps stay much closer to each other than with resampling, and, in an operation-count model with the cost of the fits included, a hybrid estimator that uses the light filter at the fine levels is projected to reduce the cost at small tolerances (Section~\ref{sec:numerics}).
\end{itemize}

The rest of the paper is organized as follows. Section~\ref{sec:general} recalls light Schr\"odinger bridges and builds them between changing manifolds, Section~\ref{sec:filter} defines the light filter and its multilevel version, Section~\ref{sec:theory} analyzes the light filter with fitted bridges and the coupling of levels, and shows why maximally coupled resampling limits the rate of particle filters, and Section~\ref{sec:numerics} reports the experiments. Section~\ref{sec:plan} discusses the results and the open questions.

\section{Light Schr\"odinger Bridges between Changing Manifolds}\label{sec:general}

The filter has to be moved, at each observation time, from a law $\rho$ on the previous manifold to a law $\eta$ on the next one. We do this with a coupling of $\rho$ and $\eta$ whose conditional law, given a point $x'$ of the source, can be sampled exactly. Drawing from it moves each sample of $\rho$ to a sample of $\eta$, with no weights. This section constructs such a coupling, independently of filtering. A key insight is that the source point enters the light Schr\"odinger bridge only through a few features, so that the bridge only has to be fitted on the target manifold: the source may lie on a manifold of any type, and consecutive manifolds may differ. We state the bridge problem, whose reference is a Gaussian transition restricted to the target manifold, and recall the light Schr\"odinger bridge in Euclidean space. We then treat a single affine subspace, where the construction reduces to the light Schr\"odinger bridge of \citep{korotin2024light}, and then products of affine subspaces and spheres. The fit and the chaining over time come last.

\subsection{The bridge problem}\label{sec:bridge}

Let $\rho$ be a probability measure on $\R^d$, the source, possibly carried by a manifold, and let $\eta$ be a probability measure on a manifold $\M\subset\R^d$, the target. Throughout, $\M$ carries a reference measure $\lambda_\M$, made explicit for each class of manifolds below. For instance, on an affine subspace $a+\operatorname{span}(V)$, on a sphere $o+r\,\mathbb S^{d-1}$ and on a product of such pieces, it is respectively Lebesgue measure in the coordinate $z$ of $x=a+Vz$, the surface measure $\varsigma$ of the unit sphere in the coordinate $u$ of $x=o+ru$, and the product of the measures of the pieces. The reference transition is a Gaussian kernel. We assume:
\begingroup\tagsleft
\begin{align}\label{A1}\tag{A1}
  f(x\,|\,x') := \mathcal N\big(x;\,m(x'),\,\Omega\big), \qquad x',x\in\R^d,
\end{align}\endgroup
with $m:\R^d\to\R^d$ measurable and $\Omega$ symmetric positive definite. The reference measure on pairs is
\[
  R(dx',dx) := \rho(dx')\, f(x\,|\,x')\,\lambda_\M(dx),
\]
that is, a source point moved by the dynamics and restricted to $\M$. The static Schr\"odinger problem is
\begin{equation}\label{eq:sb}
  P^\ast := \arg\min\big\{\KL(P\,\|\,R) : P\in\Pi(\rho,\eta)\big\},
\end{equation}
with $\Pi(\rho,\eta)$ the set of couplings of $\rho$ and $\eta$, that is the joint laws of pairs $(X',X)$ with $X'\sim\rho$ and $X\sim\eta$, and $\KL$ the relative entropy. Among all ways of moving $\rho$ to $\eta$, $P^\ast$ is the one closest to the dynamics. Equivalently, since $-\log f(x\,|\,x')$ is, up to a constant, the cost
\[
  c(x',x) := \tfrac12\big(x-m(x')\big)^\T\Omega^{-1}\big(x-m(x')\big),
\]
$P^\ast$ solves the entropic optimal transport problem with this cost: it minimizes the expected cost plus, with weight one, the relative entropy of the coupling with respect to the independent coupling $\rho\otimes\eta$, which favors spread-out couplings over deterministic maps. Moves that are unlikely under the dynamics are expensive. For $m(x')=x'$ and $\Omega=\varepsilon I$, this is entropic optimal transport with the quadratic cost $\frac12|x-x'|^2$ and regularization $\varepsilon$, the setting of \citep{korotin2024light}. We assume that moving $\rho$ to $\eta$ independently has finite cost:
\begingroup\tagsleft
\begin{align}\label{A2}\tag{A2}
  \KL(\rho\otimes\eta\,\|\,R) < \infty .
\end{align}\endgroup

\begin{proposition}[{\citep{leonard2014survey,nutz2021introduction}}]\label{prop:bridge}
Assume \eqref{A1} and \eqref{A2}. Then \eqref{eq:sb} has a unique solution, of the form
\[
  P^\ast(dx',dx) = \rho(dx')\,\frac{f(x\,|\,x')\, v^\ast(x)}{c^\ast(x')}\,\lambda_\M(dx),
  \qquad c^\ast(x') := \int_\M f(x\,|\,x')\,v^\ast(x)\,\lambda_\M(dx),
\]
for some nonnegative function $v^\ast$ on $\M$, positive $\eta$-almost everywhere, the potential.
\end{proposition}

The potential $v^\ast$ is unknown, and the conditional law $P^\ast(\cdot\,|\,x')$ requires the normalizer $c^\ast(x')$, an integral over $\M$. In Euclidean space, with $\M=\R^d$ and $f(x\,|\,x')=\mathcal N(x;x',I)$, the light Schr\"odinger bridge (LightSB) of \citep{korotin2024light,gushchin2024light} makes both explicit. It parametrizes the adjusted potential, the potential times $e^{-|x|^2/2}$, by a Gaussian mixture,
\[
  v_\theta(x):=\sum_{k=1}^K\alpha_k\,\mathcal N(x;r_k,S_k),\qquad \theta=(\alpha_k,r_k,S_k)_{k\le K}.
\] Up to a factor of $x'$ alone, which cancels in the conditional law, the normalizer is then $\sum_{k=1}^K\alpha_k\exp\{\frac12x'^\T S_kx'+x'^\T r_k\}$, and, given $x'$, the bridge draws $x$ from the mixture of the Gaussian laws $\mathcal N(r_k+S_kx',S_k)$ with weights proportional to the terms of this sum. The parameter $\theta$ is fitted from samples of $\rho$ and $\eta$, by minimizing the loss of Section~\ref{sec:fitting}, at a cost linear in the number of samples. The construction is carried over in \citep{aramayo2026light} to Riemannian manifolds such as spheres, tori, $\mathrm{SO}(3)$, $\mathrm{SE}(3)$ and their products, with geometry-specific kernels and families, but always between laws on a fixed space. On changing manifolds, the source lies on $\M_{n-1}$ and the target on $\M_n$, possibly of a different type, and the kernel is the dynamics restricted to $\M_n$. This requires the restriction of $f(\cdot\,|\,x')$ to $\M$ to have a tractable form, which we now establish, first on an affine subspace.

\subsection{A single affine subspace}\label{sec:affine}

Let $\M=a+\operatorname{span}(V)$, with $a\in\R^d$ and $V\in\R^{d\times p}$ with orthonormal columns, so that each point of $\M$ is $x=a+Vz$ with $z\in\R^p$, and let $\lambda_\M$ be Lebesgue measure in $z$. This is the manifold given by a linear observation $Y=AX$, with $\operatorname{span}(V)=\ker A$ and $Aa=y$.

The tractability of LightSB rests on one property of its Gibbs kernel. Once the factors that depend on one variable only are removed, the source and the target interact only through an inner product, $\exp\{\langle x',x\rangle/\varepsilon\}$. We look for the same structure on $\M$. The exponent of $f(a+Vz\,|\,x')$ is
\[
  -\tfrac12\,z^\T\Lambda z + z^\T V^\T\Omega^{-1}\big(m(x')-a\big) - \tfrac12\,\big(m(x')-a\big)^\T\Omega^{-1}\big(m(x')-a\big),
  \qquad \Lambda := V^\T\Omega^{-1}V,
\]
and only the middle term couples $x'$ and $z$. In the whitened coordinate $\tilde z := \Lambda^{1/2}z$, the first term becomes $-\frac12|\tilde z|^2$, which depends on $\tilde z$ only, and the coupling term becomes $\langle\tilde m(x'),\tilde z\rangle$, with
\[
  \tilde m(x') := \Lambda^{-1/2}V^\T\Omega^{-1}\big(m(x')-a\big).
\]
Since $\Lambda^{-1}V^\T\Omega^{-1}(m(x')-a)$ is the mean of the restricted Gaussian in $z$, the feature $\tilde m(x')$ is this mean written in the whitened coordinate, hence the notation. The remaining terms, in $x'$ only, are collected in
\[
  w(x') := \exp\Big\{\tfrac12|\tilde m(x')|^2-\tfrac12\big(m(x')-a\big)^\T\Omega^{-1}\big(m(x')-a\big)\Big\}.
\]

Completing the square in $\tilde z=\Lambda^{1/2}z$ in the exponent displayed above, multiplied by $-2$, yields $|\tilde z-\tilde m(x')|^2-|\tilde m(x')|^2+(m(x')-a)^\T\Omega^{-1}(m(x')-a)$. Under \eqref{A1}, for any $x'\in\R^d$ and $z\in\R^p$ we therefore have
\[
  f\big(a+Vz\,|\,x'\big) = c_0\,w(x')\,\exp\Big\{-\tfrac12\big|\Lambda^{1/2}z-\tilde m(x')\big|^2\Big\},
\]
with the Gaussian normalization $c_0:=(2\pi)^{-d/2}\det(\Omega)^{-1/2}$.

Restricted to $\M$, the transition is therefore Gaussian in the whitened coordinate, with identity covariance and mean $\tilde m(x')$, and its total mass is a constant multiple of $w(x')$,
\[
  \int_\M f(x\,|\,x')\,\lambda_\M(dx) = c_0\,(2\pi)^{p/2}\det(\Lambda)^{-1/2}\,w(x').
\] Two consequences make the bridge tractable.
\begin{itemize}
  \item The source point enters only through the feature $\tilde m(x')$. The factor $w(x')$ depends on $x'$ only and cancels between $f$ and $c^\ast$ in Proposition~\ref{prop:bridge}. Up to factors of one variable, the kernel is $\exp\{\langle\tilde m(x'),\tilde z\rangle-\frac12|\tilde z|^2\}$, the Gibbs kernel of \citep{korotin2024light} with unit regularization and input $\tilde m(x')$. The source may lie on any manifold, or on none.
  \item As in \citep{korotin2024light}, parametrize the adjusted potential, the potential times $e^{-|\tilde z|^2/2}$, which absorbs the term $-\frac12|\tilde z|^2$ of the kernel, by the Gaussian mixture family $v_\theta(\tilde z) := \sum_{k=1}^K\alpha_k\,\mathcal N(\tilde z;r_k,S_k)$, with $\alpha_k>0$, $r_k\in\R^p$ and $S_k$ symmetric positive definite. For this family, the normalizer and the conditional law are explicit \citep{korotin2024light}:
  \[
    c_\theta(x') := \int e^{\langle\tilde m(x'),\tilde z\rangle}\,v_\theta(\tilde z)\,d\tilde z = \sum_{k=1}^K\alpha_k\exp\Big\{\tfrac12\tilde m(x')^\T S_k\tilde m(x')+\tilde m(x')^\T r_k\Big\},
  \]
  and $P_\theta(\cdot\,|\,x')$ is the mixture of the Gaussian laws $\mathcal N(r_k+S_k\tilde m(x'),S_k)$ of $\tilde z$, with weights proportional to the terms of this sum. A sample on $\M$ is then $x=a+V\Lambda^{-1/2}\tilde z$.
\end{itemize}
The bridge from any source to an affine subspace is therefore LightSB in the whitened coordinate of the target, evaluated at the feature of the source. The source point enters only through its feature $\tilde m(x')$: the family $v_\theta$ lives on the target, and nothing in the bridge depends on the manifold that $x'$ lies on. The weight $w(x')$ does not enter the bridge but reappears in filtering, as the incremental weight of the optimal proposal.

\subsection{Products of affine subspaces and spheres}\label{sec:products}

A single affine subspace covers linear sensors only: one target seen by a linear sensor lies on a line, and two targets seen by linear sensors lie on a plane (Figure~\ref{fig:sets}(c)). As soon as a range is measured, this is no longer enough. A range puts a target on a circle rather than a line. Two ranges then put the pair on a torus, a circle times a circle (Figure~\ref{fig:sets}(a)), and a range and a linear sensor put it on a cylinder, a circle times a line (Figure~\ref{fig:sets}(b)). A torus or a cylinder is neither an affine subspace nor a sphere, but it is a product of pieces that are: one piece per target, a circle or a line, each in its own coordinates. A single range for the pair gives a sphere in $\R^4$ (Figure~\ref{fig:sets}(d)), a piece of its own. As the sensors switch, the filter moves between such products, a torus at one time, a cylinder at the next, and it needs the bridge from any of them to any other. We therefore treat products of affine subspaces and spheres, and show that the computation for a single affine subspace carries over piece by piece.

The bridge stays explicit whenever its two ingredients are compatible \citep[Prop.~1]{aramayo2026light}. On the target manifold, the restricted kernel times a potential component $v_{\theta,k}$ must have a closed-form normalizer $c_{\theta,k}(x')$, and their normalized product a conditional law that can be sampled directly. The mixture $v_\theta=\sum_k\alpha_kv_{\theta,k}$ then has the normalizer $c_\theta=\sum_k\alpha_kc_{\theta,k}$, and its conditional law is the mixture of the component laws with weights proportional to $\alpha_kc_{\theta,k}(x')$. The affine subspace above is one instance. We now treat spheres, then products.

\paragraph{Spheres.} Let $\M=\{o+ru:u\in\mathbb S^{d-1}\}$, a circle for $d=2$, with the surface measure $\varsigma(du)$ of the unit sphere, and let the kernel be isotropic, $\Omega=\tau^2I$. The same structure appears on a sphere, without whitening: since $|u|=1$, the quadratic term in $u$ is constant on $\M$,
\[
  |o+ru-m(x')|^2 = r^2-2r\langle m(x')-o,u\rangle+|m(x')-o|^2,
\]
so that, restricted to $\M$, $f(o+ru\,|\,x')$ is proportional, in $u$, to $e^{\langle\kappa(x'),u\rangle}$, with the feature $\kappa(x'):=r(m(x')-o)/\tau^2$. This is the von Mises--Fisher law $\mathrm{vMF}_d(\kappa(x'))$ on $\mathbb S^{d-1}$, the analog of a Gaussian law on the sphere. Its normalizer $Z_d(|\kappa(x')|)$, with $Z_d(s):=\int e^{s u_1}\varsigma(du)$, is explicit: $Z_2(s)=2\pi I_0(s)$ on the circle, with $I_0$ a modified Bessel function, and $Z_3(s)=4\pi\sinh(s)/s$ on $\mathbb S^2$. For the potential component $e^{\langle b_k,u\rangle}$, the normalizer is $Z_d(|b_k+\kappa(x')|)$ and the conditional law is $\mathrm{vMF}_d(b_k+\kappa(x'))$. This is the bridge of \citep{aramayo2026light} on the sphere, evaluated at the feature of the source. The kernel has to be isotropic: otherwise its restriction to the sphere is of Fisher--Bingham type, with no explicit normalizer.

\paragraph{Products.} With several targets, $\M$ is a product of pieces, each an affine subspace or a sphere in its own block of coordinates. For the two targets of Figure~\ref{fig:sets}, two circles give the torus, a circle and a line the cylinder, two lines the plane, and a single sphere in $\R^4$ the three-sphere. We assume that\begingroup\tagsleft\begin{align}\label{A3}\tag{A3}\parbox{0.78\textwidth}{$\Omega$ is block diagonal with respect to the blocks of the pieces of $\M$, and isotropic on the spherical pieces.}\end{align}\endgroup Then the Gaussian density factorizes over the blocks, and the computation above works in the same manner on each piece: the kernel restricted to the product is the product of its restrictions to the pieces (Lemma~\ref{lem:factor} gives the general statement). Closed forms then multiply \citep[Prop.~2]{aramayo2026light}. The potential components are products of components on the pieces, the normalizers multiply, and, given the component, the conditional law draws the pieces independently. Table~\ref{tab:pieces} summarizes the two kinds of pieces.

\begin{table}[ht]
\centering
\small
\setlength{\tabcolsep}{4pt}
\begin{tabular}{lllll}
\toprule
piece & feature of $x'$ & potential component & normalizer & conditional law\\
\midrule
affine $a+\operatorname{span}(V)$ & $\tilde m(x')$ & $\mathcal N(\tilde z;r_k,S_k)$ (adjusted) & $e^{\frac12\tilde m^\T S_k\tilde m+\tilde m^\T r_k}$ & $\mathcal N(r_k+S_k\tilde m,\,S_k)$\\
sphere $o+r\,\mathbb S^{d-1}$ & $\kappa(x')$ & $e^{\langle b_k,u\rangle}$ & $Z_d(|b_k+\kappa|)$ & $\mathrm{vMF}_d(b_k+\kappa)$\\
product & all features & product & product & independent pieces\\
\bottomrule
\end{tabular}
\vspace{0.5em}
\caption{The two kinds of pieces, with the feature of the source point, the potential component, the normalizer $c_{\theta,k}$ and the conditional law of the bridge. Products of pieces combine them.}\label{tab:pieces}
\end{table}

We write $\psi(x')$ for the collection of the features of the pieces, $w(x')$ for the product of the weights of the pieces, a constant multiple of the total mass of $f(\cdot\,|\,x')$ on $\M$, $q(\cdot\,|\,x')$ for the normalized restriction of $f(\cdot\,|\,x')$ to $\M$, and $P_\theta(\cdot\,|\,x')$ for the conditional law of the fitted bridge. Explicit formulas for general products are given in Appendix~\ref{app:products}: the factorization of the kernel (Lemma~\ref{lem:factor}), the reduction of the bridge to the features (Proposition~\ref{prop:reduction}), and the closed forms (Corollary~\ref{cor:closed}). For general curved manifolds, the restricted kernel has no conjugate family with an explicit normalizer. This case is left for further work.

\subsection{Fitting and chaining}\label{sec:fitting}

It remains to choose $\theta$ within the family of Table~\ref{tab:pieces}, written in full in \eqref{eq:family}. Assume that $\E_\rho|m(X')|^2<\infty$ and that $\eta$ has a finite second moment. By \citep[Prop.~3.1]{korotin2024light}, whose proof only uses the marginal constraints \citep{aramayo2026light}, transferred through Proposition~\ref{prop:reduction}, the loss
\begin{equation}\label{eq:loss}
  \mathcal L(\theta) := \E_{\rho}\big[\log c_\theta(X')\big] - \E_{\eta}\big[\log v_\theta(\tilde Z,U)\big]
\end{equation}
equals $\KL(P^\ast\,\|\,\rho\otimes P_\theta)$ up to an additive constant that does not depend on $\theta$, with $\rho\otimes P_\theta$ the law $\rho(dx')P_\theta(dx\,|\,x')$ of a pair. The first expectation is estimated from samples of $\rho$. In the filter, samples of $\eta$ are not available, but weighted samples are, and the second expectation is then estimated by self-normalized importance sampling. The loss is minimized with Adam, each iteration costing a number of operations linear in the number of samples.

Chaining such bridges over time transports particles from $\M_0$ to $\M_1$, $\M_2$, and so on. Each step only requires a family on its target manifold, so that consecutive manifolds may be of different types, for instance a torus followed by a three-sphere. How the fitting errors accumulate over time is studied in Section~\ref{sec:theory}.

\section{Proposed Method: Light Schr\"odinger Bridge Filters}\label{sec:filter}

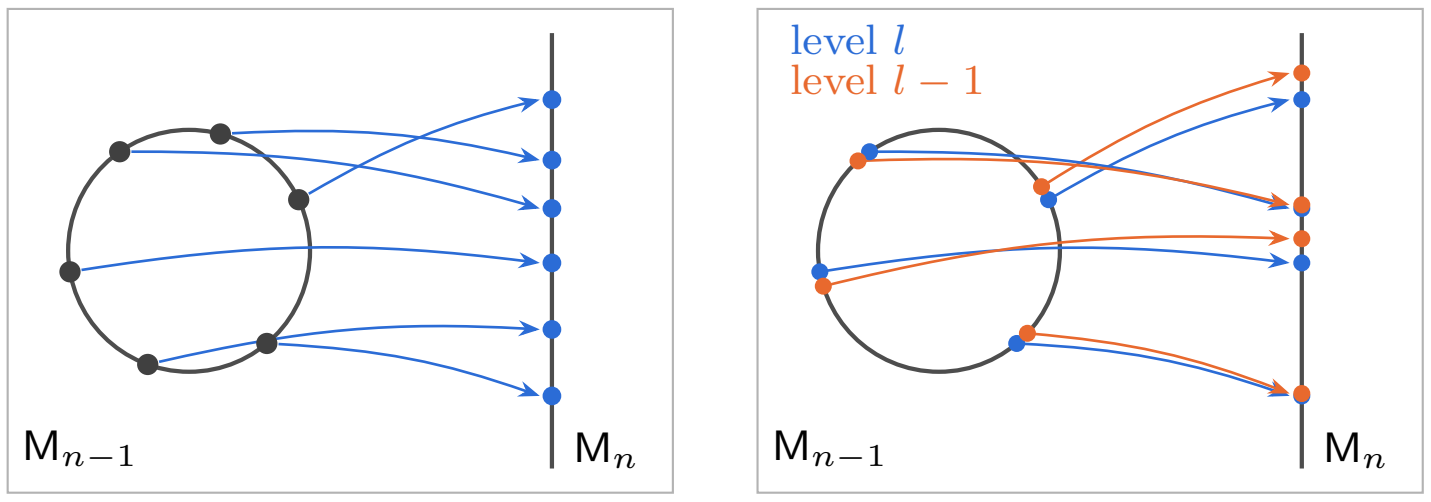
\begin{figure}[ht]
\centering
\begin{tikzpicture}[scale=0.8,
  man/.style={draw=black!70,line width=0.8pt},
  pt/.style={circle,fill=black!75,inner sep=0pt,minimum size=4pt},
  ar/.style={-{Stealth[length=4pt]},line width=0.5pt}]
\def\angles{25,75,125,190,250,310}
\def\ends{1.25,0.75,0.35,-0.1,-0.65,-1.2}
\def\frame#1{%
  \draw[gray!60,line width=0.4pt] (-1.5,-2.0) rectangle (4.0,2.0);
  \draw[man] (0,0) circle (1.0);
  \draw[man] (3.0,-1.8) -- (3.0,1.8);
  \node[font=\scriptsize] at (-0.9,-1.65) {$\M_{n-1}$};
  \node[font=\scriptsize] at (3.45,-1.65) {$\M_n$};
  \node[below,font=\small,align=center] at (1.25,-2.1) {#1};}
\begin{scope}
\frame{(a) light filter:\\equal weights, no resampling}
\foreach \a/\y in {25/1.25,75/0.75,125/0.35,190/-0.1,250/-0.65,310/-1.2}{
  \node[pt] (s) at (\a:1.0) {};
  \draw[ar,cA] (s) to[bend left=8] (2.9,\y);
  \fill[cA] (3.0,\y) circle (2.2pt);}
\end{scope}
\begin{scope}[xshift=6.2cm]
\frame{(b) two levels:\\same random numbers}
\foreach \a/\y/\d in {25/1.25/0.10,125/0.35/-0.09,190/-0.1/0.08,310/-1.2/-0.1}{
  \coordinate (s1) at (\a:1.0);
  \coordinate (s2) at (\a+7:1.0);
  \draw[ar,cA] (s1) to[bend left=8] (2.9,\y);
  \draw[ar,cB] (s2) to[bend left=8] (2.9,\y+\d+0.12);
  \fill[cA] (s1) circle (2pt); \fill[cB] (s2) circle (2pt);
  \fill[cA] (3.0,\y) circle (2pt); \fill[cB] (3.0,\y+\d+0.12) circle (2pt);}
\node[font=\scriptsize,cA,anchor=west] at (-1.4,1.75) {level $l$};
\node[font=\scriptsize,cB,anchor=west] at (-1.4,1.4) {level $l-1$};
\end{scope}
\end{tikzpicture}
\caption{One step of the light filter, from $\M_{n-1}$ (here a circle) to $\M_n$ (here a line), to be compared with Figure~\ref{fig:pf}. (a) The light filter moves each particle by a draw from a fitted bridge, which already accounts for the weights: the new particles are equally weighted. (b) In the multilevel estimator, the light filters at two levels move twin particles with the same random numbers. Without resampling, the twins stay close, so their difference has small variance.}\label{fig:scheme}
\end{figure}

We now apply the bridge to the filtering problem; we call the resulting method the light filter. It is a particle filter in which the weighting and resampling steps are replaced by a move with the fitted bridge, so that the particles keep equal weights. In what follows, ``particle filter'' without qualification refers to the standard one, which weights and resamples. We first write the filter for a model with an explicit Gaussian transition, which is the form taken by each level of the time discretization when the transition is available. We identify the bridge between consecutive filters. Section~\ref{sec:ml} then couples two levels. Figure~\ref{fig:scheme} summarizes the method.

\subsection{The filter on changing manifolds}\label{sec:fc}

Consider the state-space model
\[
  X_n = F_n(X_{n-1})+\nu_n,\quad \nu_n\sim\mathcal N_{d}(0,\Omega_n),\qquad Y_n = H_n(X_n),\qquad n\ge1,\quad X_0\sim\pi_0,
\]
which is the form of each level of the time discretization when the transition is Gaussian. Each sensor observes one block $x_{(j)}$ of the state, through a linear functional $A_{n,j}x_{(j)}$, with $A_{n,j}$ of full row rank, or through a range $|x_{(j)}-o_{n,j}|^2$ from a known position $o_{n,j}$; some blocks may be unobserved. As in Figure~\ref{fig:sets}, the manifold $\M_n$ is then a product of pieces: an affine subspace for each linear sensor (the whole block for an unobserved one), and a sphere of center $o_{n,j}$ and radius $y_{n,j}^{1/2}$ for each range. The sensors may change with $n$, and $\M_n$ with them. Linear observations alone, $Y_n=A_nX_n$, are the setting of \citep{zhumekenov2026}. We assume that $\Omega_n$ satisfies \eqref{A3} with respect to the pieces of $\M_n$: it is block diagonal, and isotropic on the range blocks. As in \citep{zhumekenov2026}, the filter then has the density (Appendix~\ref{app:coarea})
\begin{equation}\label{eq:filter}
  \pi_n(dx) \propto \int f_n\big(x\,|\,x'\big)\,\pi_{n-1}(dx')\,\lambda_{\M_n}(dx),
  \qquad f_n(\cdot\,|\,x') = \mathcal N\big(\cdot\,;F_n(x'),\Omega_n\big).
\end{equation}
This is the setting of Section~\ref{sec:general} with $\rho=\pi_{n-1}$, $\eta=\pi_n$, $m=F_n$ and $\Omega=\Omega_n$. We write $\psi_n$, $w_n$ and $q_n$ for the features, the weight and the normalized restricted kernel of Lemma~\ref{lem:factor}, and $c_{\M_n}$ for the constant of that lemma such that $\int f_n(x\,|\,x')\,\lambda_{\M_n}(dx)=c_{\M_n}w_n(x')$. The kernel $q_n(\cdot\,|\,x')$, the conditional law of $X_n$ given $X_{n-1}=x'$ and $Y_n=y_n$, is the optimal proposal, given in the linear case by \citep[Prop.~3.2]{zhumekenov2026}, and $w_n(x')$, proportional to the density of $Y_n$ at $y_n$ given $X_{n-1}=x'$, is its incremental weight. The optimal-proposal particle filter, which resamples with weights $w_n$ and moves the particles by Gaussian and von Mises--Fisher draws \citep{wood1994simulation}, is the baseline of our experiments.

\subsection{The bridge between consecutive filters}

The following lemma is the bridge of \citep[\S4.4]{bernton2019schrodinger}, written on $\M_n$. Parts (ii) and (iii) are general properties of Schr\"odinger bridges. Only (i) is specific to noiseless observations: it shows that the reference already has the right target marginal $\pi_n$, and differs from the required coupling only through its source marginal, which is $\pi_{n-1}$ reweighted by $w_n$.

\begin{lemma}[{cf.\ \citep[\S4.4]{bernton2019schrodinger}}]\label{lem:filter}
Assume that $\Omega_n$ satisfies \eqref{A3} with respect to the pieces of $\M_n$. Let $R_n(dx',dx) := \pi_{n-1}(dx')\,f_n(x\,|\,x')\,\lambda_{\M_n}(dx)$, with total mass $R_n(1):=\int R_n(dx',dx)=c_{\M_n}\pi_{n-1}(w_n)$, so that $R_n/R_n(1)$ is a probability measure, and assume that $0<\pi_{n-1}(w_n)<\infty$ and, for (ii), that \eqref{A2} holds for $(\pi_{n-1},\pi_n,R_n/R_n(1))$. Then:
\begin{enumerate}
  \item[(i)] The second marginal of $R_n/R_n(1)$ is $\pi_n$, and its first marginal is
  \[
    \pi_{n-1}(dx')\,\frac{w_n(x')}{\pi_{n-1}(w_n)}.
  \]
  \item[(ii)] The Schr\"odinger bridge between $\pi_{n-1}$ and $\pi_n$ with reference $R_n$ is
  \[
    P_n^\ast(dx',dx) = \pi_{n-1}(dx')\,\frac{q_n(dx\,|\,x')\,\bar v_n(x)}{\bar c_n(x')},
  \]
  with $\bar c_n(x'):=\int q_n(dx\,|\,x')\,\bar v_n(x)$, where $\bar v_n$ solves
  \[
    \bar v_n(x) = \frac{d\pi_n}{d\lambda_{\M_n}}(x)\Big/\int\pi_{n-1}(dx')\,\frac{1}{\bar c_n(x')}\frac{dq_n(\cdot\,|\,x')}{d\lambda_{\M_n}}(x).
  \]
  \item[(iii)] If $X'\sim\pi_{n-1}$ and $X\,|\,X'\sim P_n^\ast(dx\,|\,X')$, then $X\sim\pi_n$. Transporting equally weighted particles of $\pi_{n-1}$ through $P_n^\ast$ gives equally weighted particles of $\pi_n$, with no resampling.
  \item[(iv)] The Sinkhorn iteration for \eqref{eq:sb} rescales the reference alternately to match the first and the second marginal. Its first half-step, started at $v\equiv1$, gives the coupling $\pi_{n-1}(dx')\,q_n(dx\,|\,x')$, the moves of the optimal proposal without its weights.
\end{enumerate}
\end{lemma}

\begin{proof}
(i) follows by integrating $R_n$ in $x'$, using \eqref{eq:filter}, and in $x$, using Lemma~\ref{lem:factor}. Since $f_n(x\,|\,x')\,\lambda_{\M_n}(dx) = c_{\M_n}w_n(x')\,q_n(dx\,|\,x')$ by Lemma~\ref{lem:factor}, (ii) is the form of Proposition~\ref{prop:bridge} rewritten with $q_n$, the factor $c_{\M_n}w_n(x')$ cancelling with the normalizer, and the equation for $\bar v_n$ is the constraint on the second marginal. (iii) follows from the two marginal constraints. For (iv), take $v\equiv1$ in the update of the first potential: it equals $1/(c_{\M_n}w_n)$, which cancels the factor $c_{\M_n}w_n(x')$ of $f_n(x\,|\,x')\lambda_{\M_n}(dx)$. This ends the proof of the lemma.
\end{proof}

In other words, the bridge makes the moves of the optimal-proposal particle filter, but multiplies the proposal $q_n$ by the potential $\bar v_n$ on $\M_n$ and renormalizes it. This tilts each move toward the regions that the weights would favor, so that no weight is needed afterwards. It is the interpretation of the Sinkhorn iterations as successive reweighting (Feynman--Kac) steps \citep{delmoral2004feynman,heng2020controlled}.

The second term of the loss \eqref{eq:loss} needs samples of $\pi_n$, which we do not have. By Lemma~\ref{lem:filter}(i), draws $x^i\sim q_n(\cdot\,|\,x'^{,i})$ weighted by $w_n(x'^{,i})$ are weighted samples of $\pi_n$, and we use them instead. These weights only enter the fit: the transported particles carry equal weights.

The resulting method is Algorithm~\ref{alg:light}. When the transition density is not available, the fitting points of line~4 are replaced by weighted endpoints of diffusion bridges, as in Section~\ref{sec:num-switch}.

\begin{algorithm}[ht]
\caption{Light filter on changing manifolds}\label{alg:light}
\begin{algorithmic}[1]
\Require data $y_{1:\bar n}$, number of particles $N$, number of components $K$, fitting draws $M$ per particle
\State sample $X_0^1,\dots,X_0^N$ i.i.d.\ from $\pi_0$
\For{$n=1,\dots,\bar n$}
  \State compute the features $\psi_n(X_{n-1}^i)$ and the weights $w_n(X_{n-1}^i)$ of Lemma~\ref{lem:factor}, $i\le N$
  \State draw $M$ fitting points from $q_n(\cdot\,|\,X_{n-1}^i)$ for each $i\le N$, with weights $w_n(X_{n-1}^i)$
  \State compute $\theta_n$ by minimizing the weighted loss \eqref{eq:loss} with Adam
  \For{$i=1,\dots,N$}
    \State sample $(\tilde z^i,u^i)\sim P_{\theta_n}(\cdot\,|\,X_{n-1}^i)$ of Corollary~\ref{cor:closed}, continuously if $K\ge2$, and map it to $X_n^i\in\M_n$
  \EndFor
  \State $\widehat\pi_n^N(\varphi):=N^{-1}\sum_{i=1}^N\varphi(X_n^i)$
\EndFor
\State \Return $\widehat\pi_1^N(\varphi),\dots,\widehat\pi_{\bar n}^N(\varphi)$
\end{algorithmic}
\end{algorithm}

\subsection{Running the filter at two time steps}\label{sec:ml}

Each difference in the multilevel identity \eqref{eq:mlmc} compares the filters at two consecutive levels $l$ and $l-1$, that is, at the time steps $\Delta_l$ and $\Delta_{l-1}=2\Delta_l$. We now explain how the light filter is run at both time steps together, so that the two runs stay close.

\paragraph{Coupling the two runs.} Each step of the light filter is an explicit map of the ancestor and of uniform variables, and the fitting draws add the Gaussian variables of the diffusion bridges, if any. Two levels can therefore be coupled synchronously: both are driven by the same random variables and start from the same particles, and no coupled resampling is needed. When diffusion bridges are used, the two levels use the same ones, the coarse level evaluating the weights on every other point of the fine grid. The level enters only through the time discretization at step $\Delta_l=2^{-l}$: of the transition itself when it is Gaussian, and otherwise, for a nonlinear drift, of the weights of the fitting draws. The level-$l$ difference therefore comes from two sources: the discretization, through the weights of the fitting draws, and the difference $\theta_n^{[l]}-\theta_n^{[l-1]}$ of the fitted parameters.

\paragraph{Keeping the fits close.} For the fits to be close at neighboring levels, we start the optimizer at both levels from the same initial value of $\theta$ and use common fitting draws. At each time $n$, all levels start their fits from a common pilot fit $\theta_n^0$, computed once on an independent run of the light filter at the base level. Long fits are done only in the pilot, and each level refits briefly. With $K\ge2$, the particles are also drawn from the mixture with a sampler that is a continuous function of $\theta$, described at the end of this section.

\paragraph{Unbiasedness of the differences.} For the sum \eqref{eq:mlmc} to telescope in expectation, the level-$l$ filter must have the same expectation whether it is the fine member of the pair $(l,l-1)$ or the coarse member of the pair $(l+1,l)$. Since nothing is resampled, the particles are exchangeable, and the expectation of the level-$l$ estimate is that of one particle. Its law is fixed by the law of the fitted parameters and of the variables that drive it, and it does not depend on the number of particles averaged. The law of the fitted parameters, however, depends on the number of particles used in the fit. The sum therefore telescopes exactly when each fit at level $l$ uses the same number of particles in both pairs, for instance an auxiliary cloud of fixed size used only for the fits, and when the diffusion bridges, if any, have a law that does not depend on the level, as in Section~\ref{sec:num-switch}. Otherwise the expectations differ by terms that depend on the numbers of particles used in the fits and that we do not bound. In both cases the multilevel identity removes the discretization bias, not the fitting bias. When a single pilot fit is shared by all levels, the increments at different levels are dependent through it, and their covariances enter the variance of the multilevel estimator; an independent pilot fit for each difference restores independence. Our experiments share a single pilot. The coupled pair of filters is Algorithm~\ref{alg:coupled}.

\begin{algorithm}[H]
\caption{Synchronously coupled light filters at levels $l$ and $l-1$}\label{alg:coupled}
\begin{algorithmic}[1]
\Require level $l\ge1$, data $y_{1:\bar n}$, $N$, $K$, $M$, and pilot fits $\theta_1^0,\dots,\theta_{\bar n}^0$
\State sample $X_0^{[l],i}=X_0^{[l-1],i}$, $i\le N$, i.i.d.\ from $\pi_0$
\For{$n=1,\dots,\bar n$}
  \State draw the common random variables: those of the fitting points, of the diffusion bridges, if any, on the grid of step $\Delta_l$ (the coarse level uses every other point), and the variables $U_n^i$ of the sampler
  \For{$k\in\{l,l-1\}$}
    \State compute $\theta_n^{[k]}$ as in lines~3--5 of Algorithm~\ref{alg:light} at level $k$, started from the pilot fit $\theta_n^0$ and driven by the common variables
    \State sample $X_n^{[k],i}$ from $X_{n-1}^{[k],i}$ as in line~7 of Algorithm~\ref{alg:light}, with $\theta_n^{[k]}$ and the common variables $U_n^i$, $i\le N$
  \EndFor
\EndFor
\State \Return $N^{-1}\sum_{i=1}^N\{\varphi(X_n^{[l],i})-\varphi(X_n^{[l-1],i})\}$, $n=1,\dots,\bar n$
\end{algorithmic}
\end{algorithm}

\paragraph{The multilevel estimator.} The estimator of $\pi_n(\varphi)$ is the multilevel identity \eqref{eq:mlmc} with each term replaced by an estimate,
\begin{equation}\label{eq:ml-estimator}
  \widehat\pi_n^{\mathrm{ML}}(\varphi) := \widehat\pi_n^{[l_0],N_{l_0}}(\varphi) + \sum_{l=l_0+1}^{L}\frac1{N_l}\sum_{i=1}^{N_l}\Big\{\varphi\big(X_n^{[l],i}\big)-\varphi\big(X_n^{[l-1],i}\big)\Big\}.
\end{equation}
The first term is a single filter at the base level $l_0$, with $N_{l_0}$ particles. Each difference comes from one run of Algorithm~\ref{alg:coupled} with $N_l$ pairs of twin particles. The runs at different levels use independent random variables, and the numbers of particles $N_l$ are chosen as in Section~\ref{sec:setting-mlmc}, many at the coarse levels and few at the fine ones. We also use a hybrid estimator, in which the base level and the differences below a switch level $s$ are computed with particle filters, coupled through their resampling steps as in \citep{jasra2017multilevel}, and the differences from level $s$ on with Algorithm~\ref{alg:coupled}. Level $s-1$ then enters as a particle filter in the base term or the difference at level $s-1$ and as a light filter in the difference at level $s$, so that the hybrid estimator also carries the fitting bias of the light filter at level $s-1$. Light filters cost more per particle than particle filters, because of the fits, and this cost matters most at the coarse levels, where the Euler steps are cheap. The structure of \eqref{eq:ml-estimator} is that of multilevel particle filters \citep{jasra2017multilevel,beskos2017multilevel}; what changes is how the two filters in each difference are coupled.

\paragraph{Sampling continuously in the parameters.}\label{sec:cont-sampling} The two levels move their particles with the same uniform variables but with different fitted parameters, so the draws from the bridge of Corollary~\ref{cor:closed} must be continuous functions of the parameters. Drawing a mixture component first is not: a small change of $\theta$ makes a few particles switch component and jump by a distance of order one, so that, as with maximally coupled resampling, the variance of the increments decreases no faster than $\Delta_l$. We therefore draw the coordinates one after another, each by inverting its one-dimensional conditional distribution function, which is an explicit mixture (Appendix~\ref{app:sampler}).

\section{Bias and Variance Analysis}\label{sec:theory}

Recall from Section~\ref{sec:setting-mlmc} that, for the Euler scheme, $\alpha=\gamma=1$, so that the computing time of the multilevel estimator \eqref{eq:ml-estimator} depends on the method only through the rate $\beta$ of the variance of the level-$l$ increments, $V_l\le c_1\Delta_l^\beta$: it is of order $\epsilon^{-2}$ when $\beta>1$ and $\epsilon^{-2}|\log\epsilon|^2$ when $\beta=1$. With constant $\sigma$, two Euler schemes driven by the same Brownian path give $\beta=2$, while multilevel particle filters with maximally coupled resampling are observed at a rate of about $1$.

This section explains the difference. The light filter has no resampling step: the twins are moved by nearby maps, Lipschitz in the particle and in the fitted parameters, so that the light filter inherits the rate $\beta$ of the coupling of the dynamics, where $\beta$ is the exponent in conditions \eqref{S1}--\eqref{S2}, and its increment variance is at most of order $\Delta_l^\beta$ (Section~\ref{sec:theory-coupling}). A particle filter with maximally coupled resampling does not inherit it: for any $\beta$, the mismatches bound $N$ times its increment variance from below by a quantity of order $\Delta_l$, so that the variance rate of its increments is at most $1$ for large $N$, under a nondegeneracy condition that we prove at the first resampling in a linear Gaussian model and check numerically (Section~\ref{sec:theory-resampling}). Before these two results, Section~\ref{sec:theory-bias} bounds the bias of the light filter by its fitting errors. The bias bound concerns the population version of the light filter, with the fits taken as given, the variance bound lacks the factor $N^{-1}$ that the complexity theorem requires, and the condition on the fits is proved only in a linear Gaussian model, for a fitting procedure adapted to the proof, and checked numerically for that procedure, while for the Adam fits of our experiments the measured rates are consistent with it; we state these limits where they arise. We write $\|\rho-\rho'\|_{\mathrm{tv}}:=\sup_A|\rho(A)-\rho'(A)|$, $\operatorname{osc}(\varphi):=\sup\varphi-\inf\varphi$, so that $|\rho(\varphi)-\rho'(\varphi)|\le\operatorname{osc}(\varphi)\,\|\rho-\rho'\|_{\mathrm{tv}}$. We also write $\mathrm{D}(P):=\sup_{x,x'}\|P(x,\cdot)-P(x',\cdot)\|_{\mathrm{tv}}\le1$ for the Dobrushin coefficient of a Markov kernel $P$.

\subsection{Bias of the fitted filter}\label{sec:theory-bias}

Let $f_n$ be the transition density and
\[
  w_n(x'):=\int f_n(x\,|\,x')\,\lambda_{\M_n}(dx),\qquad q_n(dx\,|\,x'):=\frac{f_n(x\,|\,x')\,\lambda_{\M_n}(dx)}{w_n(x')};
\]
for a Gaussian transition, $q_n$ is the kernel of Lemma~\ref{lem:factor} and $w_n$ is $c_{\M_n}$ times its weight; this constant cancels in $\Phi_n$, and in Lemma~\ref{lem:filter} it amounts to dropping the factor $c_{\M_n}$ in $R_n(1)$. By \eqref{eq:filter}, the filter satisfies $\pi_n=\Phi_n(\pi_{n-1})$, with
\[
  \Phi_n(\rho)(dx) := \frac{\int\rho(dx')\,w_n(x')\,q_n(dx\,|\,x')}{\rho(w_n)}
\]
for any probability measure $\rho$ with $0<\rho(w_n)<\infty$: a Feynman--Kac step \citep{delmoral2004feynman}, which reweights $\rho$ by the potential $w_n$, normalizes it, and moves the result by the Markov kernel $q_n$. Lemma~\ref{lem:filter} holds with $\pi_{n-1}$ replaced by any such $\rho$. We write $P_n^\ast[\rho]$ for the bridge between $\rho$ and $\Phi_n(\rho)$ with the reference used in the fit, $\rho(dx')\,f_n(x\,|\,x')\,\lambda_{\M_n}(dx)$, or $\rho(dx')\,\tilde f_n(x\,|\,x')\,\lambda_{\M_n}(dx)$ with the linear kernel of Section~\ref{sec:num-switch}. Only the marginals of $P_n^\ast[\rho]$ are used in the proofs below, and they are $\rho$ and $\Phi_n(\rho)$ in both cases. The fitted filter is defined by $\widehat\pi_0:=\pi_0$ and
\[
  \widehat\pi_n(dx) := \int\widehat\pi_{n-1}(dx')\,P_{\theta_n}(dx\,|\,x')
\]
and the fitting error at time $n$ by
\[
  \delta_n := \KL\big(P_n^\ast[\widehat\pi_{n-1}]\,\big\|\,\widehat\pi_{n-1}\otimes P_{\theta_n}\big),
\]
where $\theta_n$ is the parameter fitted at time $n$, taken as given. Here $\rho\otimes P$ denotes the law $\rho(dx')P(dx\,|\,x')$ of a pair. We make the following assumption.
\begingroup\tagsleft
\begin{align}\label{F}\tag{F}
  \parbox{0.86\textwidth}{For any $n\ge1$, $0<\widehat\pi_{n-1}(w_n)<\infty$, the bridge $P_n^\ast[\widehat\pi_{n-1}]$ exists, that is \eqref{A2} holds for its marginals and its reference, and $\delta_n<\infty$.}
\end{align}\endgroup
When the reference of $P_n^\ast$ is the one whose Gibbs kernel is used in the fit, $\delta_n$ is, by \citep[Prop.~3.1]{korotin2024light}, the excess of the population loss \eqref{eq:loss}, with $\rho=\widehat\pi_{n-1}$, at $\theta_n$ over its value at the exact potential. It collects the approximation error of the family, the optimization error and, with finite samples, the statistical error. Two fits at the same time $n$ from the same source $\widehat\pi_{n-1}$ are therefore ordered by their population losses. The loss minimized in practice is an estimate of the population loss, from weighted samples, and it does not bound $\delta_n$.

A fit with a small excess loss moves the cloud almost to the right place: the error made at one step is controlled by $\delta_n$. Indeed, the first marginals of $P_n^\ast[\widehat\pi_{n-1}]$ and of $\widehat\pi_{n-1}\otimes P_{\theta_n}$ are both $\widehat\pi_{n-1}$, and their second marginals are $\Phi_n(\widehat\pi_{n-1})$, by Lemma~\ref{lem:filter}(iii), and $\widehat\pi_n$. The relative entropy does not increase under the projection on the second coordinate, so that $\KL(\Phi_n(\widehat\pi_{n-1})\,\|\,\widehat\pi_n)\le\delta_n$, and Pinsker's inequality gives, under \eqref{F}, for any $n\ge1$,
\begin{equation}\label{eq:onestep}
  \big\|\widehat\pi_n-\Phi_n(\widehat\pi_{n-1})\big\|_{\mathrm{tv}} \le (\delta_n/2)^{1/2}.
\end{equation}

An error made at time $p$ is then carried along by the later steps of the filter. These steps reweight the cloud, which can amplify the error by at most the ratio of the largest to the smallest weight, and move it by a Markov kernel, which can only shrink it. To make this precise we use the semigroup of the filter. For $n\ge1$ let $Q_n(x',dx):=w_n(x')\,q_n(dx\,|\,x')$, let $Q_{p,n}:=Q_{p+1}\cdots Q_n$ for $0\le p<n$ and $Q_{n,n}:=\mathrm{Id}$, and let
\[
  G_{p,n}:=Q_{p,n}(1),\qquad P_{p,n}(x,dy):=\frac{Q_{p,n}(x,dy)}{G_{p,n}(x)}.
\]
Then $\Phi_{p,n}:=\Phi_n\circ\cdots\circ\Phi_{p+1}$ satisfies $\Phi_{p,n}(\rho)(dy)=\rho(G_{p,n}P_{p,n}(\cdot,dy))/\rho(G_{p,n})$ \citep{delmoral2004feynman}. We assume that, for any $n\ge2$,
\begingroup\tagsleft
\begin{align}\label{W}\tag{W}
  0 < \underline w_n := \inf_{x\in\M_{n-1}}w_n(x) \le \sup_{x\in\M_{n-1}}w_n(x) =: \overline w_n < \infty,
\end{align}\endgroup
The first step is not concerned: the bound below uses $w_k$ only for $k\ge2$, and \eqref{F} gives $0<\pi_0(w_1)<\infty$, which is all that is needed at $n=1$, where $\inf w_1=0$ for a Gaussian $\pi_0$. For a Gaussian transition, $w_n$ is a continuous and positive function of $F_n(x')$ (Lemma~\ref{lem:factor}), and \eqref{W} holds when the maps $F_n$ are bounded on $\M_{n-1}$. This condition plays the role of the bounds on the likelihood in \citep[(A2)]{zhao2026sequential}. It fails in all our experiments with an unbounded manifold, including the switching test, where a localization argument would be needed.

Under \eqref{W}, a standard contraction estimate for Feynman--Kac semigroups holds on the changing manifolds \citep[Prop.~4.3.2]{delmoral2004feynman}: for any $0\le p\le n$ and any probability measures $\rho,\rho'$ carried by $\M_p$,
\begin{equation}\label{eq:lipschitz}
  \|\Phi_{p,n}(\rho)-\Phi_{p,n}(\rho')\|_{\mathrm{tv}} \le g_{p,n}\,\mathrm{D}(P_{p,n})\,\|\rho-\rho'\|_{\mathrm{tv}},
\end{equation}
with
\[
  g_{p,n}:=\frac{\sup_{\M_p}G_{p,n}}{\inf_{\M_p}G_{p,n}}\le\prod_{k=p+1}^n\frac{\overline w_k}{\underline w_k}.
\]
We give the proof in Appendix~\ref{app:proof-lipschitz}.

\begin{remark}\label{rem:ratios}
The bound on $g_{p,n}$ in \eqref{eq:lipschitz} may grow geometrically in $n-p$, but it can be made uniform in time under further conditions on the kernels. Between changing compact manifolds, as in the test on spheres of Section~\ref{sec:num-sphere}, the mixing conditions of \citep[Prop.~4.3.3 and Prop.~4.3.6]{delmoral2004feynman} apply step by step: they bound $g_{p,n}$ uniformly in time and make $\mathrm{D}(P_{p,n})$ decrease geometrically in $n-p$. Between changing noncompact manifolds, such as the affine subspaces of Section~\ref{sec:num-e4} and the plane and the cylinder of Section~\ref{sec:num-switch}, mixing fails, and a multi-step version of the drift conditions of \citep{whiteley2013stability}, which allow noncompact state spaces, would be needed.
\end{remark}

Putting \eqref{eq:onestep} and \eqref{eq:lipschitz} together, each step contributes its own fitting error, amplified or damped by the steps that follow it. This is the standard perturbation argument for Feynman--Kac flows \citep{delmoral2004feynman}; what is specific to the light filter is the one-step bound \eqref{eq:onestep}.

\begin{theorem}\label{theo:bias}
Assume \eqref{F} and \eqref{W}. Then for any $n\ge1$ and any bounded measurable function $\varphi$ we have
\[
  \big|\widehat\pi_n(\varphi)-\pi_n(\varphi)\big| \le \operatorname{osc}(\varphi)\sum_{p=1}^n g_{p,n}\,\mathrm{D}(P_{p,n})\,\Big(\frac{\delta_p}{2}\Big)^{1/2}.
\]
\end{theorem}

\begin{proof}
Since $\widehat\pi_0=\pi_0$ and $\Phi_{p-1,n}=\Phi_{p,n}\circ\Phi_p$, we have the telescoping decomposition
\[
  \widehat\pi_n-\pi_n = \sum_{p=1}^n\Big\{\Phi_{p,n}(\widehat\pi_p)-\Phi_{p,n}\big(\Phi_p(\widehat\pi_{p-1})\big)\Big\}.
\]
By the contraction estimate \eqref{eq:lipschitz} and the one-step bound \eqref{eq:onestep}, the total variation norm of the $p$-th term is at most $g_{p,n}\,\mathrm{D}(P_{p,n})\,(\delta_p/2)^{1/2}$. This ends the proof of the theorem.
\end{proof}

We end this subsection with what the theorem does and does not cover.

\paragraph{Fixed time.} The bound holds at a fixed time $n$. Using only $\mathrm{D}(P_{p,n})\le1$, the factors $g_{p,n}$ may grow geometrically in $n-p$, so that the bound may grow with $n$, as in \citep[Rem.~4.4]{zhao2026sequential}. Suppose instead that the filter forgets its initial condition, as under the conditions of Remark~\ref{rem:ratios}, that is, that the effect at time $n$ of an error made at time $p$ decreases geometrically in $n-p$: $g_{p,n}\mathrm{D}(P_{p,n})\le a\varrho^{n-p}$ for some $a>0$ and $\varrho\in]0,1[$. Then the sum is geometric, and the bias is bounded uniformly in time by $\operatorname{osc}(\varphi)\,a\,(1-\varrho)^{-1}\sup_p(\delta_p/2)^{1/2}$.

\paragraph{Populations, not particles.} The theorem is stated for the populations, with the fitted parameters given. A bound for the light filter with $N$ particles also needs a propagation of chaos estimate, which bounds the distance between the empirical law of the $N$ transported particles, which interact through the shared fit, and the population law, as in \citep[(H1)]{zhao2026sequential}.

\paragraph{Link with the experiments.} In the switching test of Section~\ref{sec:num-switch}, condition \eqref{W} fails, since the plane and the cylinder are unbounded, and the theorem does not apply. What we observe there still matches the mechanism of its proof: the bias is created at the step on $\mathbb S^3$ and carried to the later steps, as through $\Phi_{p,n}$, and a richer family on that step, which reduces its excess loss $\delta_p$, also reduces the bias. The theorem is an upper bound, and it orders the bounds, not the biases themselves.

\paragraph{Bias of the multilevel estimator.} Suppose that every level, including the base level, uses the light filter with the same procedure, and that each fit at level $l$ uses the same number of particles in both pairs that contain level $l$. Then the multilevel identity \eqref{eq:mlmc} telescopes in expectation, and the expectation of the estimator is that of the single-level light filter at the finest level $L$. Its fitting bias is that of the light filter at level $L$, whose population part, with the fitted parameters given, is bounded by Theorem~\ref{theo:bias}. It does not accumulate over the levels. If the numbers of particles differ between the two pairs, the expectations also differ by terms that depend on the numbers of particles used in the fits and that we do not bound. When a particle filter is used at the base level, or at all levels below some switch level $s$ as in the hybrid estimator of Section~\ref{sec:ml}, the telescoping is incomplete. The fitting bias of the estimator is then the difference of the fitting biases of the single-level light filters at levels $L$ and $s-1$ (Appendix~\ref{sec:num-bias}).

\subsection{Variance of the coupled increments}\label{sec:theory-coupling}

Theorem~\ref{theo:bias} concerns a single level. We now turn to the variance of the multilevel increments. At level $l$ and time $p$, the light filter moves each particle $x_{p-1}$ by one draw from the fitted bridge $P_{\theta}(\cdot\,|\,x_{p-1})$ of Corollary~\ref{cor:closed}, with $\theta=\theta_p^{[l]}$. This draw is obtained by inverting one-dimensional distribution functions at uniform random variables $U_p$, drawn independently of the fitting draws (Appendix~\ref{app:sampler}), one set $U_p^i$ for each particle $i$. We write it as
\[
  x_p=T_p^{[l]}\big(x_{p-1},\theta_p^{[l]},U_p\big),
\]
where $T_p^{[l]}(x',\theta,u)$ is the point of $\M_p$ obtained from the features of $x'$ at level $l$, the parameter $\theta$ and the uniforms $u$. The map $T_p^{[l]}$ is explicit, and it depends on the level only through the features, that is, through the level-$l$ transition. The synchronous coupling uses the same $U_p^i$ at levels $l$ and $l-1$, and $X_0^{[l],i}=X_0^{[l-1],i}$. Then $U_p^i$ is independent of $(X_{p-1}^{[l],i},X_{p-1}^{[l-1],i},\theta_p^{[l]},\theta_p^{[l-1]})$. We drop the index $i$ when it plays no role.

The idea is simple. Two particles that start at the same point and are moved with the same random variables by nearby maps stay close. Here the maps of the two levels differ only through the fitted parameters and the time discretization, and both differences are small at fine levels. Condition \eqref{S1} below says that one step is Lipschitz in the particle and in the parameter, and condition \eqref{S2} that the fits at neighboring levels differ by a quantity of order $\Delta_l^{\beta/2}$. We fix a norm $|\cdot|$ on the parameters and a compact set $\Theta$ of parameters, on which the weights $\alpha_k$ are bounded away from zero and the eigenvalues of the covariances $S_{k,j}$ are bounded above and away from zero. We consider the following conditions: there exist a function $\ell\ge0$ with $\E\,\ell(U_p)^2\le\bar\ell^{\,2}<\infty$ and constants $c_S<\infty$ and $\beta>0$ such that for any $p\ge1$, $l\ge1$, $x,\tilde x$, $\theta,\tilde\theta\in\Theta$ and $u$,
\begingroup\tagsleft
\begin{gather}
  \big|T_p^{[l]}(x,\theta,u)-T_p^{[l-1]}(\tilde x,\tilde\theta,u)\big| \le \ell(u)\big(|x-\tilde x|+(1+|x|)\big(|\theta-\tilde\theta|+\Delta_l^{\beta/2}\big)\big), \label{S1}\tag{S1}\\
  \theta_p^{[l]},\theta_p^{[l-1]}\in\Theta\ \text{a.s.},\qquad \E\big|X_{p-1}^{[l]}\big|^4\le c_S,\qquad \E\big|\theta_p^{[l]}-\theta_p^{[l-1]}\big|^4 \le c_S\,\Delta_l^{2\beta}. \label{S2}\tag{S2}
\end{gather}\endgroup

\begin{proposition}\label{prop:coupling}
Assume \eqref{S1} and \eqref{S2}. Then for any $n\ge1$ there exists a constant $\mathsf C_n<\infty$, that only depends on $(\bar\ell,c_S)$ and $n$, such that for any $l\ge1$ and $i\le N$
\[
  \E\big|X_n^{[l],i}-X_n^{[l-1],i}\big|^2 \le \mathsf C_n\,\Delta_l^{\beta},
\]
and for any Lipschitz function $\varphi$ with constant $\|\varphi\|_{\mathrm{Lip}}$ we have
\[
  \mathrm{Var}\Big(\frac1N\sum_{i=1}^N\big\{\varphi(X_n^{[l],i})-\varphi(X_n^{[l-1],i})\big\}\Big) \le \|\varphi\|_{\mathrm{Lip}}^2\,\mathsf C_n\,\Delta_l^\beta.
\]
\end{proposition}

Each step contributes a term of order $\Delta_l^\beta$, and the constant $\mathsf C_n$ may grow geometrically in $n$. In words, the light filter inherits the rate $\beta$ of the coupling of the dynamics: the twins are moved by nearby Lipschitz maps at every step, and there is no resampling step at which they could be separated. Section~\ref{sec:theory-resampling} shows that this is what fails for particle filters with maximally coupled resampling, under a nondegeneracy condition.

\begin{proof}
Let $\mathcal E_p:=\E|X_p^{[l]}-X_p^{[l-1]}|^2$ and $Z:=1+|X_{p-1}^{[l]}|$. By \eqref{S1}, the inequality $(a+b+c)^2\le3(a^2+b^2+c^2)$ for $a,b,c\in\R$, a consequence of the Cauchy--Schwarz inequality, and the independence of $U_p$,
\[
  \mathcal E_p \le 3\bar\ell^{\,2}\big(\mathcal E_{p-1}+\E\big[Z^2|\theta_p^{[l]}-\theta_p^{[l-1]}|^2\big]+\Delta_l^\beta\,\E Z^2\big).
\]
By the Cauchy--Schwarz inequality and \eqref{S2}, with $\E Z^4\le8(1+c_S)$ and $\E Z^2\le2(1+c_S^{1/2})$, we have $\mathcal E_p\le3\bar\ell^{\,2}(\mathcal E_{p-1}+c_\star\,\Delta_l^\beta)$ with $c_\star:=(8(1+c_S)c_S)^{1/2}+2(1+c_S^{1/2})$. Since $\mathcal E_0=0$, we find $\mathcal E_n\le\mathsf C_n\Delta_l^\beta$ with $\mathsf C_n:=3\bar\ell^{\,2}c_\star\sum_{k=0}^{n-1}(3\bar\ell^{\,2})^k$. The variance of the increment is at most its second moment, which, by the Cauchy--Schwarz inequality, is at most the mean over $i$ of $\E\{\varphi(X_n^{[l],i})-\varphi(X_n^{[l-1],i})\}^2\le\|\varphi\|_{\mathrm{Lip}}^2\mathcal E_n$. This ends the proof of the proposition.
\end{proof}

We end this section with a discussion of the two conditions and of what the bound misses.

\paragraph{Condition \eqref{S1}.} Consider first $K=1$ on affine blocks, for which $T_p^{[l]}(x,\theta,u)=r+S\tilde m(x)+S^{1/2}\xi(u)$ with $\xi(u)$ Gaussian. Then \eqref{S1} holds on $\Theta$ with $\ell(u)$ of order $1+|\xi(u)|$ when the features $\tilde m$ are Lipschitz: the bounds on the eigenvalues of $S$ in $\Theta$ make $S^{1/2}$ Lipschitz in $S$. The factor $1+|x|$ comes from the term $(S-\tilde S)\tilde m(x)$, and it is why \eqref{S2} uses fourth moments. On spheres, the sampler of Appendix~\ref{app:sampler} is discontinuous where the reflection taking a fixed pole to the mean direction is, and on circles and for the angles on $\mathbb S^3$ where the mean resultant vector of the mixture, which places the cut of the angle grid, vanishes. These sets depend on $x$ and $\theta$, so that \eqref{S1}, stated for all $x$, fails on spheres, and Proposition~\ref{prop:coupling} does not cover the tests of Sections~\ref{sec:num-sphere} and~\ref{sec:num-switch}. With mixtures, the continuous sampler is continuous in $\theta$, and we expect \eqref{S1} to hold on compact sets $\Theta$ that avoid these sets, with a function $\ell$ that grows as the components separate, since the inverse distribution function then varies quickly in the region of low density between them. We do not prove it. Condition \eqref{S1} fails when a component index is drawn, since a small change of $\theta$ then moves a fraction of the particles by a distance of order one. This explains the rates below one of Section~\ref{sec:num-sphere}.

\paragraph{Condition \eqref{S2}.} Condition \eqref{S2} asks that the fits at neighboring levels be close. We prove it below in the linear Gaussian setting, for a fitting procedure adapted to the proof; we check it numerically for that procedure (Appendix~\ref{app:num-fg}), and for the Adam fits of our experiments the measured rates are consistent with it. With explicit level transitions, the level enters $T_p^{[l]}$ through the features. With the bridges of Section~\ref{sec:num-switch}, the reference kernel does not depend on the level, so the level enters only the fit, through the Girsanov weights of the fitting draws, and the whole level dependence sits in \eqref{S2}. With $K=1$ the weight $\alpha_1$ cancels from the loss, which is convex in the natural parameters $(S^{-1}r,S^{-1})$, and strictly convex on compact sets of them; it is not convex in $(r,S)$. A nondegenerate minimizer then depends smoothly on the draws used in the fit. Fits at neighboring levels, whose common draws differ only through the time discretization, should therefore differ by a quantity of the order of that difference, which is what \eqref{S2} asks. With mixtures, the loss is nearly flat in the parameters of a component that carries almost no weight, and fits at neighboring levels can then differ by much more than the time discretization. This is what happens in the isolated jumps of Section~\ref{sec:num-sphere}.

\paragraph{A setting where \eqref{S2} holds.} Consider the setting of the test of Section~\ref{sec:num-e4}:
\begingroup\tagsleft
\begin{gather}
  \parbox{0.86\textwidth}{$K=1$; $\M_p=\{x:A_px=y_p\}$ with $A_p$ of orthonormal rows and $\bar y:=\max_{p\le n}|y_p|$; $\pi_0$ is Gaussian; at each level, $F^{[l]}_p=F^{[l]}$ and $\Omega^{[l]}_p=\Omega^{[l]}$, with $\hat F_l:=(\Omega^{[l]})^{-1/2}F^{[l]}(\Omega^{[l]})^{1/2}$ symmetric and $\|\hat F_l\|\le\lambda_F<1$, the eigenvalues of $\Omega^{[l]}$ in a fixed compact subset of $]0,\infty[$, and $\|F^{[l]}-F^{[l-1]}\|+\|\Omega^{[l]}-\Omega^{[l-1]}\|\le c_F\Delta_l^{\beta/2}$.}\label{LG}\tag{LG}
\end{gather}\endgroup
For the Euler scheme of the Ornstein--Uhlenbeck test, \eqref{LG} holds with $\beta=2$ and $\lambda_F=e^{-1/2}$. We take $\Theta:=\{(r,S):|S^{-1}r|\le R,\ s_-I\preceq S\preceq s_+I\}$ with $s_+\lambda_F<1$, and fit as follows:
\begingroup\tagsleft
\begin{gather}
  \parbox{0.86\textwidth}{at each time $p$, the fit uses a fresh cloud of $N_{\mathrm f}$ new draws from $\pi_0$, moved through the steps $1,\dots,p-1$ with the parameters already fitted, with the same draws and the same variables at the two levels; it runs $\bar k$ steps of projected gradient descent on the loss \eqref{eq:loss} in the natural parameters $(S^{-1}r,S^{-1})$, from a common starting point in $\Theta$, with the step $1/\mathsf L^{[l]}$ of Appendix~\ref{app:proof-S2}, computed from level $l$ alone.}\label{FG}\tag{FG}
\end{gather}\endgroup

\begin{proposition}\label{prop:S2-gauss}
Assume \eqref{LG} and \eqref{FG}. Then \eqref{S1} holds, and there exist an absolute constant $c$ and some $t_0>0$, that only depends on $(\lambda_F,s_+,\pi_0)$ and on the bounds on $\Omega^{[l]}$, such that, if $N_{\mathrm f}\ge c\,n/t_0$, condition \eqref{S2} holds for $p\le n$ with a constant $c_S$ that depends neither on $l$, $N$ nor $\bar k$. In particular the conclusions of Proposition~\ref{prop:coupling} hold.
\end{proposition}
The proof is in Appendix~\ref{app:proof-S2}. It uses the convexity of the loss in the natural parameters, and a bound on the gap between two runs of projected gradient descent from a common start that holds at every iteration, so that no convergence of the fits is needed. The fresh cloud makes the fitting particles independent given the earlier fits, which bounds the inverse moments of their mean weight. The experiments of Section~\ref{sec:numerics} run Adam on the transported cloud instead; Appendix~\ref{app:num-fg} reruns the test of Section~\ref{sec:num-e4} with \eqref{FG}. The bound on the gap between two runs extends to stochastic gradient descent with common minibatches, but not to Adam, whose normalization of the step is not a contraction. The cloud must also be large: with $\lambda_F=e^{-1/2}$ and $s_+=1.6$, $t_0\approx0.025$. The proposition thus shows that \eqref{S2} can hold for the light filter in the model of Section~\ref{sec:num-e4}. Procedure \eqref{FG} with $N_{\mathrm f}=10^3$, below the unspecified threshold of the proposition, gives a gap between the fits whose fourth moment decreases like $\Delta_l^{4.0}$, as \eqref{S2} asks with $\beta=2$ (Appendix~\ref{app:num-fg}); for the Adam fits, which we did not measure, the rates are consistent with \eqref{S2}.

\paragraph{No factor $N^{-1}$.} The bound does not decrease with $N$. The proof bounds the variance by the second moment, which contains the square of the mean increment and cannot give a factor $N^{-1}$. A factor $N^{-1}$ would follow from the decomposition of the variance into the mean of the conditional variance given the fits and the variance of the conditional mean. If the fits were computed from an independent cloud, as in \eqref{FG}, the particles would be independent given the fits, and the first term would be of order $N^{-1}\Delta_l^\beta$; the second term is of the same order only if the fluctuations of $\theta^{[l]}-\theta^{[l-1]}$, which is shared by all particles, are of order $N^{-1/2}\Delta_l^{\beta/2}$. The coupling of \citep[Thm.~4.10]{zhao2026sequential}, which couples the two levels only at the final time, obtains this factor through sample splitting.

\subsection{Why maximally coupled resampling limits the rate}\label{sec:theory-resampling}

Proposition~\ref{prop:coupling} says that the light filter inherits the rate $\beta$ of the coupling of the dynamics. We now show that a particle filter with maximally coupled resampling does not: under a nondegeneracy condition, \eqref{ND} below, $N$ times the variance of its increments is bounded below by a quantity of order $\Delta_l$, so that the variance rate of its increments is at most $1$, for any $\beta$. This is the mechanism, described in the introduction, behind the rate $\beta\approx1$ of multilevel particle filters with maximally coupled resampling. We work with the coupling used as a benchmark in Section~\ref{sec:numerics}: coupled optimal-proposal particle filters with maximally coupled resampling \citep[\S3.2]{jasra2017multilevel} and common moves. At a time $n$, write $x^{[s],j}$, $j\le N$, for the particles of level $s\in\{l,l-1\}$ before resampling, and $W^{[s],j}$ for their normalized weights. Given the history $\mathcal G$ up to this point, the resampling draws $N$ index pairs $(I^{[l]}_i,I^{[l-1]}_i)$ independently. With probability $1-\mathrm{TV}$, where
\[
  \mathrm{TV}:=\frac12\sum_{j=1}^N\big|W^{[l],j}-W^{[l-1],j}\big|,
\]
the total variation distance $\|W^{[l]}-W^{[l-1]}\|_{\mathrm{tv}}$ between the two weight vectors, seen as probability laws on $\{1,\dots,N\}$, it draws one common index from the normalized minimum $W^{[l]}\wedge W^{[l-1]}$. Otherwise it draws two independent indices from the residuals
\[
  r^{[l]}:=\frac{W^{[l]}-W^{[l]}\wedge W^{[l-1]}}{\mathrm{TV}},\qquad r^{[l-1]}:=\frac{W^{[l-1]}-W^{[l]}\wedge W^{[l-1]}}{\mathrm{TV}} ;
\]
when $\mathrm{TV}=0$, the residuals are never used.
The particles are then moved with common variables $Z_i$, independent of the indices: $X^{[s],i}=a^{[s]}(x^{[s],I^{[s]}_i},Z_i)$. The estimate is the equal-weight average after the move, and the increment is $Y^{[l]}:=N^{-1}\sum_i\{\varphi(X^{[l],i})-\varphi(X^{[l-1],i})\}$. For a probability vector $r$ and a function $g$ of the index, $\mathrm{Var}_r(g)$ denotes the variance of $j\mapsto g(j)$ under $r$.

\begin{proposition}\label{prop:resampling-lower}
For any $N\ge1$ and any measurable $\varphi$ with $\E\varphi(X^{[l],1})^2+\E\varphi(X^{[l-1],1})^2<\infty$ we have
\[
  N\,\mathrm{Var}\big(Y^{[l]}\big)\ \ge\ \E\Big[\mathrm{TV}\,\Big\{\mathrm{Var}_{r^{[l]}}\big(\varphi(a^{[l]}(x^{[l],\cdot},Z))\big)+\mathrm{Var}_{r^{[l-1]}}\big(\varphi(a^{[l-1]}(x^{[l-1],\cdot},Z))\big)\Big\}\Big],
\]
where $Z$ is an independent copy of $Z_1$ and the variances are conditional on $\mathcal G$ and $Z$.
\end{proposition}

\begin{proof}
Let $D_i:=\varphi(X^{[l],i})-\varphi(X^{[l-1],i})$. Given $\mathcal G$, the triples $(I^{[l]}_i,I^{[l-1]}_i,Z_i)$ are i.i.d., so that $\mathrm{Var}(Y^{[l]}\,|\,\mathcal G)=N^{-1}\mathrm{Var}(D_1\,|\,\mathcal G)$, and $N\,\mathrm{Var}(Y^{[l]})\ge\E[\mathrm{Var}(D_1\,|\,\mathcal G)]\ge\E[\mathrm{Var}(D_1\,|\,\mathcal G,Z_1)]$. Given $(\mathcal G,Z_1)$, let $\xi$ indicate that the two indices are drawn from the residuals, an event of probability $\mathrm{TV}$. On this event they are independent, with laws $r^{[l]}$ and $r^{[l-1]}$, so that the conditional variance of $D_1$ is the sum of the two variances in the statement. The law of total variance over $\xi$ gives $\mathrm{Var}(D_1\,|\,\mathcal G,Z_1)\ge\mathrm{TV}\cdot\mathrm{Var}(D_1\,|\,\mathcal G,Z_1,\xi=1)$. Taking expectations ends the proof of the proposition.
\end{proof}

The bound uses only the last resampling, and no constant depends on $N$. A mismatch has probability $\mathrm{TV}$, and it costs a jump whose size is set by how far apart the two residual laws put their mass. If the log-weights of twins differ by a quantity of order $\Delta_l$ that varies across the cloud, then $\mathrm{TV}$ is of order $\Delta_l$ and the residuals do not shrink with $l$. We consider the condition: there exist $c_\star>0$ and $L_0$ such that, for any $l\ge L_0$,
\begingroup\tagsleft
\begin{gather}
  \liminf_{N\to\infty}\E\Big[\mathrm{TV}\,\Big\{\mathrm{Var}_{r^{[l]}}\big(\varphi(a^{[l]}(x^{[l],\cdot},Z))\big)+\mathrm{Var}_{r^{[l-1]}}\big(\varphi(a^{[l-1]}(x^{[l-1],\cdot},Z))\big)\Big\}\Big]\ \ge\ c_\star\,\Delta_l . \label{ND}\tag{ND}
\end{gather}\endgroup

\begin{corollary}\label{cor:resampling-lower}
Under \eqref{ND}, $\liminf_NN\,\mathrm{Var}(Y^{[l]})\ge c_\star\Delta_l$ for any $l\ge L_0$. In particular the variance rate of maximally coupled resampling is at most $1$ for large $N$, for any rate $\beta$ of the coupling of the dynamics.
\end{corollary}

With constant $\sigma$ the Euler scheme has $\beta=2$, so that the rate is halved, as the upper bound $\Delta_l^{\beta/2}$ of \citep{jasra2017multilevel} allows. That bound is for the bootstrap particle filter with a likelihood bounded above and below, so our model is outside its assumptions. The halving is also described in \citep{ballesio2020wasserstein} and \citep[\S5.3]{jasra2018clt}, as an upper bound and from numerical evidence; we are not aware of a proof of a lower bound. The next proposition proves \eqref{ND} at the first resampling in a linear Gaussian model, where both levels start from the same draws and only the weights differ.

\begin{proposition}\label{prop:resampling-gauss}
Let $dX_t=-BX_t\,dt+dW_t$ in $\R^d$, with $B$ symmetric positive definite, be discretized by the Euler scheme, with $X_0\sim\mathcal N(\mu_0,\Sigma_0)$, $\Sigma_0\succ0$, one scalar noiseless observation $y_1=a^\T X_1$, $|a|=1$, the optimal proposal and $\varphi(x)=c^\T x$. Let $Y^{[l]}$ be the increment after the first resampling and move, and let $p^{[s]}=\mathcal N(\mu^{[s]},\Sigma^{[s]})$ be the law of $X_0$ given $y_1$ at level $s$. Assume that
\begin{enumerate}
  \item[(i)] $(\mu',\Sigma')\neq(0,0)$, where $\mu^{[l-1]}-\mu^{[l]}=\Delta_l\mu'+o(\Delta_l)$ and $\Sigma^{[l-1]}-\Sigma^{[l]}=\Delta_l\Sigma'+o(\Delta_l)$;
  \item[(ii)] $c$ is not parallel to $a$, that is, $\varphi$ is not constant on $\M_1$.
\end{enumerate}
Then there exists $c_\star(y_1)>0$ such that
\[
  \liminf_{l\to\infty}\,\liminf_{N\to\infty}\,\frac{N\,\mathrm{Var}\big(Y^{[l]}\big)}{\Delta_l}\ \ge\ c_\star(y_1).
\]
\end{proposition}
The proof is in Appendix~\ref{app:proof-gauss}.

\begin{remark}\label{rem:weighted}
Proposition~\ref{prop:resampling-lower} takes the estimate after the resampling and the move. In the order of Section~\ref{sec:setting-pf}, the particles are moved, weighted and averaged before they are resampled, so that the estimate is a ratio and the last resampling that matters at time $n$ is the one at time $n-1$. The same argument then bounds $N\,\mathrm{Var}(Y_n^{[l]})$ from below by the mismatch term $\mathrm{TV}_{n-1}V_{\mathrm{mm}}$ of that resampling, with $V_{\mathrm{mm}}$ the residual variance of Proposition~\ref{prop:resampling-lower} for the weighted estimate, up to the remainder of the linearization of the ratio, which vanishes under a remainder condition \eqref{Rem} (Lemma~\ref{lem:resampling-weighted} in Appendix~\ref{app:proof-weighted}). The rate is then at most $1$ under \eqref{Rem} and the analog of \eqref{ND} for the mismatch term. This is the order of the switching-sensor test of Section~\ref{sec:num-switch}, where we check the mismatch term numerically but do not verify \eqref{Rem} for the Girsanov weights.
\end{remark}

\paragraph{Scope.} The bound concerns couplings of resampling that keep the twins together or redraw them independently from the residuals. It says nothing against couplings of transport type, which move the mass of one cloud onto the other: the sorting of \citep{ballesio2020wasserstein} in dimension one, the ensemble transform filters \citep{reich2013nonparametric,gregory2016multilevel}, and the light filter. It bounds $N$ times the variance of the particle filter's increment from below; the variance of the light filter's increment is not of this form, since its fits are shared by all particles (Section~\ref{sec:theory-coupling}).

\paragraph{Summary.} The two couplings compare as follows.
\begin{itemize}
  \item The light filter inherits the rate of the coupling of the dynamics: under \eqref{S1}--\eqref{S2}, the variance of its increments is at most of order $\Delta_l^\beta$ (Proposition~\ref{prop:coupling}).
  \item A particle filter with maximally coupled resampling does not: under \eqref{ND}, $N$ times the variance of its increments is at least of order $\Delta_l$ as $N\to\infty$, for any $\beta$ (Corollary~\ref{cor:resampling-lower}). Condition \eqref{ND} is proved at the first resampling in a linear Gaussian model (Proposition~\ref{prop:resampling-gauss}) and checked numerically on the examples of Sections~\ref{sec:num-e4} and~\ref{sec:num-sphere} (Table~\ref{tab:mismatch}). For the switching-sensor test, Lemma~\ref{lem:resampling-weighted} gives the same conclusion under \eqref{Rem}, which we do not verify, and the analog of \eqref{ND} for its mismatch term, which holds numerically: the measured mismatches are again of order $\Delta_l$ (Section~\ref{sec:num-switch}).
\end{itemize}
With constant $\sigma$, we expect $\beta=2$ on affine manifolds, and the two variance rates then differ by one order in $\Delta_l$; this is the difference that Figures~\ref{fig:rates} and~\ref{fig:manifolds}(a)--(b) show, with $\hat\beta\approx2$ for the light coupling and $\hat\beta\approx1$ for the maximal coupling, and on the switching-sensor test the maximal coupling stays at or below $1$ (Section~\ref{sec:num-switch}). The first bound rests on \eqref{S2}, which we prove only in the setting of Proposition~\ref{prop:S2-gauss}, and Proposition~\ref{prop:coupling} does not cover the spheres, where the light coupling reaches the same rate outside its assumptions. The two bounds also concern different scalings in $N$, as explained in the scope paragraph above.

\section{Numerical Results}\label{sec:numerics}

We now test the light filter and its multilevel coupling. Section~\ref{sec:num-e1} checks the fitted light filter against the exact filter in a linear Gaussian model. Sections~\ref{sec:num-e4} and~\ref{sec:num-cost} then ask, for linear observations and explicit level transitions, whether the synchronous coupling gives increment variances of order $\Delta_l^2$, at what cost, and with what bias. Sections~\ref{sec:num-sphere} and~\ref{sec:num-switch} ask whether this carries over to spheres and to manifolds that change type, with nonlinear dynamics. Section~\ref{sec:theory} gives the two outcomes to compare: the variance of the increments of the light coupling decreases like $\Delta_l^2$ if \eqref{S1}--\eqref{S2} hold with $\beta=2$ (Proposition~\ref{prop:coupling}), and that of the maximal coupling no faster than $\Delta_l$, under \eqref{ND} or, for the switching-sensor test, under \eqref{Rem} and the analog of \eqref{ND} for Lemma~\ref{lem:resampling-weighted} (Section~\ref{sec:theory-resampling}).

All computations use double precision. On affine manifolds we use the LightSB code of \citep{korotin2024light}, unchanged, with $\varepsilon=1$ in the whitened coordinate. The spherical and product families are our own implementation. The potential is initialized at $v_\theta\equiv1$ up to a small perturbation ($r_k\approx0$, $S_k=I$, $\alpha_k=1/K$), so that the fit starts from the moves of the optimal proposal without its weights (Lemma~\ref{lem:filter}(iv)). Unless stated otherwise, the loss \eqref{eq:loss} is estimated with $M=8$ weighted draws from $q_n$ per particle and minimized with Adam at learning rate $2\times10^{-2}$.

\subsection{A correctness check}\label{sec:num-e1}

We first check that the fitted light filter targets the right filter, in a case where the filter is known exactly. With $d_x=10$, $F_n(x)=0.9x$, $\Omega_n=I$, $n=1,\dots,20$, a new random $A_n$ at each time and $d_y\in\{1,5,9\}$, the filters are Gaussian, given by the Kalman filter. The potential $\bar v_n$ of Lemma~\ref{lem:filter} is log-quadratic and $K=1$ is exact. The light filter with $N=10^3$ and $K=1$ has an error at the level of the Monte Carlo error of the optimal-proposal particle filter with $N=10^3$. Its squared error of the mean is $2.0\times10^{-1}$, $3.9\times10^{-2}$ and $1.9\times10^{-3}$ for $d_y=1,5,9$, averaged over $n$ and ten data sets, with at most $500$ iterations, against $2.8\times10^{-1}$, $3.3\times10^{-2}$ and $1.5\times10^{-3}$ for the particle filter. Using $K=5$ only adds estimation error. At a single level the light filter is not competitive: it takes $3$ to $20$ seconds for $20$ steps, against $0.02$ to $0.04$ seconds for a particle filter with $N=10^4$. That particle filter is about $5$ to $9$ times more accurate. The experiments below are about the coupling of levels.

\subsection{Multilevel coupling}\label{sec:num-e4}

We now turn to the coupling of levels, first in a linear model where each level-$l$ filter is known exactly. The model is the Ornstein--Uhlenbeck diffusion
\[
  dX_t = -BX_t\,dt + dW_t \quad\text{in } \R^{d_x},
\]
with $B = Q\operatorname{diag}(\lambda_1,\dots,\lambda_{d_x})Q^\T$, $\lambda_i$ equally spaced in $[0.5,1.5]$ and $Q$ a random orthogonal matrix, started from $X_0\sim\mathcal N(0,I_{d_x})$ and observed without noise at times $n=1,\dots,10$ through $Y_n=A_nX_n$ with $d_y=1$. Either $A_n = d_x^{-1/2}(1,\dots,1)$ for all $n$, so that the manifolds $\M_n$ are parallel hyperplanes, or $A_n$ is a new random unit vector at each $n$, so that the manifolds rotate. The data are simulated with the Euler scheme at step $2^{-12}$. Level $l$ is the Euler scheme at step $\Delta_l=2^{-l}$ on each unit interval. Since the drift is linear, the level-$l$ transition over a unit interval is an explicit Gaussian kernel. The level-$l$ filter $\pi_n^{[l]}$ is therefore given by a Kalman filter, both particle methods can use the exact optimal proposal, and no bridge between observations is needed. The experiment thus isolates the coupling of the levels. The test function is $\varphi(x)=x_1$.

Three couplings of the filters at levels $l$ and $l-1$ are compared, all with $N=10^3$:
\begin{itemize}
  \item maximal coupling, for short: coupled optimal-proposal particle filters with maximally coupled resampling \citep{jasra2017multilevel} and common Gaussian moves;
  \item Wasserstein coupling: the same, with the resampling coupled through the sorted cumulative weights \citep{ballesio2020wasserstein}. Since the move depends on the ancestor only through $\tilde m_n$, the two clouds are sorted by $\tilde m_n$. This coupling requires $\M_n$ of dimension one, that is $d_x=2$;
  \item light coupling: the light filter at both levels, with $K=1$, without resampling. Both fits start from the same point and use common fitting draws, with $100$ iterations and a cosine decay of the learning rate to $10^{-3}$ of its initial value. The particles are moved with common uniform variables and common Gaussian variables.
\end{itemize}

\begin{figure}[ht]
\centering
\includegraphics[width=\textwidth]{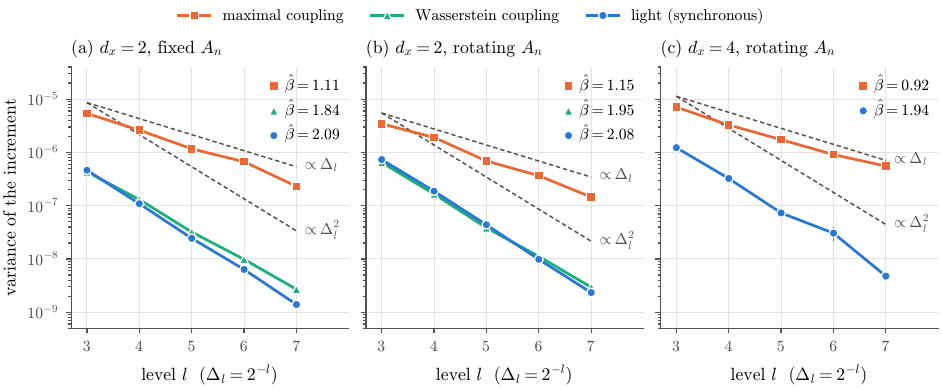}
\caption{Variance of the coupled increments against the level, averaged over $n=1,\dots,10$, with $N=10^3$ particles and $50$ independent runs per level: (a) $d_x=2$, fixed $A_n$; (b) $d_x=2$, rotating $A_n$; (c) $d_x=4$, rotating $A_n$. Bars are $\pm$ one standard error. The dashed lines have slopes $\Delta_l$ and $\Delta_l^2$.}\label{fig:rates}
\end{figure}

Figure~\ref{fig:rates} shows the variance over $50$ independent runs of the increment $\pi_n^{[l],N}(\varphi)-\pi_n^{[l-1],N}(\varphi)$, with $\pi_n^{[l],N}$ the level-$l$ filter with $N$ particles, averaged over $n$. Since $\Delta_l=2^{-l}$, the points of $\log_2V_l$ against $l$ lie near a line of slope $-\beta$, and we report the fitted rate $\hat\beta$, on levels $3$ to $7$. The maximal coupling gives $\hat\beta$ between $0.92$ and $1.15$. This is consistent with the bound of Section~\ref{sec:theory-resampling}, which concerns large $N$ and fine levels, while $\hat\beta$ is fitted at $N=10^3$ on levels $3$ to $7$; Table~\ref{tab:mismatch} checks the bound itself at $N=10^3$. The Wasserstein coupling recovers $\hat\beta\approx2$ when $\M_n$ is one-dimensional, and the light coupling matches it there, with a slightly smaller variance at the finest level. With $d_x=4$ and rotating manifolds, where no sorting is available, the light coupling keeps $\hat\beta=1.94$, against $0.92$ for the maximal coupling. This is the rate that Proposition~\ref{prop:coupling} gives if \eqref{S1}--\eqref{S2} hold with $\beta=2$; here $K=1$ on affine blocks, for which \eqref{S1} holds, and \eqref{S2} is proved for the fitting procedure of Proposition~\ref{prop:S2-gauss}, but not for the Adam fits used here (Section~\ref{sec:theory-coupling}). With that procedure and $N_{\mathrm f}=10^3$, the same test gives $\hat\beta=2.05\pm0.04$, against $1.94\pm0.07$ for the Adam fits and $0.92\pm0.06$ for the maximal coupling (bootstrap standard errors over runs), with a fitting bias not distinguishable from Monte Carlo error (Appendix~\ref{app:num-fg}). At $l=7$, the increment variance of the light coupling is about $120$ times smaller than that of the maximal coupling.

How the two fits are coupled matters. Starting both fits from the same initial parameter with the same fitting draws reaches $\hat\beta\approx2$ with $100$ iterations ($\hat\beta=2.09$ for $d_x=4$ with fixed $A_n$), whereas starting the fine fit from the converged coarse fit needs about $400$ iterations to reach it (Appendix~\ref{app:num-fits}).

\subsection{Cost against error}\label{sec:num-cost}

A faster decrease of the increment variances only pays if it outweighs the cost of the fits, which we now quantify. We follow the standard multilevel calculation \citep{giles2008multilevel}. For a tolerance $\epsilon_{\mathrm{rms}}$, the finest level $L$ is the smallest with squared bias at most $\epsilon_{\mathrm{rms}}^2/2$. The bias of each level is computed from the Kalman filters, and the limit $\Delta_l\to0$ is estimated by Richardson extrapolation from levels $6$ and $7$, that is, by $2\pi_n^{[7]}(\varphi)-\pi_n^{[6]}(\varphi)$, which removes the leading term of a bias proportional to $\Delta_l$. The sample sizes $N_l\propto\epsilon_{\mathrm{rms}}^{-2}(\mathrm{Var}_l/C_l)^{1/2}$, with $N_l\ge100$, make the variance at most $\epsilon_{\mathrm{rms}}^2/2$. Here $C_l$ is the cost per particle of a pair of filters and $\mathrm{Var}_l$ is $N$ times the variance of the increment measured above at $N=10^3$, extrapolated beyond $l=7$ with the fitted rate. The unit of cost is one Euler step for one particle. A pair costs $2^l+2^{l-1}$ per particle and time step, and a fit costs $100\times(M+1)\times K=900$ gradient evaluations per particle, each counted as one unit, with two fits per pair. The base level $l_0=2$ is a particle filter. Since a fit costs more than the Euler steps at coarse levels, we also consider the hybrid estimator of Section~\ref{sec:ml}, with the maximal coupling below a switch level and the light coupling from it. The switch level is chosen to minimize the cost.

\begin{table}[ht]
\centering
\begin{tabular}{rrrrrrr}
\toprule
$\epsilon_{\mathrm{rms}}$ & $L$ & single level & maximal & hybrid & gain & light, fits free\\
\midrule
$10^{-2}$ & 4 & $2.0\times10^{6}$ & $1.0\times10^{6}$ & $1.0\times10^{6}$ & 1.0 & $6.3\times10^{5}$\\
$10^{-3}$ & 8 & $3.3\times10^{9}$ & $2.5\times10^{8}$ & $1.7\times10^{8}$ & 1.4 & $7.5\times10^{7}$\\
$10^{-4}$ & 11 & $2.6\times10^{12}$ & $4.2\times10^{10}$ & $1.8\times10^{10}$ & 2.3 & $7.8\times10^{9}$\\
$10^{-5}$ & 14 & $2.1\times10^{15}$ & $6.6\times10^{12}$ & $1.8\times10^{12}$ & 3.6 & $7.9\times10^{11}$\\
\bottomrule
\end{tabular}
\vspace{0.5em}
\caption{Cost (in Euler steps of one particle) to reach a root-mean-square error $\epsilon_{\mathrm{rms}}$, $d_x=4$, rotating $A_n$. The hybrid uses the light coupling from level $6$ on for $\epsilon_{\mathrm{rms}}\le10^{-3}$. The gain is the ratio of the costs of the maximal-coupling and hybrid estimators. The last column ignores the cost of the fits. The variances are extrapolated beyond $l=7$ and assumed to scale like $N^{-1}$. For the light coupling they decrease more slowly in $N$, so the gains at small tolerances are optimistic. The fits are costed on the transported particles, whose number $N_l$ varies with the level, so that the estimator costed does not telescope exactly (Section~\ref{sec:ml}).}\label{tab:cost}
\end{table}

With $\gamma=1$, the maximal coupling has $\beta\approx\gamma$, the critical case in which every level contributes equally and the computing time is of order $\epsilon^{-2}|\log\epsilon|^2$, while the light coupling has $\beta\approx2>\gamma$, for which it is of order $\epsilon^{-2}$. With the fits costed and a particle filter at the base level, the light coupling at every level $l\ge3$ costs $4.9$ and $1.8$ times as much as the maximal coupling at $10^{-2}$ and $10^{-3}$, about the same ($1.1$) at $10^{-4}$, and $0.7$ times as much at $10^{-5}$. The hybrid is cheaper at small tolerances, by a factor $1.4$ at $10^{-3}$ and $3.6$ at $10^{-5}$ (at $10^{-2}$ the finest level is below the switch level).

Two caveats apply. First, counting one gradient evaluation as one Euler step is a convention: in wall-clock time, a pair of light filters takes $4.3$ seconds against a few milliseconds for a pair of particle filters. Second, the variances are assumed to scale like $N^{-1}$. The light increment variances decrease more slowly, like $N^{-0.7}$ at $l=5$ (Appendix~\ref{app:num-N}), so the gains at $10^{-4}$ and $10^{-5}$, which use many particles at fine levels, are optimistic.

Finally, the hybrid estimator inherits the difference $b_L-b_{s-1}$, where $b_l$ now denotes the fitting bias of the single-level light filter at level $l$, relative to the exact level-$l$ filter, at the finest level $L$ and at the level $s-1$ just below the switch level. A check against the Kalman filters (Appendix~\ref{sec:num-bias}) detects no bias with $100$ iterations and $M=8$, whereas $20$ iterations and $M=1$ bias the filter by $7$ to $9\times10^{-3}$ at every level without changing the rate. The check can only detect biases larger than about $2\times10^{-3}$, well above the tolerances $10^{-4}$ and $10^{-5}$ of Table~\ref{tab:cost}, so it does not establish that the bias is below those tolerances.

\subsection{Changing spheres}\label{sec:num-sphere}

We now leave affine manifolds for spheres, where the restricted kernel is of von Mises--Fisher type (Lemma~\ref{lem:factor}) and the center and radius of $\M_n$ are given by the data. At a single level with small noise, the light filter is less accurate than the optimal-proposal particle filter on spheres (Appendix~\ref{app:num-sphere}). We focus on the coupling of levels.

The state follows $dX_t=\lambda(x_\star-X_t)\,dt+dW_t$ in $\R^3$, $\lambda=0.5$, $x_\star=(2,0,0)$, $X_0\sim\mathcal N(x_\star,0.25\,I)$, observed through $Y_n=|X_n-o_n|^2$, $n=1,\dots,10$, with either $o_n=0$ or a beacon moving on a circle, $o_n=2(\cos0.7n,\sin0.7n,0)$. In the second case both the center and the radius of $\M_n$ change at every step. As in the affine test, the level-$l$ transition over a unit interval is an explicit Gaussian kernel, here isotropic, so that Lemma~\ref{lem:factor} applies at every level and the experiment isolates the coupling. The maximal coupling uses the von Mises--Fisher optimal proposal, sampled by inversion with common uniform variables. The light coupling uses $K=1$, $200$ iterations, $M=8$, both fits started from the same point with common fitting draws. Each level has a reference particle filter with $2\times10^5$ particles; $N=10^3$ and $50$ runs per level.

\begin{figure}[ht]
\centering
\includegraphics[width=\textwidth]{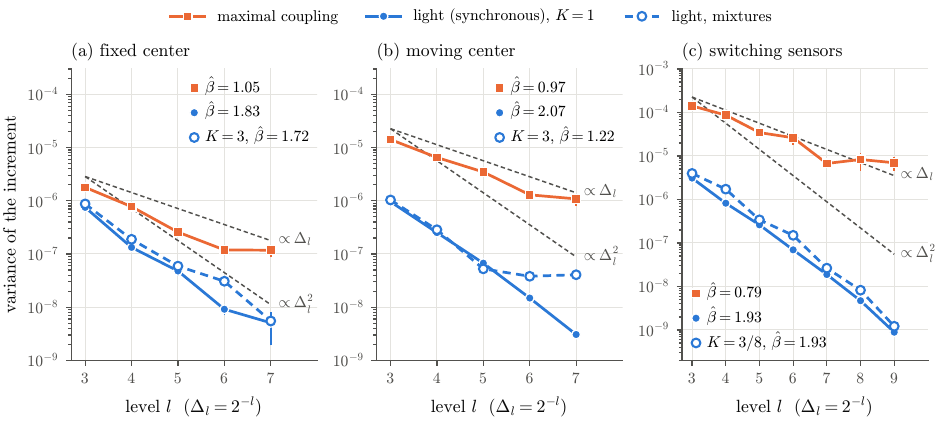}
\caption{Variance of the coupled increments on Riemannian manifolds, $N=10^3$, $50$ runs per level. (a), (b) Changing spheres with Ornstein--Uhlenbeck dynamics, fixed center and moving beacon; hollow markers: light coupling with $K=3$ components, continuous sampling and a common pilot fit. (c) Switching sensors with nonlinear dynamics (Section~\ref{sec:num-switch}), levels $3$ to $9$; hollow markers: $K=8$ at the steps on $\mathbb S^3$ with a pilot fit and short refits, $K=3$ elsewhere. The legends give $\hat\beta$ over all the levels shown.}\label{fig:manifolds}
\end{figure}

Figure~\ref{fig:manifolds}(a)--(b) gives $\hat\beta=1.83$ and $2.07$ for the light coupling against $1.05$ and $0.97$ for the maximal coupling. The rates of the maximal coupling are close to the rate $1$ that Section~\ref{sec:theory-resampling} allows for large $N$, and Table~\ref{tab:mismatch} checks the bound itself. Proposition~\ref{prop:coupling} does not cover the spheres, since \eqref{S1}, stated for all $x$, fails on the exceptional sets of $(x,\theta)$ where the sampler is discontinuous (Section~\ref{sec:theory-coupling}), but the light coupling still has the variance rate $\Delta_l^2$. At $l=7$ the light increments have $23$ and $350$ times smaller variance. With the fixed center, the decrease from $l=6$ to $l=7$ is within the Monte Carlo error. With $K=3$ components, a component index drawn for each particle and fits started from nearly identical components, the light coupling gives $\hat\beta=0.76$ and $0.74$, below those of the maximal coupling. This is because particles change component between the two levels: when a component index is drawn, a small change of $\theta$ moves a fraction of the particles by a distance of order one, so that \eqref{S1} fails, and the light coupling no longer inherits $\beta$ (Section~\ref{sec:theory-coupling}).

With the continuous sampler of Appendix~\ref{app:sampler} and a common pilot fit with $200$ refitting iterations at learning rate $2\times10^{-2}$, the rate is restored with the fixed center. There $\hat\beta=1.72$, with increment variances close to those of $K=1$ (hollow markers in Figure~\ref{fig:manifolds}(a)). With the moving center, $\hat\beta=1.22$ (Figure~\ref{fig:manifolds}(b)), because a few runs carry isolated large increments (Appendix~\ref{app:num-sphere}). These jumps are a failure of \eqref{S2}: at the step $n=4$, the fits put almost all the weight on one component, the parameters of the other components are then almost undetermined by the loss, and the fits at neighboring levels can move them far apart (Appendix~\ref{app:num-sphere}). Here the filters are unimodal and a single component already represents them well, so that mixtures are not needed on the spheres. They are needed in the next section.

\paragraph{Mismatches under maximal coupling.} We check \eqref{ND} on the affine test with rotating $A_n$ of Section~\ref{sec:num-e4} and on the two sphere tests, by computing, at every resampling of the maximal coupling, the expectation in Proposition~\ref{prop:resampling-lower}, averaged over $16$ independent copies of the move variables, with $N=10^3$, $\varphi(x)=x_1$ and $50$ runs per level. The coupled filters are those of Sections~\ref{sec:num-e4} and~\ref{sec:num-sphere}, run with the same code; the additional draws use a separate random stream, so that the filters themselves are unchanged. Table~\ref{tab:mismatch} gives the bound and $N$ times the measured variance of the increments, both averaged over $n$ and divided by $\Delta_l$.

\begin{table}[ht]
\centering
\small
\begin{tabular}{lcccccc}
\toprule
 & \multicolumn{2}{c}{affine, rotating $A_n$} & \multicolumn{2}{c}{spheres, fixed center} & \multicolumn{2}{c}{spheres, moving beacon}\\
\cmidrule(lr){2-3}\cmidrule(lr){4-5}\cmidrule(lr){6-7}
$l$ & $N\mathrm{Var}/\Delta_l$ & bound$/\Delta_l$ & $N\mathrm{Var}/\Delta_l$ & bound$/\Delta_l$ & $N\mathrm{Var}/\Delta_l$ & bound$/\Delta_l$\\
\midrule
3 & $5.7\times10^{-2}$ & $2.49\times10^{-2}$ & $1.4\times10^{-2}$ & $4.25\times10^{-3}$ & $1.1\times10^{-1}$ & $2.51\times10^{-2}$\\
4 & $5.2\times10^{-2}$ & $2.58\times10^{-2}$ & $1.2\times10^{-2}$ & $3.98\times10^{-3}$ & $1.0\times10^{-1}$ & $2.58\times10^{-2}$\\
5 & $5.6\times10^{-2}$ & $2.63\times10^{-2}$ & $8.2\times10^{-3}$ & $3.88\times10^{-3}$ & $1.1\times10^{-1}$ & $2.64\times10^{-2}$\\
6 & $5.8\times10^{-2}$ & $2.61\times10^{-2}$ & $7.6\times10^{-3}$ & $3.76\times10^{-3}$ & $8.3\times10^{-2}$ & $2.63\times10^{-2}$\\
7 & $7.1\times10^{-2}$ & $2.66\times10^{-2}$ & $1.5\times10^{-2}$ & $3.79\times10^{-3}$ & $1.4\times10^{-1}$ & $2.55\times10^{-2}$\\
\midrule
rate & $0.92$ & $0.98$ & $1.05$ & $1.04$ & $0.97$ & $0.99$\\
\bottomrule
\end{tabular}
\vspace{0.5em}
\caption{Maximal coupling: $N$ times the variance of the increments and the lower bound of Proposition~\ref{prop:resampling-lower}, both divided by $\Delta_l$ and averaged over $n=1,\dots,10$; rates fitted over levels $3$ to $7$. $N=10^3$, $50$ runs per level.}\label{tab:mismatch}
\end{table}

On the three tests the bound divided by $\Delta_l$ is flat in $l$, so that \eqref{ND} holds numerically, and it accounts for a fifth to a half of the measured variance: much of the variance of the maximal coupling comes from the mismatches of a single resampling step, as the introduction describes.

\subsection{Switching sensors: manifolds of changing type}\label{sec:num-switch}

The last test combines affine and spherical factors, with manifolds that change type at every step and nonlinear dynamics. It is a stress test, with sensors chosen so that the manifold takes each of the types of Figure~\ref{fig:sets}, rather than a model of a particular application. Two targets move in the plane, $x=(x_{(1)},x_{(2)})\in\R^4$, with $dX_t=\mu(X_t)\,dt+dW_t$ and
\[
  \mu(x)_{(i)} = \omega Jx_{(i)}-\zeta x_{(i)}+\zeta\rho_0\,\frac{x_{(i)}}{(|x_{(i)}|^2+1)^{1/2}},\qquad i\in\{1,2\},
\]
$J$ the rotation by $\pi/2$, $\zeta=1$, $\rho_0=2.5$, $\omega=1$: each target rotates and is attracted to a circle, with a nonlinear Lipschitz drift.

At time $n$ each target is observed either by a range sensor, $|x_{(i)}-o_i(n)|^2$ with a beacon $o_i(n)=3(\cos\vartheta_{i,n},\sin\vartheta_{i,n})$, $\vartheta_{i,n}=0.5(n-1)+\pi(i-1)$, or by a linear sensor at a fixed position $p_i$. The linear sensor reports that the target lies on a given line through $p_i$: $\langle e_{i,n},x_{(i)}-p_i\rangle=0$. The unit normal $e_{i,n}$ is known in advance. In the simulations, $e_{i,n}$ is generated from the simulated state and then treated as a design. This is not a bearing. Conditioning on a noiseless bearing angle, or on the unoriented line through $p_i$ given by the data, would add the non-constant co-area factor $|x_{(i)}-p_i|$, which is outside the setting of this paper. When the targets are not resolved, the total range $|x-(o_1(n),o_2(n))|^2$ is observed. The schedule $\mathrm{RR},\mathrm{RL},\mathrm{LL},\mathrm{U},\mathrm{LR}$, repeated over $n=1,\dots,10$, makes $\M_n$ a torus, a cylinder, a plane, a three-sphere and a cylinder.

\paragraph{Transition density.} The drift is nonlinear, so the transition density has no closed form, and a path simulated freely between two observations ends on $\M_n$ with probability zero. We use the standard representation of a transition density as an expectation over bridges of a simpler process \citep{beskos2006exact,heng2024unbiased}. Let $d\tilde X_t=B\tilde X_t\,dt+dW_t$ with $B=\omega J-0.3\,I$, a linear diffusion whose transition density over one interval is the Gaussian $\tilde f_n(z\,|\,x')=\mathcal N(z;\,e^{B}x',\,\tilde\Sigma)$, with $\tilde\Sigma:=\int_0^1e^{Bs}e^{B^\T s}\,ds=\frac{1-e^{-0.6}}{0.6}\,I$. It is isotropic, so that Lemma~\ref{lem:factor} and Corollary~\ref{cor:closed} apply in every mode. By Girsanov's theorem and conditioning on the endpoint \citep[Lemma~1]{beskos2006exact},
\begin{equation}\label{eq:girsanov}
  f_n(z\,|\,x') = \tilde f_n(z\,|\,x')\,\E\big[G(\tilde X)\,\big|\,\tilde X_0=x',\,\tilde X_1=z\big],
\end{equation}
where the expectation is over the diffusion bridge of $\tilde X$ from $x'$ to $z$, that is, $\tilde X$ conditioned to start at $x'$ and end at $z$, a Gaussian process that can be simulated exactly, and $G$ is the density of the law of $X$ with respect to that of $\tilde X$ on the paths over $[0,1]$ \citep[\S2.1]{heng2024unbiased}. Here $\mu(x)-Bx=\nabla A(x)$, with $A(x)=\sum_{i=1}^2\{(0.3-\zeta)|x_{(i)}|^2/2+\zeta\rho_0(|x_{(i)}|^2+1)^{1/2}\}$, the rotations cancelling, and Itô's formula leaves only time integrals in $G$ \citep[Cor.~1]{beskos2006exact}:
\begin{equation}\label{eq:girsanov-ito}
  \log G(X) = A(X_1)-A(X_0)-\int_0^1\Big\{\frac{|\mu(X_t)|^2-|BX_t|^2}{2}+\frac12\,\Delta A(X_t)\Big\}\,dt .
\end{equation}
At level $l$, the time integral is replaced by a sum on the grid of step $\Delta_l$, along the diffusion bridge simulated exactly on that grid. In the light filter, this only changes the fitting draws, in line~4 of Algorithm~\ref{alg:light}: a draw $z$ from the normalized restriction $\tilde q_n(\cdot\,|\,x')$ of $\tilde f_n(\cdot\,|\,x')$ to $\M_n$, followed by a diffusion bridge from $x'$ to $z$ and weighted by $\tilde w_n(x')\,G$, is a weighted draw of $\pi_n$, and the bridge is fitted with the reference $\pi_{n-1}(dx')\,\tilde f_n(x\,|\,x')$. The particle filters use the same weighted draws as proposals. At the two levels of a pair, the diffusion bridges are the same, the coarse level summing on every other point of the fine grid.

The maximal coupling uses common uniforms and bridge variables and maximally coupled resampling. The light coupling uses common fitting draws, bridge variables and uniforms, and all fits start from a pilot fit at the base level. The reference at each level is a particle filter with the same bridges and $10^5$ particles; $N=10^3$, and $50$ runs per level at the levels $3$ to $9$.

\begin{table}[ht]
\centering
\small
\setlength{\tabcolsep}{3.0pt}
\begin{tabular}{lcccccccccc}
\toprule
 & \multicolumn{2}{c}{$\hat\beta$} & & & \multicolumn{2}{c}{$l=3$} & \multicolumn{2}{c}{$l=5$} & \multicolumn{2}{c}{$l=7$}\\
\cmidrule(lr){2-3}\cmidrule(lr){6-7}\cmidrule(lr){8-9}\cmidrule(lr){10-11}
method & $3$--$9$ & $7$--$9$ & var.\ $l=9$ & s/pair & bias & $t_n^2$ & bias & $t_n^2$ & bias & $t_n^2$\\
\midrule
maximal coupling & 0.79 & $-0.03$ & $7.0\times10^{-6}$ & 2.0 & 7.9 & 3.60 & 6.7 & 0.78 & 5.9 & 0.97\\
light, $K=1$ & 1.93 & 2.20 & $8.9\times10^{-10}$ & 14 & 16.5 & 14.3 & 16.6 & 14.3 & 18.2 & 19.4\\
light, $K=3/8$, pilot & 1.93 & 2.22 & $1.2\times10^{-9}$ & 26 & 7.5 & 5.83 & 7.5 & 7.39 & 7.1 & 5.86\\
\bottomrule
\end{tabular}
\vspace{0.5em}
\caption{Switching-sensor test, $N=10^3$. Rate $\hat\beta$ of the coupled increments over levels $3$ to $9$ and over levels $7$ to $9$, their variance at $l=9$ and the time per pair of filters at $l=9$ ($50$ runs per level). Root-mean-square bias over $n=1,\dots,10$ (in units of $10^{-3}$) and mean of $t_n^2$ of single-level particle filters ($100$ runs, five particle filters with $2\times10^5$ particles as reference), where $t_n$ is the difference between the mean of the runs and the reference at time $n$, divided by its standard error, so that the mean of $t_n^2$ is close to one without bias. ``$K=3/8$'': $K=8$ and $M=8$ at the steps $\mathrm U$, with a pilot fit of $2000$ iterations and refits of $300$ iterations, and $K=3$, $200$ iterations and $M=4$ elsewhere.}\label{tab:switch-bias}
\end{table}

The light coupling gives the rate $\Delta_l^2$ on these manifolds of changing type (Table~\ref{tab:switch-bias} and Figure~\ref{fig:manifolds}(c)), the rate that Proposition~\ref{prop:coupling} gives with $\beta=2$, although the proposition does not cover this test, since \eqref{S1} fails on the circles and on $\mathbb S^3$ (Section~\ref{sec:theory-coupling}). With $K=1$, $\hat\beta=1.93$ over levels $3$ to $9$ and $2.20$ over levels $7$ to $9$, and the rate is between $1.9$ and $2.1$ at each of the ten time steps separately. The maximal coupling gives $\hat\beta=0.79$, and its increment variance stops decreasing from $l=7$ on, at about $7\times10^{-6}$. At $l=7$, before this plateau, the increment variance of the light coupling is about $350$ times smaller, and at $l=9$ about $8000$ times smaller. We have not identified the cause of the plateau of the maximal coupling. The mismatches of the resampling at $n-1$ are its leading term at $N=10^3$, but they do not account for the plateau. With the same runs, the expectation in Lemma~\ref{lem:resampling-weighted} without the remainder, $\E[\mathrm{TV}_{n-1}V_{\mathrm{mm}}]$, divided by $\Delta_l$ and averaged over $n=2,\dots,10$, is $0.78$, $0.76$, $0.74$, $0.92$, $0.80$, $1.14$ and $1.40$ at $l=3,\dots,9$ (sixteen copies of the move variables, the additional draws on a separate random stream; the filters are unchanged), so that the analog of \eqref{ND} holds numerically, with $\mathrm{TV}_{n-1}$ of order $\Delta_l$. It lies below $N$ times the measured variance at every level and accounts for $35$ to $83\%$ of it. From $l=7$ to $l=9$, however, it decreases by a factor of about $2.3$ while the variance stays flat. Its median over runs and times is flat, $0.22$ to $0.28$ in the same units, and its largest values carry a growing share of the mean at $l=8$ and $9$. Mixtures keep the rate: with $K=8$ on the steps $\mathrm U$ and $K=3$ elsewhere, $\hat\beta=1.93$ over levels $3$ to $9$, and the variance at $l=9$ is about $6000$ times smaller than with the maximal coupling.

The rate relies on two details of the construction. First, the weights must be computed in the form \eqref{eq:girsanov-ito}: with the stochastic integral of $G$, the light coupling gives only $\hat\beta\approx1$, and with a Brownian reference, $B=0$, the weights degenerate. Second, the angles on the circles and on $\mathbb S^3$ must be drawn on grids whose cut lies opposite the mean direction; with a fixed cut, $\hat\beta\approx1$ (Appendix~\ref{app:sampler}).

The fit, however, leaves a bias. With one component, the single-level light filter is biased by about $1.7\times10^{-2}$ at every level, with a mean $t_n^2$ of $14$ to $19$ (Table~\ref{tab:switch-bias}). The particle filter has a mean $t_n^2$ close to one at $l=5$ and $l=7$, so that its root-mean-square bias, about $6\times10^{-3}$, is the Monte Carlo resolution of the check. The bias of the light filter does not decrease with the level: it is a fitting error, which the multilevel identity does not remove. By time step, it is small up to the first step $\mathrm U$, where $\M_n$ is a three-sphere, and appears at the steps that follow, so that the family with one component misrepresents the filter on the three-sphere. With $K=8$ on the steps $\mathrm U$ and $K=3$ elsewhere, the bias falls to about $7\times10^{-3}$ at every level, close to the resolution of the check but still detected (mean $t_n^2$ of $6$ to $7$), and the rate is unchanged.

Each fit also needs $M$ bridges of $2^l$ steps per particle, which the cost model of Section~\ref{sec:num-cost} does not include.

\section{Discussion}\label{sec:plan}

At a single level, the light filter is about as accurate as the optimal-proposal particle filter with the same number of particles on affine manifolds, less accurate on spheres with small noise, and biased by its fits on manifolds that change type; it is always much more expensive. Its value lies in the coupling of levels. Because the transported particles carry equal weights, two levels can be driven by the same random variables at every step, with no resampling to couple. The increment variances then decrease like $\Delta_l^2$ instead of $\Delta_l$, on rotating affine manifolds and on spheres with a moving center, where couplings based on sorting do not apply. In an operation-count model with the fits costed, a hybrid estimator that uses the light coupling at the fine levels only is projected to be cheaper at small tolerances (Section~\ref{sec:num-cost}); no multilevel estimator was run to a tolerance.

The experiments also show what the coupling requires: fits started from a common point with common draws, sampling that is continuous in the parameters, and fits that are both accurate and stable across levels. On manifolds that change type, with a nonlinear drift, two further details matter: the Girsanov weight must contain only time integrals, and the angles must be drawn on grids whose cut lies opposite the mean direction. With both, one component and mixtures keep the rate $\Delta_l^2$. One component leaves a fitting bias, made at the steps on the three-sphere, and mixtures on those steps reduce it to close to the resolution of our check without losing the rate.

Our analysis explains the mechanism behind these observations, at the level of the populations: the bias is bounded through the excess losses of the fits, when the weights are bounded, and the rate follows from the continuity of the sampler and the stability of the fits, \eqref{S1}--\eqref{S2}, which on spheres hold at best away from the discontinuities of the sampler. It also explains why the maximal coupling of Section~\ref{sec:numerics} does not reach this rate: with maximally coupled resampling, the mismatches of the resampling bound $N$ times the variance of the increments from below by a quantity of order $\Delta_l$, for large $N$ and for any rate $\beta$ of the coupling of the dynamics, under the nondegeneracy condition \eqref{ND}, or, for the switching test, under \eqref{Rem} and the analog of \eqref{ND} for Lemma~\ref{lem:resampling-weighted}. The bound on the bias does not cover the switching test, where the weights are unbounded, and a localization argument is needed there. Several questions remain. Condition \eqref{ND} is proved only at the first resampling in a linear Gaussian model, and \eqref{Rem} is not verified for the Girsanov weights of the switching test, where the plateau of the variance of the maximal coupling is also not explained by the mismatches. Condition \eqref{S2} is proved only for $K=1$ in the linear Gaussian model, for projected gradient descent on a fresh cloud (Proposition~\ref{prop:S2-gauss}); it is not proved for Adam on the transported cloud, which our experiments use, and for mixtures the role of the pilot fit has to be understood. The complexity theorem of multilevel Monte Carlo needs a bound of order $N^{-1}\Delta_l^{\beta}$ on the variance of the increment with $N$ particles, whereas Proposition~\ref{prop:coupling} gives $\Delta_l^{\beta}$ without the factor $N^{-1}$. A bound for the light filter with $N$ particles, rather than for its population version, also needs a propagation of chaos estimate, as in \citep{zhao2026sequential}.

The construction extends in several directions. Nonlinear observations and diffusions with a state-dependent diffusion coefficient lead to curved manifolds, on which the restricted kernel has no closed form. Since the source point enters the bridge only through its features, each step only needs a tractable family on its target manifold, and the families of \citep{aramayo2026light} on tori, $\mathrm{SO}(3)$ and $\mathrm{SE}(3)$ suggest that such families exist beyond spheres. Low-noise observations $Y_n=A_nX_n+h\,\xi_n$ can be handled by lifting to an affine manifold in $(x,\xi)$, on which the kernel is again Gaussian \citep{zhumekenov2026}. The multilevel filter of \citep{zhao2026sequential} for noisy observations couples the two levels only at the final time, which does not need the fits to be stable across levels. Combining such a terminal correction with the synchronous coupling, where the latter degrades, remains to be studied. Finally, the fitted bridge has an explicit density, so that the fitting bias could be removed by weighting each transported particle by the ratio of the restricted dynamics to this density, at the price of weights that are no longer equal.

\appendix

\section*{Appendix}

\section{Notation}\label{app:notation}

\begin{center}
\small
\begin{tabular}{@{}p{0.2\textwidth}p{0.62\textwidth}l@{}}
\toprule
Symbol & Meaning & Where\\
\midrule
\multicolumn{3}{@{}l}{\emph{Model and filter}}\\
$X_t$, $X_n$ & hidden state in $\R^d$, at time $t$ and at observation time $n$ & \S\ref{sec:setting-filter}\\
$\mu$, $\sigma$ & drift and constant diffusion coefficient of the signal & \S\ref{sec:setting-filter}\\
$H_n$, $Y_n$, $y_n$ & observation map, noiseless observation, observed value & \S\ref{sec:setting-filter}\\
$\M_n$ & manifold $\{x:H_n(x)=y_n\}$ on which $X_n$ lies & \S\ref{sec:setting-manifolds}\\
$\lambda_{\M_n}$ & Lebesgue measure in $z$, or surface measure of the unit sphere in $u$, on $\M_n$ & \S\ref{sec:bridge}\\
$\pi_n$, $\pi_n(\varphi)$ & filter (law of $X_n$ given $y_{1:n}$) and its mean of $\varphi$ & \S\ref{sec:setting-filter}\\
$f_n(x\,|\,x')$ & transition density from time $n-1$ to time $n$ & \S\ref{sec:setting-filter}\\
$F_n$, $\Omega_n$ & mean map and covariance of a Gaussian transition & \S\ref{sec:fc}\\
$A_n$, $o_n$ & linear observation matrix, position of a range sensor & \S\ref{sec:fc}\\
\midrule
\multicolumn{3}{@{}l}{\emph{Levels}}\\
$l$, $L$, $\Delta_l$ & level, finest level, time step $2^{-l}$ & \S\ref{sec:setting-disc}\\
$f_n^{[l]}$, $\pi_n^{[l]}$ & level-$l$ transition density and filter & \S\ref{sec:setting-disc}\\
$\alpha$, $\beta$, $\gamma$ & rates of the bias, of the increment variances and of the cost & \S\ref{sec:setting-mlmc}\\
$N$ & number of particles & Alg.~\ref{alg:light}\\
\midrule
\multicolumn{3}{@{}l}{\emph{Bridge}}\\
$\rho$, $\eta$ & source and target laws of a bridge & \S\ref{sec:bridge}\\
$w_n(x')$ & mass of $f_n(\cdot\,|\,x')$ on $\M_n$, the incremental weight & \S\ref{sec:fc}\\
$q_n(\cdot\,|\,x')$ & $f_n(\cdot\,|\,x')$ restricted to $\M_n$ and normalized, the optimal proposal & \S\ref{sec:fc}\\
$\psi_n(x')$ & features of the source point: $\tilde m$ on affine pieces, $\kappa$ on spheres & \S\ref{sec:products}\\
$\mathrm{vMF}_d(\kappa)$, $Z_d$ & von Mises--Fisher law on $\mathbb S^{d-1}$ and its normalizer & \S\ref{sec:products}\\
$v_\theta$, $\theta$, $K$ & fitted (adjusted) potential, its parameters, number of mixture components & \S\ref{sec:fitting}\\
$\mathcal L(\theta)$ & loss of the fit & \S\ref{sec:fitting}\\
$P_\theta(\cdot\,|\,x')$ & conditional law of the fitted bridge, used to move a particle & \S\ref{sec:fitting}\\
$M$ & fitting draws per particle & Alg.~\ref{alg:light}\\
$\theta_n^{[l]}$, $\theta_n^0$ & fit at time $n$ and level $l$, common pilot fit & \S\ref{sec:ml}\\
\midrule
\multicolumn{3}{@{}l}{\emph{Analysis}}\\
$\Phi_n$, $\Phi_{p,n}$ & one step and several steps of the exact filter & \S\ref{sec:theory-bias}\\
$\widehat\pi_n$, $\delta_n$ & fitted filter and excess loss of the fit at time $n$ & \S\ref{sec:theory-bias}\\
$\mathrm D(P)$, $\operatorname{osc}(\varphi)$ & Dobrushin coefficient, oscillation & \S\ref{sec:theory}\\
$T_p^{[l]}$, $U_p$ & one step of the light filter at level $l$, as a map of the ancestor, the fit and the uniforms $U_p$ & \S\ref{sec:theory-coupling}\\
\eqref{S1}, \eqref{S2} & Lipschitz condition on one step, closeness of the fits at neighboring levels & \S\ref{sec:theory-coupling}\\
\eqref{LG}, \eqref{FG}, $N_{\mathrm f}$ & linear Gaussian setting, fits by projected gradient descent on a fresh cloud of size $N_{\mathrm f}$ & \S\ref{sec:theory-coupling}\\
$\mathrm{TV}$, $r^{[s]}$ & total variation between the weights of two levels, residuals of maximally coupled resampling & \S\ref{sec:theory-resampling}\\
\eqref{ND} & nondegeneracy of the mismatches & \S\ref{sec:theory-resampling}\\
\eqref{Rem}, $V_{\mathrm{mm}}$ & remainder of a ratio estimator, mismatch variance & App.~\ref{app:proof-weighted}\\
\bottomrule
\end{tabular}
\end{center}

\section{Bridges on Products of Affine Subspaces and Spheres}\label{app:products}

\subsection{Closed forms on products}\label{app:closed}

We split the coordinates of $x\in\R^d$ into disjoint blocks $x_{(j)}\in\R^{d_j}$, $j\le J$, each of one of two types. A block is either an affine block $j\in\mathcal A$, given by $a_j\in\R^{d_j}$ and $V_j\in\R^{d_j\times p_j}$ with orthonormal columns as above, or a spherical block $j\in\mathcal S$, given by a center $o_j\in\R^{d_j}$ and a radius $r_j>0$, with $d_j\ge2$. We consider the manifold
\[
  \M := \prod_{j\in\mathcal A}\big\{a_j+\operatorname{span}(V_j)\big\}\times\prod_{j\in\mathcal S}\big\{o_j+r_j\,\mathbb S^{d_j-1}\big\},
\]
with points $x_{(j)}=a_j+V_jz_j$, $z_j\in\R^{p_j}$, for $j\in\mathcal A$, and $x_{(j)}=o_j+r_ju_j$, $u_j\in\mathbb S^{d_j-1}$, for $j\in\mathcal S$. The measure $\lambda_\M$ is the product of Lebesgue measure in the coordinates $z_j$ and of the surface measures $\varsigma(du_j)$ of the unit spheres. It differs from the Riemannian measure by the constant factor $\prod_{j\in\mathcal S}r_j^{d_j-1}$.

We assume that $\Omega$ is block diagonal with respect to the blocks, with blocks $\Omega_{(j)}$, and isotropic on the spherical blocks, $\Omega_{(j)}=\tau_j^2I$ with $\tau_j>0$ for $j\in\mathcal S$. An anisotropic Gaussian kernel on a sphere is of Fisher--Bingham type, with no explicit normalizer. On an affine block, the computation of Section~\ref{sec:affine} applies, with $\Lambda_j:=V_j^\T\Omega_{(j)}^{-1}V_j$ and $\tilde z_j:=\Lambda_j^{1/2}z_j$. On a spherical block, $|u_j|=1$ removes the quadratic term in $u_j$, and the source and target interact through $\langle\kappa_j(x'),u_j\rangle$. We set
\[
  \tilde m_j(x') := \Lambda_j^{-1/2}V_j^\T\Omega_{(j)}^{-1}\big(m_{(j)}(x')-a_j\big),\quad j\in\mathcal A,\qquad
  \kappa_j(x') := \frac{r_j\,(m_{(j)}(x')-o_j)}{\tau_j^2},\quad j\in\mathcal S,
\]
and $w(x') := \prod_{j\in\mathcal A}w_j(x')\prod_{j\in\mathcal S}w_j(x')$ with $w_j$ as in Section~\ref{sec:affine} for $j\in\mathcal A$ and
\[
  w_j(x') := Z_{d_j}\big(|\kappa_j(x')|\big)\exp\Big\{-\frac{|m_{(j)}(x')-o_j|^2}{2\tau_j^2}\Big\},\quad j\in\mathcal S,
\]
where $Z_p(s) := \int_{\mathbb S^{p-1}} e^{s\,u_1}\,\varsigma(du) = (2\pi)^{p/2}s^{1-p/2}I_{p/2-1}(s)$ and $I_\nu$ is the modified Bessel function of the first kind. In particular, $Z_2(s)=2\pi I_0(s)$, $Z_3(s)=4\pi\sinh(s)/s$ and $Z_4(s)=4\pi^2I_1(s)/s$. We write $\mathrm{vMF}_p(du;\kappa) := e^{\langle\kappa,u\rangle}\varsigma(du)/Z_p(|\kappa|)$ for the von Mises--Fisher law on $\mathbb S^{p-1}$ with natural parameter $\kappa\in\R^p$. We write $\psi(x') := \big((\tilde m_j(x'))_{j\in\mathcal A},(\kappa_j(x'))_{j\in\mathcal S}\big)$ for the features of a source point.

\begin{lemma}\label{lem:factor}
Assume \eqref{A1} and \eqref{A3}. Then there exists a constant $c_\M'>0$ that only depends on $\Omega$, $(p_j)_{j\in\mathcal A}$ and $(r_j)_{j\in\mathcal S}$ such that for any $x'\in\R^d$ and any $x\in\M$ we have
\[
  f(x\,|\,x') = c_\M'\,w(x')\prod_{j\in\mathcal A}\mathcal N\big(\tilde z_j;\,\tilde m_j(x'),\,I\big)\prod_{j\in\mathcal S}\frac{e^{\langle\kappa_j(x'),u_j\rangle}}{Z_{d_j}(|\kappa_j(x')|)}.
\]
In particular, the restriction of $f(\cdot\,|\,x')$ to $\M$, normalized, is the law $q(\cdot\,|\,x')$ under which the coordinates are independent, $\tilde z_j\sim\mathcal N(\tilde m_j(x'),I)$ for $j\in\mathcal A$ and $u_j\sim\mathrm{vMF}_{d_j}(\kappa_j(x'))$ for $j\in\mathcal S$, and its total mass is $\int_\M f(x\,|\,x')\,\lambda_\M(dx)=c_\M w(x')$, with $c_\M:=c_\M'\prod_{j\in\mathcal A}\det(\Lambda_j)^{-1/2}$.
\end{lemma}

\begin{proof}
The Gaussian density factorizes over the blocks. On an affine block, this is the computation of Section~\ref{sec:affine}. On a spherical block, since $|r_ju_j|=r_j$,
\[
  |o_j+r_ju_j-m_{(j)}(x')|^2 = r_j^2-2r_j\langle m_{(j)}(x')-o_j,u_j\rangle+|m_{(j)}(x')-o_j|^2 .
\]
The factors $e^{-r_j^2/(2\tau_j^2)}$, the Gaussian normalization and the factors $(2\pi)^{p_j/2}$ form $c_\M'$. Since $\lambda_\M$ is Lebesgue measure in $z_j$ and $\tilde z_j=\Lambda_j^{1/2}z_j$, the Gaussian factor in $\tilde z_j$ integrates to $\det(\Lambda_j)^{-1/2}$, and the von Mises--Fisher factors integrate to one. This ends the proof of the lemma.
\end{proof}

On affine blocks the kernel is thus the Gibbs kernel of LightSB with unit regularization in the whitened coordinate. On spherical blocks it is the kernel $\exp\{\langle x',x\rangle/\varepsilon\}$ of \citep{aramayo2026light} on the sphere, with input $\kappa_j(x')$ and $\varepsilon=1$. As in Section~\ref{sec:affine}, the bridge only sees the source through its features.

\begin{proposition}\label{prop:reduction}
Assume \eqref{A1}, \eqref{A2} and \eqref{A3}, and that $\E_\rho|m(X')|^2<\infty$. Then the conditional law $P^\ast(\cdot\,|\,x')$ of Proposition~\ref{prop:bridge} depends on $x'$ only through $\psi(x')$. In the coordinates $(\tilde z,u)$ it coincides with the conditional law of the static Schr\"odinger bridge between the image measure $\psi_\#\rho$ and $\eta$ with Gibbs kernel
\[
  \exp\Big\{\sum_{j\in\mathcal A}\Big(\langle\tilde m_j,\tilde z_j\rangle-\frac12|\tilde z_j|^2\Big)+\sum_{j\in\mathcal S}\langle\kappa_j,u_j\rangle\Big\},
\]
that is, LightSB with unit regularization on the affine blocks and the bridge of \citep{aramayo2026light} with unit regularization on the spherical blocks.
\end{proposition}

\begin{proof}
By Lemma~\ref{lem:factor}, the factors $c_\M'\,w(x')$ and the normalizers of the factors of $q(\cdot\,|\,x')$, which depend on $x'$ only, cancel between $f(x\,|\,x')$ and $c^\ast(x')$ in Proposition~\ref{prop:bridge}. Hence $P^\ast(d\tilde z,du\,|\,x')$ is proportional to the Gibbs kernel above, evaluated at $\psi(x')$, times $v^\ast$, and depends on $x'$ only through $\psi(x')$. The image of $P^\ast$ under $(x',x)\mapsto(\psi(x'),x)$ is therefore a coupling of $\psi_\#\rho$ and $\eta$ whose density with respect to the reference $\tilde R$, the law $\psi_\#\rho$ times the Gibbs kernel above, is a product of a function of the source and a function of the target. The reduced problem with reference $\tilde R$ satisfies \eqref{A2}. Indeed, by Lemma~\ref{lem:factor}, $dR/d\tilde R(x',x)=h(x')$ with $h(x'):=c_h'\,w(x')\prod_{j\in\mathcal A}e^{-|\tilde m_j(x')|^2/2}\prod_{j\in\mathcal S}Z_{d_j}(|\kappa_j(x')|)^{-1}$ for a constant $c_h'>0$, and on each block $-c_h(1+|m(x')|^2)\le\log h(x')\le c_h$ for some constant $c_h$. Hence the relative entropies with respect to $R$ and $\tilde R$ of a coupling with first marginal $\rho$ differ by $\E_\rho[\log h(X')]$, which is finite when $\E_\rho|m(X')|^2<\infty$. In particular, the image of $P^\ast$ has finite relative entropy with respect to $\tilde R$, and its density is of product form. By the characterization of the Schr\"odinger bridge as the unique coupling of product form with finite relative entropy \citep{nutz2021introduction}, this image is the bridge between $\psi_\#\rho$ and $\eta$. This ends the proof of the proposition.
\end{proof}

As in LightSB, on the affine blocks we parametrize the adjusted potential, the potential times $e^{-|\tilde z_j|^2/2}$, so that the Gibbs kernel reduces to $\exp\{\sum_{\mathcal A}\langle\tilde m_j,\tilde z_j\rangle+\sum_{\mathcal S}\langle\kappa_j,u_j\rangle\}$. On the spherical blocks no adjustment is needed, since $|u_j|=1$. Take the family
\begin{equation}\label{eq:family}
  v_\theta(\tilde z,u) := \sum_{k=1}^K\alpha_k\prod_{j\in\mathcal A}\mathcal N\big(\tilde z_j;\,r_{k,j},\,S_{k,j}\big)\prod_{j\in\mathcal S}e^{\langle b_{k,j},u_j\rangle},
\end{equation}
with $\theta := (\alpha_k,(r_{k,j},S_{k,j})_{j\in\mathcal A},(b_{k,j})_{j\in\mathcal S})_{k\le K}$, $\alpha_k>0$, $r_{k,j}\in\R^{p_j}$, $S_{k,j}$ symmetric positive definite and $b_{k,j}\in\R^{d_j}$. For a single affine block, this is the family of Section~\ref{sec:affine}. Let
\[
  c_\theta(x') := \int\exp\Big\{\sum_{j\in\mathcal A}\langle\tilde m_j(x'),\tilde z_j\rangle+\sum_{j\in\mathcal S}\langle\kappa_j(x'),u_j\rangle\Big\}\,v_\theta(\tilde z,u)\,d\tilde z\,\varsigma(du),
\]
and let $P_\theta(d\tilde z,du\,|\,x')$ be the normalized integrand.

\begin{corollary}[{\citep{korotin2024light}, \citep[Prop.~2]{aramayo2026light}}]\label{cor:closed}
For the family \eqref{eq:family} and any $x'$ we have
\[
  c_\theta(x') = \sum_{k=1}^K\alpha_k\prod_{j\in\mathcal A}\exp\Big\{\tfrac12\tilde m_j(x')^\T S_{k,j}\tilde m_j(x')+\tilde m_j(x')^\T r_{k,j}\Big\}\prod_{j\in\mathcal S}Z_{d_j}\big(|b_{k,j}+\kappa_j(x')|\big),
\]
and $P_\theta(\cdot\,|\,x')$ is the mixture, with weights $\tilde\alpha_k(x')$ proportional to the terms of this sum, of the laws under which the coordinates are independent, with
\[
  \tilde z_j\sim\mathcal N\big(r_{k,j}+S_{k,j}\tilde m_j(x'),\,S_{k,j}\big),\quad j\in\mathcal A,\qquad u_j\sim\mathrm{vMF}_{d_j}\big(b_{k,j}+\kappa_j(x')\big),\quad j\in\mathcal S.
\]
\end{corollary}

\begin{proof}
These are the normalizer and the conditional law of LightSB on the affine blocks and of \citep{aramayo2026light} on the spherical blocks. They are combined by the closure under products of \citep[Prop.~2]{aramayo2026light} and evaluated at the features $\psi(x')$ instead of $x'/\varepsilon$. This ends the proof of the corollary.
\end{proof}

\subsection{The measure on the manifolds}\label{app:coarea}

For the observations of Section~\ref{sec:fc}, the conditional law of $X_n$ given $H_n(X_n)=y_n$ has a density with respect to the Riemannian measure of $\M_n$ proportional to the predicted density times the co-area factor $\det(\partial H_n\,\partial H_n^\T)^{-1/2}$. With $\mathcal A$ and $\mathcal S$ the sets of linear and range blocks, this factor is constant on $\M_n$, equal to
\begin{equation}\label{eq:coarea-sphere}
  \prod_{j\in\mathcal A}\det(A_{n,j}A_{n,j}^\T)^{-1/2}\prod_{j\in\mathcal S}\frac{1}{2r_{n,j}},\qquad r_{n,j}:=y_{n,j}^{1/2},
\end{equation}
and the Riemannian measure of $\M_n$ differs from $\lambda_{\M_n}$ by a constant factor. Hence the constant can be absorbed, and the filter has the density \eqref{eq:filter} with respect to $\lambda_{\M_n}$.

\section{Continuous Sampling from the Fitted Bridge}\label{app:sampler}

The coupling of Section~\ref{sec:ml} needs draws from the bridge of Corollary~\ref{cor:closed} that are continuous functions of the fitted parameters. For one component $\mathrm{vMF}_p(\kappa)$, with mean direction $\kappa/|\kappa|$, we sample by inversion: exactly on $\mathbb S^2$, through the explicit distribution function of $\langle u,\kappa/|\kappa|\rangle$, and on a fine grid on $\mathbb S^1$ and $\mathbb S^3$, with the distribution function interpolated linearly between the grid points. The draws are then continuous in the parameters and the uniform variables, as the coupling requires, except where the reflection taking a fixed pole to the mean direction is discontinuous, that is where the two coincide.

For $K\ge2$, drawing a component index with probabilities $\tilde\alpha_k(x')$ and then a draw from that component is a discontinuous function of $\theta$: a fraction of the particles of the order of $|\theta^{[l]}-\theta^{[l-1]}|$ changes component between the two levels. We therefore draw the coordinates one after another, each from its conditional law under $P_\theta(\cdot\,|\,x')$ given the coordinates already drawn. This conditional law is again a mixture of the same families, with the weights $\tilde\alpha_k(x')$ multiplied by the component densities of the coordinates already drawn. We draw it by inverting its one-dimensional distribution function: the angle on $\mathbb S^1$, the coordinates one after another on an affine factor, and on $\mathbb S^3$ the coordinates of Lemma~\ref{lem:hopf} below. Each draw is then a continuous function of $\theta$, of the features and of the uniform variables, except at the discontinuities of the reflection above and of the cut below. The normalizers and the densities stay explicit, and only the inversion is numerical.

On a circle, and for the two angles of the coordinates of $\mathbb S^3$ below, the inversion uses a grid of angles that starts and ends at a cut. We place the cut opposite the direction of the mean resultant vector of the mixture, $\sum_k\tilde\alpha_k(x')\,\mathrm{mean}_k$, with $\mathrm{mean}_k$ the mean resultant vector $\E_k[u]$ of component $k$, where the density is typically small. This direction is undefined, and the sampler discontinuous, where the mean resultant vector vanishes. With a fixed cut, say at $\pm\pi$, a component whose mode lies near the cut puts mass at both ends of the grid. A small change of the parameters then moves mass across the cut, and the particles whose uniform variable falls in the low-density region between the two ends jump around the circle. The fraction of such particles is proportional to the change of the parameters, so that the variance of the coupled increments decreases only like $\Delta_l$. In the switching test of Section~\ref{sec:num-switch}, a fixed cut gives $\hat\beta\approx1$, and the cut opposite the mean direction gives $\hat\beta\approx2$.

\begin{lemma}\label{lem:hopf}
Let $\kappa=(\kappa_{12},\kappa_{34})\in\R^2\times\R^2$ and let $u\sim\mathrm{vMF}_4(\kappa)$ on $\mathbb S^3$. Write $u=(\cos\chi\,(\cos t_1,\sin t_1),\,\sin\chi\,(\cos t_2,\sin t_2))$ with $\chi\in\,]0,\pi/2[$ and $t_1,t_2\in\,]-\pi,\pi]$. Then $\chi$ has a density proportional to
\[
  I_0\big(|\kappa_{12}|\cos\chi\big)\,I_0\big(|\kappa_{34}|\sin\chi\big)\cos\chi\,\sin\chi ,
\]
and, given $\chi$, the angles $t_1$ and $t_2$ are independent, with von Mises laws of natural parameters $\cos\chi\,\kappa_{12}$ and $\sin\chi\,\kappa_{34}$.
\end{lemma}

\begin{proof}
In these coordinates the surface measure of $\mathbb S^3$ is $\varsigma(du)=\cos\chi\sin\chi\,d\chi\,dt_1\,dt_2$, and $\langle\kappa,u\rangle=\cos\chi\,\langle\kappa_{12},(\cos t_1,\sin t_1)\rangle+\sin\chi\,\langle\kappa_{34},(\cos t_2,\sin t_2)\rangle$. Integrating $e^{\langle\kappa,u\rangle}$ in $t_1$ and $t_2$ gives $(2\pi)^2I_0(|\kappa_{12}|\cos\chi)\,I_0(|\kappa_{34}|\sin\chi)$, and the conditional law of $(t_1,t_2)$ given $\chi$ factorizes. This ends the proof of the lemma.
\end{proof}

For a mixture of von Mises--Fisher laws on $\mathbb S^3$, the law of $\chi$ is the mixture, with the weights $\tilde\alpha_k(x')$, of these densities normalized,
\[
  (2\pi)^2\,I_0\big(|\kappa_{k,12}|\cos\chi\big)\,I_0\big(|\kappa_{k,34}|\sin\chi\big)\cos\chi\,\sin\chi\,/\,Z_4(|\kappa_k|).
\] The weights are updated after $\chi$ and after $t_1$, so that the draw takes three one-dimensional inversions. On $\mathbb S^2$, with natural parameters $\kappa_k:=b_k+\kappa(x')$, the polar coordinate $t$ about a fixed axis has a density proportional to $\sum_k\alpha_k e^{\kappa_{k,3}t}I_0(|\kappa_{k,12}|(1-t^2)^{1/2})$, with the weights $\alpha_k$ of $\theta$, multiplied, in a product of blocks, by the normalizers of the blocks not yet drawn, and the azimuth given $t$ is a mixture of von Mises laws.

\section{\texorpdfstring{Proofs for Section~\ref{sec:theory}}{Proofs for Section 4}}\label{app:proofs}

\subsection{\texorpdfstring{The contraction estimate \eqref{eq:lipschitz}}{The contraction estimate}}\label{app:proof-lipschitz}
The proof follows Lemma~4.3.1 and Proposition~4.3.2 of \citep{delmoral2004feynman}, on the changing manifolds.

\begin{proof}[Proof of \eqref{eq:lipschitz}]
Write $\Gamma_G(\rho)(dx):=G(x)\rho(dx)/\rho(G)$, so that $\Phi_{p,n}(\rho)=\Gamma_{G_{p,n}}(\rho)P_{p,n}$. For a measurable set $A$ and $h:=G\,(1_A-\Gamma_G(\rho')(A))$, we have $\rho'(h)=0$ and $\Gamma_G(\rho)(A)-\Gamma_G(\rho')(A)=(\rho-\rho')(h)/\rho(G)$. Since $h$ takes its values in $[-c_A\sup G,(1-c_A)\sup G]$ with $c_A:=\Gamma_G(\rho')(A)\in[0,1]$, we have $\operatorname{osc}(h)\le\sup G$, hence
\[
  \|\Gamma_G(\rho)-\Gamma_G(\rho')\|_{\mathrm{tv}} \le \frac{\sup G}{\inf G}\,\|\rho-\rho'\|_{\mathrm{tv}}.
\]
The Markov kernel $P_{p,n}$ contracts the total variation by $\mathrm{D}(P_{p,n})$. Finally,
\[
  G_{p,n}(x)=w_{p+1}(x)\int q_{p+1}(dy\,|\,x)\,G_{p+1,n}(y),
\]
with $q_{p+1}(\cdot\,|\,x)$ carried by $\M_{p+1}$, so that $g_{p,n}\le(\overline w_{p+1}/\underline w_{p+1})\,g_{p+1,n}$ and $g_{n,n}=1$. This ends the proof of \eqref{eq:lipschitz}.
\end{proof}

\subsection{\texorpdfstring{Proof of Proposition~\ref{prop:resampling-gauss}}{Proof of the linear Gaussian lower bound}}\label{app:proof-gauss}

\begin{proof}[Proof of Proposition~\ref{prop:resampling-gauss}]
At level $s$, write $(F_s,\Omega_s)$ for the transition, $\sigma_s^2:=a^\T\Omega_sa$, $K_s:=\Omega_sa/\sigma_s^2$ and $G_s:=(I-K_sa^\T)F_s$. The weight is $w_s(x)\propto\sigma_s^{-1}\exp\{-\frac12(y_1-a^\T F_sx)^2/\sigma_s^2\}$, and the move is $G_sx+K_sy_1+\Xi_sZ$, with $Z\sim\mathcal N(0,I)$ common to the two levels and $\Xi_s$ a square root of the conditional covariance $\Omega_s-\Omega_saa^\T\Omega_s/\sigma_s^2$. Since $\varphi$ is linear, the terms that do not depend on the index drop out of each residual variance, and, with $A_j:=c^\T G_lx_j$, $B_j:=c^\T G_{l-1}x_j$ and $W^{[s],j}=w_s(x_j)/\sum_kw_s(x_k)$, Proposition~\ref{prop:resampling-lower} reads $N\,\mathrm{Var}(Y^{[l]})\ge\E\,\mathcal R_N$ with
\[
  \mathcal R_N:=\min_m\sum_j\big(W^{[l],j}-W^{[l-1],j}\big)_+(A_j-m)^2+\min_m\sum_j\big(W^{[l-1],j}-W^{[l],j}\big)_+(B_j-m)^2,
\]
since $\mathrm{TV}\,r^{[l],j}=(W^{[l],j}-W^{[l-1],j})_+$. The $x_j$ are i.i.d.\ from $\pi_0$, the weights are bounded and $t\mapsto t_+$ is $1$-Lipschitz, so that the strong law of large numbers gives $\mathcal R_N\to R_\infty(l)$ almost surely, where $R_\infty(l)$ is obtained by replacing the sums by integrals against $(p^{[l]}-p^{[l-1]})_+$ and $(p^{[l-1]}-p^{[l]})_+$. Since $\mathcal R_N\ge0$, Fatou's lemma gives $\liminf_N\E\,\mathcal R_N\ge R_\infty(l)$. The Euler matrices are smooth in the step, with $F_{l-1}-F_l$ and $\Omega_{l-1}-\Omega_l$ of order $\Delta_l$, so that $p^{[l-1]}=p^{[l]}\{1+\Delta_l\mathsf h+o(\Delta_l)\}$, where $\mathsf h$ is the derivative of the Gaussian log-density in the direction $(\mu',\Sigma')$, a polynomial of degree at most $2$ with mean zero. Hence $R_\infty(l)/\Delta_l\to c_\star(y_1):=\min_m\int p\,\mathsf h_-(g^\T x-m)^2dx+\min_m\int p\,\mathsf h_+(g^\T x-m)^2dx$, with $p$ the limit of $p^{[l]}$ and $g:=G^\T c$. Under (i), $\mathsf h\neq0$, so that $\mathsf h_+$ and $\mathsf h_-$ are positive on open sets. Since $g=F^\T(c-a\,K^\T c)$ with $F$ invertible and $K^\T a=1$, $g=0$ if and only if $c$ is parallel to $a$, so that under (ii) $x\mapsto g^\T x$ is not constant on any open set. Both minima are therefore positive. This ends the proof of the proposition.
\end{proof}

\subsection{The lower bound when the estimate precedes resampling}\label{app:proof-weighted}

We give the statement behind Remark~\ref{rem:weighted}. In Proposition~\ref{prop:resampling-lower}, the estimate is the equal-weight average taken after the resampling and the move. With the three parts of a step in the order of Section~\ref{sec:setting-pf}, the step runs differently: the resampled particles are moved with common variables $U$, then weighted, the estimate is the weighted average, and only then are the particles resampled. The last resampling that matters for the estimate at time $n$ is then the one at time $n-1$, and the estimate is a ratio. Write $\mathcal G$ for the history up to the weights at time $n-1$, with particles $x^{[s],j}$, weights $W^{[s],j}$, total variation $\mathrm{TV}_{n-1}$ and residuals $r^{[s]}$ as in Section~\ref{sec:theory-resampling}. For an ancestor $j$ and move variables $U$, write $Z^{[s]}(j,U)$ for the moved particle and $\omega_s(j,U)>0$ for its unnormalized weight at time $n$. The increment is
\[
  Y_n^{[l]}:=\frac{\sum_i\omega_l\,\varphi(Z^{[l]})}{\sum_i\omega_l}-\frac{\sum_i\omega_{l-1}\,\varphi(Z^{[l-1]})}{\sum_i\omega_{l-1}},
\]
with the terms evaluated at $(I^{[s]}_i,U_i)$. Let $b_s:=\E[\omega_s(I^{[s]},U)\,|\,\mathcal G]$, $\mu_s:=\E[\omega_s\varphi(Z^{[s]})\,|\,\mathcal G]/b_s$ and $d_s(j,U):=\omega_s(j,U)\{\varphi(Z^{[s]}(j,U))-\mu_s\}/b_s$. With $\bar A_s:=N^{-1}\sum_i\omega_s\varphi(Z^{[s]})$ and $\bar B_s:=N^{-1}\sum_i\omega_s$, let $R_N:=(\bar A_l-\mu_l\bar B_l)(\bar B_l^{-1}-b_l^{-1})-(\bar A_{l-1}-\mu_{l-1}\bar B_{l-1})(\bar B_{l-1}^{-1}-b_{l-1}^{-1})$ be the remainder of the linearization of the two ratios, and consider the condition
\begingroup\tagsleft
\begin{gather}
  \lim_{N\to\infty}N\,\E\big[R_N^2\big]=0\qquad\text{and}\qquad\sup_{N\ge1}\E\big[\mathrm{TV}_{n-1}V_{\mathrm{mm}}\big]<\infty, \label{Rem}\tag{R}
\end{gather}\endgroup
where
\[
  V_{\mathrm{mm}}:=\E_U\Big[\mathrm{Var}_{r^{[l]}}\big(d_l(\cdot,U)\big)+\mathrm{Var}_{r^{[l-1]}}\big(d_{l-1}(\cdot,U)\big)\Big].
\]

\begin{lemma}\label{lem:resampling-weighted}
For any $N\ge1$,
\[
  N\,\mathrm{Var}\big(Y_n^{[l]}\big)\ \ge\ \E\Big[\Big(\big(\mathrm{TV}_{n-1}V_{\mathrm{mm}}\big)^{1/2}-\big(N\,\E[R_N^2\,|\,\mathcal G]\big)^{1/2}\Big)_+^2\Big].
\]
Assume \eqref{Rem}. Then $\liminf_NN\,\mathrm{Var}(Y_n^{[l]})\ge\liminf_N\E[\mathrm{TV}_{n-1}V_{\mathrm{mm}}]$.
\end{lemma}

Condition \eqref{Rem} is the usual $O(N^{-1})$ bound on the remainder of a ratio estimator. It holds, for instance, when the weights are bounded and $\bar B_s^{-1}$ has bounded fourth moments. This is the order of the switching-sensor test of Section~\ref{sec:num-switch}, where $U$ contains the endpoint uniforms and the bridge increments; we do not verify \eqref{Rem} for its Girsanov weights.

\begin{proof}[Proof of Lemma~\ref{lem:resampling-weighted}]
Given $\mathcal G$, the triples $(I^{[l]}_i,I^{[l-1]}_i,U_i)$ are i.i.d. The identity $\bar A/\bar B-\mu=(\bar A-\mu\bar B)/\bar B$ gives $Y_n^{[l]}-(\mu_l-\mu_{l-1})=N^{-1}\sum_iD_i+R_N$, with $D_i:=d_l(I^{[l]}_i,U_i)-d_{l-1}(I^{[l-1]}_i,U_i)$, which has mean zero given $\mathcal G$. We have $\mathrm{Var}(Y_n^{[l]})\ge\E[\mathrm{Var}(Y_n^{[l]}\,|\,\mathcal G)]$, and, by the triangle inequality in $L^2$, $\mathrm{Var}(Y_n^{[l]}\,|\,\mathcal G)^{1/2}\ge\mathrm{Var}(N^{-1}\sum_iD_i\,|\,\mathcal G)^{1/2}-\E[R_N^2\,|\,\mathcal G]^{1/2}$. The variance $\mathrm{Var}(N^{-1}\sum_iD_i\,|\,\mathcal G)$ equals $\mathrm{Var}(D_1\,|\,\mathcal G)/N$, and, as in the proof of Proposition~\ref{prop:resampling-lower}, conditioning on $U_1$ and on the event that the indices are drawn from the residuals gives $\mathrm{Var}(D_1\,|\,\mathcal G)\ge\mathrm{TV}_{n-1}V_{\mathrm{mm}}$. This gives the first bound. For the second, $\E[(a^{1/2}-b^{1/2})_+^2]\ge\E a-2(\E a\,\E b)^{1/2}$ for nonnegative $a,b$, by the Cauchy--Schwarz inequality, and \eqref{Rem} gives $\E b=N\E[R_N^2]\to0$ with $\E a$ bounded. This ends the proof of the lemma.
\end{proof}

\subsection{\texorpdfstring{Proof of Proposition~\ref{prop:S2-gauss}}{Proof of the closeness of the fits in the linear Gaussian case}}\label{app:proof-S2}

Fix a time $p$ and a level $l$, and drop $p$ from the notation when it plays no role. The manifold $\M_p$ is affine of dimension $k$, and we work in its whitened coordinate $\tilde z\in\R^k$ (Section~\ref{sec:general}). With $K=1$, the adjusted potential is $v_\theta(\tilde z)=\mathcal N(\tilde z;r,S)$, and the weight $\alpha_1$ cancels. We use the natural parameters $h:=S^{-1}r$ and $\mathsf H:=S^{-1}$, write $\vartheta:=(h,\mathsf H)$, with the norm $|\vartheta|^2:=|h|^2+|\mathsf H|_F^2$, and let
\[
  \mathsf A(\zeta,\mathsf H):=\log\int_{\R^k}\exp\big\{\zeta^\T z-\tfrac12z^\T\mathsf Hz\big\}\,dz=\tfrac12\zeta^\T\mathsf H^{-1}\zeta-\tfrac12\log\det\mathsf H+\tfrac k2\log2\pi
\]
be the Gaussian log-partition function. In these parameters, $\Theta=\{|h|\le R\}\times\{\lambda_-I\preceq\mathsf H\preceq\lambda_+I\}$ with $\lambda_\pm:=s_\mp^{-1}$. It is convex, and its Euclidean projection $\Pi_\Theta$ projects $h$ on the ball and clips the eigenvalues of $\mathsf H$ to $[\lambda_-,\lambda_+]$. Constants $C$ may change from line to line; they do not depend on $l$, $N$, $N_{\mathrm f}$ or $\bar k$.

\begin{lemma}\label{lem:natural}
For $K=1$, the loss \eqref{eq:loss} is
\[
  \mathcal L(\vartheta)=\E_\rho\big[\mathsf A\big(h+\tilde m(X'),\mathsf H\big)\big]-\big\langle h,\E_\eta\tilde Z\big\rangle+\tfrac12\big\langle\mathsf H,\E_\eta\tilde Z\tilde Z^\T\big\rangle .
\]
It is convex in $\vartheta$.
\end{lemma}

\begin{proof}
With $r=Sh$ and $\tilde m=\tilde m(x')$, we have $\log c_\theta(x')=\tilde m^\T r+\tfrac12\tilde m^\T S\tilde m=\mathsf A(h+\tilde m,\mathsf H)-\tfrac12h^\T Sh+\tfrac12\log\det\mathsf H-\tfrac k2\log2\pi$. On the other side, $-\log v_\theta(z)=\tfrac12z^\T\mathsf Hz-h^\T z+\tfrac12h^\T Sh-\tfrac12\log\det\mathsf H+\tfrac k2\log2\pi$. Adding the two expressions, all the terms other than $\mathsf A$, $-h^\T z$ and $\tfrac12z^\T\mathsf Hz$ cancel. Since $\mathsf A$ is a log-partition function, it is jointly convex in $(\zeta,\mathsf H)$, and the other two terms are linear in $\vartheta$. This ends the proof of the lemma.
\end{proof}

\begin{lemma}\label{lem:hessian}
For any $(\zeta,\mathsf H)$ with $\lambda_-I\preceq\mathsf H\preceq\lambda_+I$, let $\bar z:=\mathsf H^{-1}\zeta$. Then for any direction $(b,B)$, with $B$ symmetric, we have
\[
  \frac{c_\Theta\,(|b|^2+|B|_F^2)}{1+4|\bar z|^2}\;\le\;D^2\mathsf A(\zeta,\mathsf H)[(b,B),(b,B)]\;\le\;C_\Theta\,(1+|\bar z|^2)\,(|b|^2+|B|_F^2),
\]
with $c_\Theta:=\min(s_-/4,s_-^2/2)$ and $C_\Theta:=2s_++s_+^2/2$.
\end{lemma}

\begin{proof}
$\mathsf A$ is the log-partition function of the statistic $(z,-\frac12zz^\T)$, so that, with $Z\sim\mathcal N(\bar z,S)$, $D^2\mathsf A[(b,B),(b,B)]=\mathrm{Var}(b^\T Z-\tfrac12Z^\T BZ)$. Writing $Z=\bar z+\varepsilon$, the variable equals a constant plus $(b-B\bar z)^\T\varepsilon-\frac12\varepsilon^\T B\varepsilon$. The linear and quadratic parts of a centered Gaussian vector are uncorrelated, and $\mathrm{Var}(\varepsilon^\T B\varepsilon)=2\operatorname{tr}(BSBS)$, so that
\[
  D^2\mathsf A[(b,B),(b,B)]=(b-B\bar z)^\T S(b-B\bar z)+\tfrac12\operatorname{tr}(BSBS),
\]
with $s_-^2|B|_F^2\le\operatorname{tr}(BSBS)\le s_+^2|B|_F^2$. For the upper bound, use $|b-B\bar z|^2\le2|b|^2+2|B|_F^2|\bar z|^2$. For the lower bound, if $|b|\le2|B|_F|\bar z|$, then $|b|^2+|B|_F^2\le(1+4|\bar z|^2)|B|_F^2$ and the second term alone gives the bound; otherwise $|b-B\bar z|\ge|b|/2$, and the sum is at least $\frac{s_-}4|b|^2+\frac{s_-^2}2|B|_F^2$. This ends the proof of the lemma.
\end{proof}

\begin{lemma}\label{lem:gd}
Let $f,g$ be $C^2$ on a neighborhood of $\Theta$, with $\mathsf mI\preceq\nabla^2f\preceq\mathsf LI$ on $\Theta$, $0<\mathsf m\le\mathsf L$. Let $\vartheta_0=\vartheta'_0\in\Theta$, and for $k\ge0$ let $\vartheta_{k+1}:=\Pi_\Theta(\vartheta_k-\mathsf s\nabla f(\vartheta_k))$ and $\vartheta'_{k+1}:=\Pi_\Theta(\vartheta'_k-\mathsf s'\nabla g(\vartheta'_k))$, with $\mathsf s:=1/\mathsf L$ and $\mathsf s'>0$. Then for any $k\ge0$ we have
\[
  |\vartheta_k-\vartheta'_k|\le\frac1{\mathsf m}\sup_\Theta|\nabla f-\nabla g|+\frac{\mathsf L}{\mathsf m}\,|\mathsf s-\mathsf s'|\,\sup_\Theta|\nabla g| .
\]
\end{lemma}

\begin{proof}
Let $e_k:=|\vartheta_k-\vartheta'_k|$. The projection is nonexpansive. Since $\Theta$ is convex, $\nabla f(\vartheta_k)-\nabla f(\vartheta'_k)=\bar H(\vartheta_k-\vartheta'_k)$, with $\bar H$ the average of $\nabla^2f$ on the segment between them, whose eigenvalues lie in $[\mathsf m,\mathsf L]$. Hence, with $\mathcal D:=\sup_\Theta|\nabla f-\nabla g|$ and $\mathcal B:=\sup_\Theta|\nabla g|$,
\[
  e_{k+1}\le\|I-\mathsf s\bar H\|\,e_k+\mathsf s\,|\nabla f(\vartheta'_k)-\nabla g(\vartheta'_k)|+|\mathsf s-\mathsf s'|\,|\nabla g(\vartheta'_k)|\le\Big(1-\frac{\mathsf m}{\mathsf L}\Big)e_k+\mathsf s\mathcal D+|\mathsf s-\mathsf s'|\,\mathcal B .
\]
Since $e_0=0$, induction gives $e_k\le(\mathsf L/\mathsf m)(\mathsf s\mathcal D+|\mathsf s-\mathsf s'|\mathcal B)$ for any $k$. This ends the proof of the lemma.
\end{proof}

The bound does not depend on the number of iterations, and it says nothing about convergence to a minimizer. The same proof applies to losses that change with $k$ and satisfy the bounds for every $k$, that is, to stochastic gradient descent with common minibatches.

\paragraph{The fits.} At time $p$, the fresh cloud of level $l$ consists of the particles $\check X^{[l],i}_{p-1}$, $i\le N_{\mathrm f}$, with features $\tilde m_i^{[l]}:=\tilde m^{[l]}_p(\check X^{[l],i}_{p-1})$ and weights $w_i^{[l]}:=w_p^{[l]}(\check X^{[l],i}_{p-1})$. With common Gaussian variables $\xi_{ij}\sim\mathcal N(0,I_k)$, $j\le M$, the fitting points are $\tilde z^{[l]}_{ij}=\tilde m_i^{[l]}+\xi_{ij}$ (Section~\ref{sec:general}). Let
\[
  \bar w^{[l]}:=\frac1{N_{\mathrm f}}\sum_iw_i^{[l]},\qquad \hat{\mathsf g}^{[l]}:=\frac{1}{N_{\mathrm f}\bar w^{[l]}}\sum_iw_i^{[l]}\,\bar g_i^{[l]},\qquad \bar g_i^{[l]}:=\frac1M\sum_j\big(\tilde z^{[l]}_{ij},\,\tilde z^{[l]}_{ij}\tilde z^{[l]\T}_{ij}\big),
\]
and let $\mathcal L^{[l]}(\vartheta):=N_{\mathrm f}^{-1}\sum_i\mathsf A(h+\tilde m^{[l]}_i,\mathsf H)-\langle h,\hat{\mathsf g}^{[l]}_1\rangle+\frac12\langle\mathsf H,\hat{\mathsf g}^{[l]}_2\rangle$ be the loss of Lemma~\ref{lem:natural} with $\rho$ the empirical law of the cloud and $\eta$ the weighted empirical law of the fitting points. On $\Theta$, $|\mathsf H^{-1}(h+\tilde m_i)|\le s_+(R+|\tilde m_i|)$. With
\[
  \mathsf F^{[l]}:=\frac1{N_{\mathrm f}}\sum_i|\tilde m_i^{[l]}|^2,\qquad \mathsf m^{[l]}:=\frac{c_\Theta}{1+8s_+^2(R^2+\mathsf F^{[l]})},\qquad \mathsf L^{[l]}:=C_\Theta\big(1+2s_+^2(R^2+\mathsf F^{[l]})\big),
\]
Lemma~\ref{lem:hessian} and Jensen's inequality for the convex function $t\mapsto1/(1+4t)$ give $\mathsf m^{[l]}I\preceq\nabla^2\mathcal L^{[l]}\preceq\mathsf L^{[l]}I$ on $\Theta$, and $\mathsf L^{[l]}/\mathsf m^{[l]}\le C(1+\mathsf F^{[l]})^2$. In \eqref{FG}, the step is $\mathsf s^{[l]}:=1/\mathsf L^{[l]}$, which is computed from level $l$ alone, so that the fit of level $l$ has the same law in the pairs $(l,l-1)$ and $(l+1,l)$.

\paragraph{The maps.} Under \eqref{LG}, write $\hat x:=\Omega_l^{-1/2}x$ with $\Omega_l:=\Omega^{[l]}$, let $O_p^{[l]}:=\Omega_l^{-1/2}V\Lambda^{-1/2}$, which has orthonormal columns, and let $x^{\star[l]}_p$ be the point of $\hat\M_p:=\Omega_l^{-1/2}\M_p$ closest to $0$, with $|x^{\star[l]}_p|\le r^\star:=\bar y\,\sup_l\lambda_{\min}(\Omega_l)^{-1/2}$. Taking $\Omega_l^{1/2}x^{\star[l]}_p$ as the base point of $\M_p$ in Section~\ref{sec:general}, we have $\tilde m^{[l]}_p(x)=O_p^{[l]\T}(\hat F_l\hat x-x^{\star[l]}_p)$, $\log w^{[l]}_p(x)=-\frac12\operatorname{dist}(\hat F_l\hat x,\hat\M_p)^2$, so that $w\le1$, and one step reads, in the coordinate $\hat x$,
\[
  \hat x_p=x^{\star[l]}_p+O^{[l]}_p\big\{S\big(h+O^{[l]\T}_p(\hat F_l\hat x_{p-1}-x^{\star[l]}_p)\big)+S^{1/2}\xi(u)\big\},
\]
with $\xi(u)\sim\mathcal N(0,I_k)$ when $u$ is uniform.

\begin{lemma}\label{lem:S1}
Assume \eqref{LG}. Then \eqref{S1} holds with $\ell(u)=C(1+|\xi(u)|)$, and for any $x,\tilde x$,
\begin{gather}
  |\tilde m^{[l]}_p(x)|\le C(1+|x|),\qquad |\tilde m^{[l]}_p(x)-\tilde m^{[l-1]}_p(\tilde x)|\le C\big(|x-\tilde x|+(1+|x|)\Delta_l^{\beta/2}\big),\label{G1}\tag{G1}\\
  |\log w^{[l]}_p(x)-\log w^{[l-1]}_p(\tilde x)|\le C(1+|x|+|\tilde x|)\big(|x-\tilde x|+(1+|x|)\Delta_l^{\beta/2}\big).\notag
\end{gather}
\end{lemma}

\begin{proof}
Every matrix and vector in the expression of one step, written back in the coordinate $x$, is a smooth function of $(F^{[l]},\Omega^{[l]})$ on the compact set of \eqref{LG}, so that they differ between levels by at most $C\Delta_l^{\beta/2}$. On $\Theta$, the maps $\vartheta\mapsto S$, $S^{1/2}$ and $Sh$ are Lipschitz, since the eigenvalues of $S$ are bounded away from $0$. The expression is affine in $x$ and in $\xi$, so that $|T^{[l]}_p(x,\theta,u)-T^{[l-1]}_p(\tilde x,\tilde\theta,u)|\le C(|x-\tilde x|+(1+|x|+|\xi(u)|)(|\theta-\tilde\theta|+\Delta_l^{\beta/2}))$, which is \eqref{S1} with $\ell(u)=C(1+|\xi(u)|)$, since $1+|x|+|\xi|\le(1+|x|)(1+|\xi|)$. The same expansion gives \eqref{G1}, since $\tilde m$ is affine and $\log w$ is quadratic in $x$, with coefficients smooth in $(F^{[l]},\Omega^{[l]})$. This ends the proof of the lemma.
\end{proof}

\begin{lemma}\label{lem:gauss}
Assume \eqref{LG}, and let $\theta_1,\theta_2,\dots\in\Theta$ be fixed. A particle drawn from $\pi_0$ and moved by the light filter of level $l$ with these parameters has, at time $p$, a Gaussian law $\mathcal N(\mu_p,\Sigma_p)$ in the coordinate $\hat x$, with
\[
  |\mu_p|\le\bar\mu:=|\hat\mu_0|\vee\frac{(1+s_+)r^\star+s_+R}{1-s_+\lambda_F},\qquad \|\Sigma_p\|\le\bar\Sigma:=\|\hat\Sigma_0\|\vee\frac{s_+}{1-(s_+\lambda_F)^2},
\]
where $(\hat\mu_0,\hat\Sigma_0)$ are the mean and covariance of $\pi_0$ in the coordinate $\hat x$, bounded uniformly in $l$. In particular, for any $q$, $\sup_l\E|\check X^{[l],i}_p|^q<\infty$, and the same holds for the transported particles.
\end{lemma}

\begin{proof}
The step is affine in $\hat x_{p-1}$, with linear part $\mathsf G_p:=O_pSO_p^\T\hat F_l$ of norm at most $s_+\lambda_F$, and with Gaussian noise of covariance $O_pSO_p^\T\preceq s_+I$. Hence $\mu_p=\mathsf G_p\mu_{p-1}+\mathsf c_p$ with $|\mathsf c_p|\le(1+s_+)r^\star+s_+R$, and $\Sigma_p=\mathsf G_p\Sigma_{p-1}\mathsf G_p^\T+O_pSO_p^\T$, so that $\|\Sigma_p\|\le(s_+\lambda_F)^2\|\Sigma_{p-1}\|+s_+$. Both recursions stay below the stated bounds by induction. For random parameters in $\Theta$, the conditional law given the parameters satisfies the same bounds, which gives the moments. This ends the proof of the lemma.
\end{proof}

\begin{lemma}\label{lem:G3}
Assume \eqref{LG} and \eqref{FG}, and let $t_0:=(4\lambda_F^2\bar\Sigma)^{-1}$. Then for any $q\le t_0N_{\mathrm f}$, $l\ge0$ and $p\ge1$,
\begin{equation}\label{G2}\tag{G2}
  \E\big[(\bar w^{[l]}_p)^{-q}\big]\le\exp\big\{q\,\big(r^{\star2}+2d\,\lambda_F^2\bar\Sigma+2\lambda_F^2\bar\mu^2\big)\big\}.
\end{equation}
\end{lemma}

\begin{proof}
We have $\phi:=-\log w_p=\operatorname{dist}(\hat F_l\hat x,\hat\M_p)^2/2\le\lambda_F^2|\hat x|^2+r^{\star2}$. By Jensen's inequality, $\bar w\ge\exp\{-N_{\mathrm f}^{-1}\sum_i\phi(\check X^i)\}$. Given $\theta_{1:p-1}$, the fresh particles are i.i.d., so that, with $t:=q/N_{\mathrm f}\le t_0$, $\E[\bar w^{-q}\,|\,\theta_{1:p-1}]\le(\E[e^{t\phi(\check X)}\,|\,\theta_{1:p-1}])^{N_{\mathrm f}}$. For $\hat X\sim\mathcal N(\mu,\Sigma)$ and $2t'\|\Sigma\|\le\frac12$, $\E e^{t'|\hat X|^2}\le(1-2t'\|\Sigma\|)^{-d/2}e^{2t'|\mu|^2}$. Apply this with $t'=t\lambda_F^2$, use Lemma~\ref{lem:gauss} and $(1-u)^{-1}\le e^{2u}$ for $u\le\frac12$, and raise to the power $N_{\mathrm f}$. The bound is uniform in $\theta_{1:p-1}\in\Theta$. This ends the proof of the lemma.
\end{proof}

\begin{proof}[Proof of Proposition~\ref{prop:S2-gauss}]
Condition \eqref{S1} holds by Lemma~\ref{lem:S1}, the fits lie in $\Theta$ since they are projected, and the moments of $X^{[l]}_{p-1}$ are bounded by Lemma~\ref{lem:gauss}. It remains to bound the gap between the fits. We prove, by induction on $p$, the statement $(\mathrm I_p)$: for any $q\in[1,4\mathsf p^{3(n-p)}]$, $\E|\check X^{[l],i}_p-\check X^{[l-1],i}_p|^q+\E|X^{[l],i}_p-X^{[l-1],i}_p|^q\le C_{p,q}\Delta_l^{q\beta/2}$, together with $\E|\vartheta^{[l]}_p-\vartheta^{[l-1]}_p|^q\le C_{p,q}\Delta_l^{q\beta/2}$ for any $q\in[1,4\mathsf p^{3(n-p)+1}]$. Since $\check X^{[l]}_0=\check X^{[l-1]}_0$ and $X^{[l]}_0=X^{[l-1]}_0$, $(\mathrm I_0)$ holds.

Fix $p\le n$ and assume $(\mathrm I_{p-1})$. Write $e_i:=|\check X^{[l],i}_{p-1}-\check X^{[l-1],i}_{p-1}|$ and $\epsilon_i:=e_i+(1+|\check X^{[l],i}_{p-1}|)\Delta_l^{\beta/2}$, and let $\mathcal P_i$ be a generic polynomial in $|\check X^{[l],i}_{p-1}|$, $|\check X^{[l-1],i}_{p-1}|$ and $M^{-1}\sum_j|\xi_{ij}|$, which may change from line to line. By Lemma~\ref{lem:gauss}, $\E\mathcal P_1^q<\infty$ for any $q$, and since the fresh particles are exchangeable, $\E[(N_{\mathrm f}^{-1}\sum_ia_i)^q]\le\E a_1^q$ for exchangeable $a_i\ge0$.

\emph{Gradient gap.} We have $\nabla_\zeta\mathsf A=\mathsf H^{-1}\zeta$ and $\nabla_{\mathsf H}\mathsf A=-\frac12(S+S\zeta\zeta^\T S)$, so that, uniformly on $\Theta$, $|\nabla\mathsf A(h+\tilde m,\mathsf H)-\nabla\mathsf A(h+\tilde m',\mathsf H)|\le C(1+|\tilde m|+|\tilde m'|)|\tilde m-\tilde m'|$. Since the fitting points share the $\xi_{ij}$, $\tilde z^{[l]}_{ij}-\tilde z^{[l-1]}_{ij}=\tilde m^{[l]}_i-\tilde m^{[l-1]}_i$, so that $|\bar g^{[l]}_i-\bar g^{[l-1]}_i|\le\mathcal P_i\epsilon_i$ by \eqref{G1}. Since the weights lie in $]0,1]$, $|w_i^{[l]}-w_i^{[l-1]}|\le|\log w_i^{[l]}-\log w_i^{[l-1]}|\le\mathcal P_i\epsilon_i$. Writing
\[
  \hat{\mathsf g}^{[l]}-\hat{\mathsf g}^{[l-1]}=\frac{N_{\mathrm f}^{-1}\sum_i\big(w_i^{[l]}\bar g_i^{[l]}-w_i^{[l-1]}\bar g_i^{[l-1]}\big)}{\bar w^{[l]}}+\hat{\mathsf g}^{[l-1]}\,\frac{\bar w^{[l-1]}-\bar w^{[l]}}{\bar w^{[l]}},
\]
with $|\hat{\mathsf g}^{[l-1]}|\le N_{\mathrm f}^{-1}\sum_i|\bar g_i^{[l-1]}|/\bar w^{[l-1]}$,
we obtain
\[
  \mathcal D:=\sup_\Theta|\nabla\mathcal L^{[l]}-\nabla\mathcal L^{[l-1]}|\le\mathcal Y\cdot\frac1{N_{\mathrm f}}\sum_i\mathcal P_i\epsilon_i,\quad \mathcal Y:=C\Big(1+\frac1{\bar w^{[l]}}\Big)\Big(1+\frac1{\bar w^{[l-1]}}\Big)\Big(1+\frac1{N_{\mathrm f}}\sum_i\mathcal P_i\Big).
\]

\emph{Steps.} Since $\mathsf L^{[l]}\ge C_\Theta$, $|\mathsf s^{[l]}-\mathsf s^{[l-1]}|\le C|\mathsf F^{[l]}-\mathsf F^{[l-1]}|\le N_{\mathrm f}^{-1}\sum_i\mathcal P_i\epsilon_i$. Moreover $\sup_\Theta|\nabla\mathcal L^{[l-1]}|\le C(1+\mathsf F^{[l-1]})+|\hat{\mathsf g}^{[l-1]}|$.

\emph{Fits.} By Lemma~\ref{lem:gd} with $f=\mathcal L^{[l]}$ and $g=\mathcal L^{[l-1]}$, $|\vartheta^{[l]}_p-\vartheta^{[l-1]}_p|\le\mathcal Y'\cdot N_{\mathrm f}^{-1}\sum_i\mathcal P_i\epsilon_i$, with $\mathcal Y'$ a product of $\mathcal Y$, of $1/\mathsf m^{[l]}$ and of polynomials in $\mathsf F^{[l]}$, $\mathsf F^{[l-1]}$ and $N_{\mathrm f}^{-1}\sum_i\mathcal P_i$. By \eqref{G2}, Lemma~\ref{lem:gauss} and H\"older's inequality, $\E\mathcal Y'^q<\infty$ as long as the order used in \eqref{G2} is at most $t_0N_{\mathrm f}$. Fix $\mathsf p:=1+1/(3n)$ and its conjugate exponent $\mathsf p':=1+3n$. By H\"older's inequality, twice, exchangeability and $(\mathrm I_{p-1})$,
\[
  \E|\vartheta^{[l]}_p-\vartheta^{[l-1]}_p|^q\le(\E\mathcal Y'^{q\mathsf p'})^{1/\mathsf p'}\big(\E\mathcal P_1^{q\mathsf p\mathsf p'}\big)^{1/(\mathsf p\mathsf p')}\big(\E\epsilon_1^{q\mathsf p^2}\big)^{1/\mathsf p^2}\le C\Delta_l^{q\beta/2}.
\]

\emph{Particles.} The fresh and the transported particles are moved by the same maps, with $U_p$ independent of the other variables. By \eqref{S1}, $|X^{[l]}_p-X^{[l-1]}_p|\le\ell(U_p)\big(e+(1+|X^{[l]}_{p-1}|)(|\theta^{[l]}_p-\theta^{[l-1]}_p|+\Delta_l^{\beta/2})\big)$ with $e:=|X^{[l]}_{p-1}-X^{[l-1]}_{p-1}|$, and the same holds for the fresh particles, with $e_i$ in place of $e$. The map from the natural parameters to $(r,S)$ is Lipschitz on $\Theta$. H\"older's inequality with the same exponents, the bound on the fits and $(\mathrm I_{p-1})$ give $(\mathrm I_p)$.

Each step of the induction raises the order by the factor $\mathsf p^3$, so that the orders of the gaps stay below $4\mathsf p^{3n}\le4\mathrm e$, and $(\mathrm I_n)$ gives \eqref{S2}, which uses $q=4$. The orders of the moments of $\mathcal Y'$, and hence those used in \eqref{G2}, are then at most $c\,n$ for an absolute constant $c$, so that Lemma~\ref{lem:G3} applies at every step as soon as $c\,n\le t_0N_{\mathrm f}$. No constant depends on $l$, $N$ or $\bar k$. This ends the proof of the proposition.
\end{proof}

\section{Additional Numerical Results}\label{app:numerics}

\subsection{Coupling the fits}\label{app:num-fits}

\begin{figure}[ht]
\centering
\includegraphics[width=0.72\textwidth]{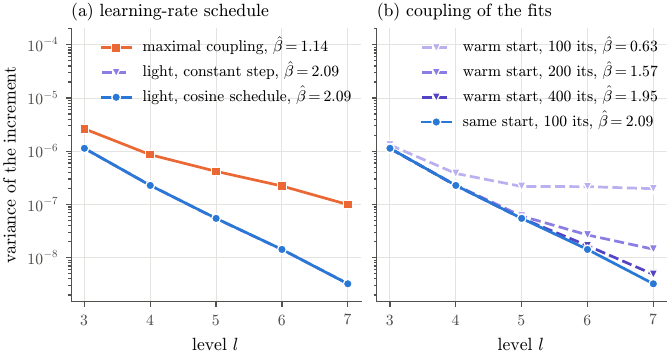}
\caption{Effect of the fit on the variance of the light increments, $d_x=4$, fixed $A_n$. (a) Fits started from the same point, $100$ iterations, with a constant learning rate and with a cosine schedule; the two light curves overlap. (b) Cosine schedule: fine fit warm-started from the coarse fit with $100$, $200$ and $400$ iterations, and both fits started from the same point with $100$ iterations.}\label{fig:ablation}
\end{figure}

The rate depends on how the two fits are coupled (Figure~\ref{fig:ablation}, $d_x=4$, fixed $A_n$, Section~\ref{sec:num-e4}). When the fine fit is warm-started from the converged coarse fit, the remaining optimization error is not common to the two levels, and the rate $\hat\beta\approx2$ requires about $400$ iterations. When both fits start from the same point with the same fitting draws, the two optimization paths differ only through the level-dependent inputs, by a quantity of order $\Delta_l$ even before convergence. In this case, $100$ iterations suffice ($\hat\beta=2.09$, $4.3$ seconds per pair of filters, against $17$ seconds for warm starts with $400$ iterations). With the same start, a constant learning rate gives the same rate as the cosine schedule.

\subsection{\texorpdfstring{Fits by projected gradient descent on a fresh cloud}{Fits by projected gradient descent on a fresh cloud}}\label{app:num-fg}

We rerun the test of Section~\ref{sec:num-e4} with $d_x=4$ and rotating $A_n$, with the light coupling fitted by procedure \eqref{FG} of Section~\ref{sec:theory-coupling} instead of Adam: a fresh cloud of $N_{\mathrm f}=10^3$ draws, $M=8$ fitting draws per particle, $2000$ steps of projected gradient descent with the step $1/\mathsf L^{[l]}$ of Appendix~\ref{app:proof-S2}, and $\Theta$ with $R=3$, $s_-=0.1$ and $s_+=1.6$, so that $s_+\lambda_F<1$. The data, the levels, the $N=10^3$ transported particles and the $50$ runs per level are those of Figure~\ref{fig:rates}(c). Table~\ref{tab:fg} gives the variance of the increment, the gap $|\vartheta^{[l]}-\vartheta^{[l-1]}|$ between the fits in the natural parameters, and the bias against the level-$l$ Kalman filter, averaged over $n=1,\dots,10$. The variance decreases at the rate of the Adam fits, and the gap like $\Delta_l$; its fourth moment, which \eqref{S2} bounds, decreases like $\Delta_l^{4.01}$, and like $\Delta_l^{3.9}$ to $\Delta_l^{4.2}$ at each time step. The bias is of the same size as with Adam, and not distinguishable from Monte Carlo error. No fit ends on the boundary of $\Theta$, and the norm of the last gradient is at most $3.4\times10^{-3}$, so that the constraints do not shape these results.

\begin{table}[htbp]
\centering
\small
\begin{tabular}{rcccc}
\toprule
$l$ & variance, \eqref{FG} & variance, Adam & mean fit gap, \eqref{FG} & bias, \eqref{FG} (standard error)\\
\midrule
$3$ & $1.20\times10^{-6}$ & $1.23\times10^{-6}$ & $9.0\times10^{-3}$ & $2.4\times10^{-3}$ ($3.2\times10^{-3}$)\\
$4$ & $2.99\times10^{-7}$ & $3.25\times10^{-7}$ & $4.4\times10^{-3}$ & $1.9\times10^{-3}$ ($3.3\times10^{-3}$)\\
$5$ & $7.68\times10^{-8}$ & $7.31\times10^{-8}$ & $2.2\times10^{-3}$ & $2.0\times10^{-3}$ ($3.2\times10^{-3}$)\\
$6$ & $1.66\times10^{-8}$ & $3.05\times10^{-8}$ & $1.1\times10^{-3}$ & $3.8\times10^{-3}$ ($3.2\times10^{-3}$)\\
$7$ & $4.17\times10^{-9}$ & $4.77\times10^{-9}$ & $5.5\times10^{-4}$ & $2.9\times10^{-3}$ ($2.9\times10^{-3}$)\\
\midrule
rate & $2.05\pm0.04$ & $1.94\pm0.07$ & $1.01$ & \\
\bottomrule
\end{tabular}
\caption{The test of Section~\ref{sec:num-e4} ($d_x=4$, rotating $A_n$) with fits by procedure \eqref{FG} and with Adam; $50$ runs per level, rates fitted on levels $3$ to $7$, with bootstrap standard errors over runs.}
\label{tab:fg}
\end{table}

\subsection{Dependence on the number of particles}\label{app:num-N}

The multilevel calculation of Section~\ref{sec:num-cost} assumes that the variances scale like $N^{-1}$, which is not guaranteed for the light coupling (Section~\ref{sec:theory-coupling}). A check with $N\in\{250,10^3,4\cdot10^3\}$ ($20$ runs) gives light increment variances decreasing like $N^{-0.94}$ at $l=7$ but only like $N^{-0.7}$ at $l=5$. The maximal coupling scales like $N^{-1}$ within the Monte Carlo error, and the rate in $\Delta_l$ of the light coupling is similar at the three values of $N$ ($1.9$ to $2.3$ on levels $5$ to $7$).

\subsection{Bias check}\label{sec:num-bias}

Write $b_l := \E\,\pi_n^{[l],N}(\varphi) - \pi_n^{[l]}(\varphi)$ for the bias of the single-level light filter at level $l$. The hybrid estimator with switch level $s$ has bias $b_L-b_{s-1}$ relative to $\pi_n^{[L]}(\varphi)$, besides the $O(N_l^{-1})$ terms common to particle methods, and $b_L$ with a light filter also at the base level. We run $100$ single-level filters with $N=10^3$ at levels $l\in\{3,5,7\}$ and compare their mean with the level-$l$ Kalman filter at each $n$ through $t_n$, the difference divided by its standard error. Without bias, the mean of $t_n^2$ over the ten times is close to $1$. The value $1.83$, the $95\%$ quantile of $\chi^2_{10}/10$, is only an indicative threshold, since the $t_n$ come from the same runs and are correlated over $n$.

\begin{table}[ht]
\centering
\begin{tabular}{lcccccc}
\toprule
 & \multicolumn{2}{c}{$l=3$} & \multicolumn{2}{c}{$l=5$} & \multicolumn{2}{c}{$l=7$}\\
\cmidrule(lr){2-3}\cmidrule(lr){4-5}\cmidrule(lr){6-7}
method & bias & $t_n^2$ & bias & $t_n^2$ & bias & $t_n^2$\\
\midrule
particle filter & $2.0\times10^{-3}$ & 0.52 & $2.7\times10^{-3}$ & 1.31 & $3.5\times10^{-3}$ & 1.59\\
light, $100$ iterations, $M=8$ & $2.1\times10^{-3}$ & 0.71 & $1.7\times10^{-3}$ & 0.70 & $2.1\times10^{-3}$ & 0.83\\
light, $20$ iterations, $M=1$ & $6.7\times10^{-3}$ & 2.79 & $8.1\times10^{-3}$ & 3.80 & $8.6\times10^{-3}$ & 4.05\\
\bottomrule
\end{tabular}
\vspace{0.5em}
\caption{Root-mean-square bias over $n$ and mean of $t_n^2$ for single-level filters, $N=10^3$, $100$ runs, $d_x=4$, rotating $A_n$.}\label{tab:bias}
\end{table}

With $100$ iterations and $M=8$, no bias is detected (Table~\ref{tab:bias}). With $20$ iterations and $M=1$, the coupling still gives $\hat\beta=2.16$, but the filter is biased by $7$ to $9\times10^{-3}$ at every level. The resolution of the check, about $2\times10^{-3}$, is well above the tolerances of Table~\ref{tab:cost}, so the experiment does not establish that $b_L-b_{s-1}$ is small at those tolerances.

\subsection{Spheres}\label{app:num-sphere}

At a single level, for $X_n = F(X_{n-1})+\tau\,\nu_n$ in $\R^3$ with a nonlinear $F$ (a rotation and a componentwise sine) observed through $Y_n=|X_n|^2$, the light filter with $N=10^3$ matches the optimal-proposal particle filter for $\tau=1$. With the best of $K\in\{1,3,10\}$, however, its error is about $3$ times larger at $\tau=0.5$ and $12$ times larger at $\tau=0.25$. A mixture of linear tilts $e^{\langle b_k,u\rangle}$ with fixed $b_k$ cannot rescale the concentration $|\kappa(x')|$ uniformly over source points, unlike the Gaussian family on affine blocks, and Bingham-type potentials have no explicit normalizer.

With the moving center and $K=3$, the increments are heavy-tailed: four runs at $l=6$ and three at $l=7$ carry increments of more than $10$ and up to about $150$ times the scaled median absolute deviation, all at the steps $n=3$ and $n=4$. Without these runs, the variance at $l=7$ is $3.8\times10^{-9}$, close to the $3.1\times10^{-9}$ of $K=1$. The jumps do not come from the cut of the azimuth grid, which is placed opposite the mean direction as in Appendix~\ref{app:sampler}: a fixed cut gives the same rate and the same jumps. Replaying three runs with jumps and three without, with the same random numbers, locates them at the step $n=4$, where both fits put a weight of about $0.99$ on one component. In the runs with jumps, the squared distance between the parameters of the fine and coarse fits at that step is $0.1$ to $9$, against $10^{-4}$ to $2\times10^{-3}$ in the other runs and $10^{-5}$ to $10^{-3}$ at the other steps. The distance is not reduced by relabeling the components, so the fits do not swap components: they move the parameters of the components with almost no weight, which the loss barely constrains. Moving the parameters a small fraction of the way between the two fits moves the particles proportionally, so the sampler itself stays smooth.

\end{document}